\documentclass[11pt,english]{article}
\input epsf.tex
\usepackage{amssymb}
\usepackage{pstcol}

\usepackage{babel}
\makeatletter
\@addtoreset{equation}{section}
\makeatother

\def\dalemb#1#2{{\vbox{\hrule height .#2pt
        \hbox{\vrule width.#2pt height#1pt \kern#1pt
                \vrule width.#2pt}
        \hrule height.#2pt}}}

\def\co{{\cal O}}

\def\0{{\sst{(0)}}}
\def\1{{\sst{(1)}}}
\def\2{{\sst{(2)}}}
\def\3{{\sst{(3)}}}
\def\4{{\sst{(4)}}}
\def\5{{\sst{(5)}}}
\def\6{{\sst{(6)}}}
\def\7{{\sst{(7)}}}
\def\8{{\sst{(8)}}}
\def\n{{\sst{(n)}}}

\def\ep{\epsilon}
\def\td{\tilde}

\def\half{{\textstyle{1\over2}}}

\def\qu{{\textstyle{1\over 4}}}

\let\a=\alpha \let\b=\beta \let\g=\gamma \let\d=\delta \let\e=\epsilon
  \let\q=\theta  
\let\l=\lambda \let\m=\mu \let\n=\nu  \let\r=\rho
\let\s=\sigma \let\t=\tau  \let\f=\phi \let\c=\chi 
  \let\D=\Delta  \let\L=\Lambda
   \let\F=\Phi 
 \let\W=\Omega   \let\G=\Gamma
\let\la=\label  
  
\def\nn{\nonumber} \def\bd{\begin{document}} \def\ed{\end{document}}
\def\ds{\documentstyle} \let\fr=\frac \let\bl=\bigl \let\br=\bigr
\let\Br=\Bigr \let\Bl=\Bigl
\let\bm=\bibitem
\let\na=\nabla
\let\pa=\partial \let\ov=\overline
\newcommand{\be}{\begin{equation}}
\newcommand{\ee}{\end{equation}}
\def\ba{\begin{array}}
\def\ea{\end{array}}
\def\ft#1#2{{\textstyle{{\scriptstyle #1}\over {\scriptstyle #2}}}}
\def\fft#1#2{{#1 \over #2}}
\def\del{\partial}
\def\sst#1{{\scriptscriptstyle #1}}
 \def\oneone{\rlap 1\mkern4mu{\rm l}}
\def\ie{{\it i.e.\ }}
\def\via{{\it via}}
\def\semi{{\ltimes}}
\def\str{{\rm str}}
\def\Dm{{{D_{\sst{max}}}}}
\def\vac{ \left | 0 \right \rangle }
\def\kvac{ \left | k \right \rangle }

\def\sp{\; \; \;}

\def\bol{ \left | B (p^+) \right \rangle}
\def\bo1{ \left | B^0 (p^+) \right \rangle}

\def\bolt{ \left | B (p^+) \right \rangle_{\t}}

\def\boxl{ \left | B (x^-) \right \rangle}

\def\<{ \langle }
\def\>{ \rangle }

\def\vf{\varphi}

\def\ls{{(l,0)}}
\def\lv{{(l,\pm1)}}
\def\lt{{(l,\pm2)}}

\def\lse#1{{(l_{#1},0)}}
\def\lve#1{{(l_{#1},\pm1)}}
\def\lte#1{{(l_{#1},\pm2)}}

\def\lsg#1{{5(l_{#1},0)}}
\def\lvg#1{{5(l_{#1},\pm1)}}
\def\ltg#1{{5(l_{#1},\pm2)}}

\def\lsi#1{{5{(#1,0)}}}
\def\lvi#1{{5{(#1,\pm1)}}}
\def\lti#1{{5{(#1,\pm2)}}}

\def\lsr#1{{1{(#1,0)}}}
\def\lvr#1{{1{(#1,\pm1)}}}
\def\ltr#1{{1{(#1,\pm2)}}}

\def\cn{{\cal N}}
\def\cao{{\cal O}}
\def\cD{{\cal D}}
\def\cE{{\cal E}}
\def\cF{{\cal F}}
\def\cG{{\cal G}}
\def\cH{{\cal H}}
\def\cK{{\cal K}}
\def\cO{{\cal O}}
\def\cP{{\cal P}}
\def\cQ{{\cal Q}}
\def\cR{{\cal R}}
\def\cS{{\cal S}}
\def\cT{{\cal T}}
\def\cU{{\cal U}}
\def\cV{{\cal V}}
\def\cW{{\cal W}}
\newcommand{\nono}{\nonumber}
\newcommand{\eqref}[1]{(\ref{#1})}
\newcommand{\dtilde}[1]{\tilde{\tilde{#1}}}
\newcommand{\hatb}[1]{\hat{\ov{#1}}}
\newcommand{\hatt}[1]{\hat{\tilde{#1}}}
\newcommand{\emnr}{{e_\m}^{\n\r}}

\newcommand{\hsp}{\hspace{0.5cm}}

\newcommand{\ho}[1]{$\, ^{#1}$}
\newcommand{\hoch}[1]{$\, ^{#1}$}
\newcommand{\bea}{\begin{eqnarray}}
\newcommand{\eea}{\end{eqnarray}}
\newcommand{\ra}{\rightarrow}
\newcommand{\lra}{\longrightarrow}
\newcommand{\Lra}{\Leftrightarrow}
\newcommand{\ap}{\alpha^\prime}
\newcommand{\bp}{\tilde \beta^\prime}
\newcommand{\tr}{{\rm tr} }
\newcommand{\Tr}{{\rm Tr} }
\newcommand{\NP}{Nucl. Phys. }

\newcommand{\ams}{{\it Institute for Theoretical Physics,
University of Amsterdam, \\
Valckenierstraat 65, 1018XE Amsterdam, The Netherlands} \\
{\tt skenderi, taylor@science.uva.nl}}
\newcommand{\auth}{Kostas Skenderis and Marika Taylor}

\def\red{\color{red}}

\vspace{10pt}

\begin{document}
\begin{flushright}
\hfill{ITFA-2007-17}
\end{flushright}

\vspace{25pt}

\begin{center}

{\Large \bf Anatomy of bubbling solutions}

\vspace{20pt}

\auth

\vspace{15pt}

\vspace{8pt}

{\ams}

\vspace{20pt}

\underline{ABSTRACT}
\end{center}

We present a comprehensive analysis of holography for
the bubbling solutions of Lin-Lunin-Maldacena.
These solutions are uniquely determined by a
coloring of a 2-plane, which was argued to correspond
to the phase space of free fermions. We show that
in general this phase space distribution does not determine
fully the 1/2 BPS state of $N=4$ SYM that the gravitational
solution is dual to, but it does determine it enough so that vevs
of all single trace 1/2 BPS operators in that state
are uniquely determined to leading order in the large $N$ limit.
These are precisely the vevs encoded in the asymptotics
of the LLM solutions. We extract these vevs for operators
up to dimension 4 using holographic
renormalization and KK holography and show exact agreement
with the field theory expressions.

\pagebreak

{\small \setcounter{tocdepth}{2}
\tableofcontents }
\pagebreak

\section{Introduction}

Supersymmetric supergravity solutions play an important role in
developing our understanding of holographic dualities. Such solutions
provide valuable examples where one can carry out detailed
computations and, using non-renormalization properties, make
quantitative tests of gravity/gauge dualities at and away from conformal
fixed points. Given the precise holographic dictionary available in
such cases, one may also understand in detail how the
spacetime is reconstructed from gauge theory data. One would hope to
take from these examples generally applicable methods and principles,
along with insight into the inner workings of holography.

Supergravity solutions that asymptote to $AdS_5 \times S^5$  describe
either a deformation of $N=4$ SYM or the theory in a non-trivial
state. The most supersymmetric non-trivial vacua of $N=4$ SYM theory
preserve 16 supersymmetries. In this context it is interesting to consider
the SYM theory on both Minkowski spacetime
$R^{(1,3)}$, in which case we have $N=4$ on
the Coulomb branch, and the theory on $R \times S^3$. These
two cases are equivalent in the conformal
vacuum since the two backgrounds are mapped to each other by a Weyl
transformation but differ on a generic half supersymmetric state (which
spontaneously breaks the conformal invariance of the $N=4$ SYM theory).
Of course, the two theories are still
related in the decompactification limit of $S^3$.

Using standard D-brane  physics, one expects that the holographic dual of
$N=4$ SYM on the
Coulomb branch is the near-horizon limit of multi-center D3 brane
solutions \cite{Kraus:1998hv}. The Coulomb branch may be parametrized
by the vevs of chiral primary operators and the gravity/gauge
theory duality together with non-renormalization theorems
imply that the vevs computed at weak coupling are non-renormalized and
must therefore also be reproduced by the
holographic computation. In \cite{Skenderis:2006di},  building on
 \cite{Klebanov:1999tb,Skenderis:2006uy},
we indeed succeeded in
extracting these vevs from a generic multi-center solution,
showing exact agreement with field theory. This provides
a highly non-trivial test of the correspondence away from the conformal point
-- an infinite number of vevs was quantitatively matched --
and also illustrates the maturity of holographic methods as
it shows that one can go beyond qualitative matching,
performing precise quantitative computations.

Supergravity solutions corresponding to $N=4$ SYM on $R \times S^3$
were recently constructed in \cite{Lin:2004nb}. The solutions of
\cite{Lin:2004nb}
preserve an $R \times SO(4) \times SO(4)$ bosonic symmetry and
16 supersymmetries. These ``bubbling solutions''
are uniquely determined by a coloring of the 2-plane into black and white
regions. Based on earlier work relating 1/2 BPS states to
free fermions \cite{Corley:2001zk,Berenstein:2004kk},
this distribution was argued to map to the phase space distribution
of free fermions and supporting evidence was provided in
\cite{Mandal:2005wv,Grant:2005qc,Takayama:2005yq, Balasubramanian:2005mg};
see also \cite{Ghodsi:2005ks,Donos:2005vm,Berenstein:2005aa,Silva:2005fa,
Yamada:2005um,Yoneya:2005si}.

It is often stated in the literature that the gauge theory dual
of the bubbling solution is the matrix model associated with
free fermions. One of the aims of this work is to understand
to which extent this assertion is valid by applying standard
AdS/CFT methods. That is, we will address the question of whether the
matrix model captures the entire vacuum structure.
In the AdS/CFT correspondence the
asymptotics of the supergravity solution encode QFT data.
In particular the vacuum structure of the dual QFT
can be extracted from the near boundary asymptotics of the solution.
In the first part of this
paper we will use holographic renormalization \cite{Skenderis:2002wp}
and KK holography \cite{Skenderis:2006uy}
in order to extract the vevs from the solutions of
\cite{Lin:2004nb}. Any proposal for the field theory dual
must reproduce these results\footnote{Here we assume that
the 16 supercharges protect the vevs from acquiring
quantum corrections, as in the case of $N=4$ SYM on $R^{(1,3)}$.}.

Let us now discuss the QFT side.
By the operator-state correspondence, all 1/2 BPS states
of $N=4$ SYM can be obtained by acting with 1/2 BPS operators
on the conformal vacuum. The operators constructed from
the 6 scalars $X^m$ %,\ m{=}1,{\ldots}, 6,$
of $N=4$ SYM lie in the
$(0,l,0)$ representation
of the $SU(4)$ R-symmetry (see, for example, the review
\cite{D'Hoker:2002aw}). Up to a $U(3) \subset SU(4)$ rotation,
every such operator can be represented holomorphically
using a single complex combination of the scalars $Z=X^1+i X^2$.
Thus gauge invariant operators built from
these scalars preserve an $SO(4)$ part
of the $SU(4)$ R-symmetry and have a definite $SO(2)$ charge $j$
under rotations in the $X^1{-}X^2$ plane. A convenient basis for these
operators
is the Schur polynomial basis \cite{Corley:2001zk}
and an arbitrary half BPS state $|\Phi \rangle$
preserving $SO(4)$ R symmetry can be written as a superposition of states
\be \label{state}
| \Phi \rangle = \sum_{R} a_{R} \chi_{R}(Z) | \Omega \rangle
\equiv {\cal O}_\F | \Omega \rangle
\ee
for suitable complex coefficients $a_R$, where $\chi_{R}(Z)$ is the Schur
polynomial associated with the\footnote{We are really interested
in $SU(N)$ gauge theory but the difference between $U(N)$ and $SU(N)$
is subleading in the large $N$ limit.}  $U(N)$ representation $R$
and $| \Omega \rangle$ is the conformal vacuum. The representation
$R$ may be labeled by a Young tableau and the associated
Schur polynomial $\chi_{R}(Z)$ has degree equal to the number of
boxes $n$ and in general involves both single and multi-trace contributions.
Thus the operator $\co_\Phi$ is equal to a sum of terms
each of which has dimension equal to charge, $\D=j=n$, for any $n>1$.
It follows that in order to specify the theory we need to supply the
coefficients $a_R$ and any gravitational dual should
encode these coefficients.

When the field theory is formulated on $R \times S^3$ one may
reduce over the $S^3$ to obtain a one-dimensional
model involving an infinite number of fields (KK modes).
Given the large amount
of supersymmetry, however, one might anticipate that the vacuum
structure, i.e. the coefficients $a_R$ in (\ref{state}),
may be encoded, at least to leading order in the
large $N$ limit, in the truncation of the $S^3$ reduction
to only the $s$-mode of the complex scalar $Z$.
We will take this as a working assumption in this paper.
It would be
interesting to investigate the validity of this assertion
in general.
We should emphasize that one should be very cautious about
using properties of this matrix model which are subleading in $N$
(to infer properties of the dual spacetimes etc.) since these
are likely to be different from the true $1/N$ corrections
of $N=4$ SYM on $R \times S^3$.

Standard arguments map the matrix model to free
fermions, whose phase space distribution is meant to correspond
to the coloring of the 2-plane that determines the LLM solution.
In the fermion picture, we have $N$ free fermions with
the ground state being the completely filled Dirac sea.
This state corresponds to $AdS_5 \times S^5$. Excited states
are in one to one correspondence with the Schur polynomials,
the excitation numbers being directly determined from the
the length of the rows of the associated
Young tableaux. A generic excited state
in the free fermion picture is then in direct correspondence
with the 1/2 BPS state (\ref{state}) of $N=4$ SYM.
The crucial question is then: does a phase space distribution
for the fermions uniquely determine the state $| \F \rangle$?
We show that this is {\em not} the case, i.e.
the phase space distribution does not determine
all coefficients $a_R$, but it does
determine the state enough so that the vevs of {\em all} (single trace)
{\it chiral primaries} in this state are uniquely determined!
These are precisely the vevs encoded in the asymptotics
of the supergravity solution.

The results of the holographic computation show that the LLM
solutions (generically) encode vevs of {\it all} $SO(4)$ singlet
operators; such operators can be labeled by their $SO(2)$ charge $j$.
So to check the correspondence one should compute these
vevs in the field theory. Here we face the first
obstacle. While maximally charged operators (those whose $SO(2)$
charge is equal in magnitude to the dimension) involve
only the $Z$ field and so can be implemented
in the matrix model \cite{Takayama:2005yq}, all other operators involve all
six scalars and thus appear to involve fields not included in the
matrix model.

To see how one can deal with this issue, recall that the 1-point function
of an operator ${\cal O}$
in the state $|\Phi\rangle$ is equivalent to the 3-point function
in the conformal vacuum  between ${\cal O}$, the operator that
creates $|\Phi\rangle$ from $|\Omega\rangle$ and its conjugate,
\be \label{1to3}
\langle \co \rangle_\F = \langle \Omega | \co_\F^\dagger \co \co_\F
|\Omega \rangle.
\ee
Suppose this correlator is computed in free field theory. Since $\co_\F$
is constructed only from $Z$ the 3-point function
receives contributions only from part of $\co$ that contains $Z$.
Thus for the free field computation of the 1-point functions
in (\ref{1to3}) one may set to zero
all fields but $Z$ (and $\bar{Z}$) in the chiral primary operators.
We emphasize, however, that this truncation would in general give incorrect
answers if used for different computations, e.g. two point functions.
If the free field computation were to be renormalized, then fields
apart from $Z$ would of course contribute in loops.
However, three-point functions of single trace chiral primary operators
of $N=4$ SYM are known not to renormalize \cite{LMRS} and it is
believed that three point functions of protected multi-trace operators
are similarly non-renormalized \cite{Skiba}. We indeed find that the vevs
computed using free field results for multi trace operators do agree
with those extracted holographically, thus confirming the expectation
of non-renormalization.

The truncated operators can therefore be implemented in the
matrix model. We do this explicitly for  all
operators up to dimension four (whose vevs we also extract from the gravity
solutions). In particular, we show that each of these operators
can be expressed as linear combinations of
bilinears of fermion creation and annihilation
operators. The coefficients in the linear combinations
are fixed by demanding that the operators
have zero expectation values at the conformal vacuum, 3-point functions
with single trace operators are correctly reproduced,
and the vev of the operators
have the correct limit in the decompactification limit of the $S^3$.
In this limit the phase space distribution maps to the distribution
of eigenvalues of the scalars in the Coulomb branch of $N=4$ SYM.
Having implemented the operators in the matrix model it is then
straightforward to compute their vevs in a general state
$|\F\rangle$ and we find exact agreement with the holographic
computations!

\bigskip

This paper is organized as follows. In the next section we
summarize how to extract holographic data from asymptotically
$AdS_5 \times S^5$ solutions. The resulting
expressions for the holographic vevs
in terms of supergravity field asymptotics are applicable not just to
the bubbling solutions of interest in this paper, but to more general
$1/4$ and $1/8$ BPS bubbling solutions.
In section \ref{bub_sec} we review the LLM solutions and
extract the vevs of all maximally charged operators and
of all operators with any charge up to dimension four.
In section \ref{dual_qft} we discuss the dual description of the
bubbling solutions. We show what information about the state is
captured by the distribution and hence the
gravity solution; we reproduce the holographic vevs in section 5
and we explicitly match certain specific
symmetric distributions with 1/2 BPS states in section 6.
In section \ref{disc} we discuss our results.

Appendices A and B
review relevant properties of spherical harmonics and of scalar chiral
primary operators in ${\cal N} = 4$ SYM whilst appendix C discusses
the large N scaling of three point functions. Appendix D is rather tangential
to the focus of the paper: we discuss the Killing spinors for the
LLM supergravity solutions. These were discussed in \cite{Lin:2004nb}
but only half of them were correctly identified and they are
missing local phase factors which (drop out of the fermion bilinears
used to construct the supergravity solution but which) are needed to
solve the Killing spinor equations.

\section{Extracting holographic data}

In this section we will give a self-contained summary of the method of
Kaluza-Klein holography, developed in \cite{Skenderis:2006uy}, which
allows the computation of all 1-point functions from any
asymptotically $AdS _p\times X_q$ supergravity solution.

The basic steps in this method are the following. First one expresses
the deviation of the supergravity solution from $AdS _p\times X_q$ in
terms of the complete basis of harmonics of the compact manifold
$X_q$; let the expansion coefficients be denoted collectively as
$\psi^{\cal I}$. Now one forms gauge invariant combinations of these
fluctuations, $\hat{\psi}^{\cal I}$, that satisfy field equations which can be
expanded perturbatively in the number of fluctuation
fields. Schematically these field equations may be written
\be
{\cal L}_{\cal I} \hat{\psi}^{\cal I} = {\cal L}_{ {\cal I  J K} }
\hat{\psi}^{\cal J} \hat{\psi}^{\cal K}
+ {\cal L}_{ {\cal I  J K L}}
\hat{\psi}^{\cal J} \hat{\psi}^{\cal K} \hat{\psi}^{\cal L} + \cdots,
\ee
where $ {\cal L}_{{\cal I}_1 \cdots {\cal I}_n}$ is an appropriate differential
operator. Since $ {\cal L}_{{\cal I}_1 \cdots {\cal I}_n}$ involves
derivatives, the set of field equations cannot generically be
integrated into an action. However, one can always define $p$-dimensional
fields $\Psi^{\cal I}$ by a non-linear Kaluza-Klein reduction map of
the fields $\psi^{\cal I}$:
\be
\Psi^{\cal I} = \psi^{\cal I} + {\cal K}^{I}_{ {\cal J K } }
\psi^{\cal J} \psi^{\cal K} + \cdots,
\ee
where ${\cal K}^I_{\cal J K}$ contains appropriate derivatives. The
reduction map is such that the fields $\Psi^{\cal I}$
do satisfy field equations which can be integrated into an action.
Given this $p$-dimensional action, it is then straightforward to obtain the
one point functions of operators in terms of the asymptotics of the
fields $\Psi^{\cal I}$, using the well-developed techniques of
holographic renormalization
\cite{Henningson:1998gx,Balasubramanian:1999re,deHaro:2000xn,Skenderis:2000in,BFS1,BFS2,PS1,PS2};
for a review, see \cite{Skenderis:2002wp}.
We will now give the details of each step in the case of interest.

\bigskip

Let us consider any asymptotically $AdS_5 \times S^5$ solution of type
IIB, which involves only the metric and 5-form field
strength. (It is straightforward to include all other fields of type
IIB, but unnecessary for our application here to the LLM bubbling
solutions.) The IIB SUGRA field equations\footnote{The field strength differs by a
factor of 4 from the conventions in \cite{Polchinski:1998rr}.}
for the metric and 5-form field strength are given by:
\be
R_{MN} = \frac{1}{6} F_{MPQRS} F_N{}^{PQRS}, \qquad
F=*F.
\ee
These equations admit an $AdS_5 \times S^5$ solution
\bea
ds_o^2 &=& \frac{dz^2}{z^2} + \frac{1}{z^2} d x_{||}^2
+ d\q^2 + \sin^2 \q d \W_3^2 + \cos^2 \q d \f^2 \\
&&F^{o}_{\m \n \r \s \t} = \e_{\m \n \r \s \t}, \qquad
F^{o}_{abcde} = \e_{abcde}. \nn
\eea
where $(\m,\n)$ and $(a,b)$ denote $AdS_5$ and $S^5$ indices
respectively; $M,N,...$ are $10d$ indices whilst
$x$ denotes AdS coordinates and $y$ denotes $S^5$ coordinates.
We will consider here solutions that are deformations of
$AdS_5 \times S^5$ such that
\bea
g_{MN} &=& g^o_{MN} + h_{MN}, \\
F_{MNPQR} &=&  F^o_{MNPQR} + f_{MNPQR}.\nn
\eea
These fluctuations can be expanded in spherical harmonics as:
\bea \label{fluct_h}
h_{\m \n}(x,y) &=& \sum {h}^{I_1}_{\m \n}(x) Y^{I_1}(y); \nonumber \\
h_{\m a} (x,y)&=&
\sum ({B}^{I_5}_{(v)\m}(x) Y_a^{I_5}(y)
+ {B}^{I_1}_{(s)\mu}(x) D_a Y^{I_1}(y)); \nonumber \\
h_{(ab)}(x,y)
&=& \sum (\hat{\phi}_{(t)}^{I_{14}}(x) Y_{(ab)}^{I_{14}}(y)
+ \phi^{I_5}_{(v)}(x) D_{(a} Y^{I_5}_{b)}(y)
+ \phi^{I_1}_{(s)}(x) D_{(a} D_{b)} Y^{I_1}(y) ); \nonumber \\
h_{a}^a(x,y) &=& \sum {\pi}^{I_1}(x) Y^{I_1}(y),
\eea
and
\bea
f_{\m \n \r \s \t}(x,y) &=& \sum
5 D_{[\m} b^{I_1}_{\n \r \s \t]}(x) Y^{I_1}(y); \\
f_{a \m \n \r \s}(x,y) &=& \sum (b^{I_1}_{\m \n \r \s}(x) D_a Y^{I_1}(y)
+ 4 D_{[\m} b^{I_5}_{\n \r \s]}(x) Y_a^{I_5}(y)); \nonumber \\
f_{ab \m \n \r}(x,y) &=& \sum
(3 D_{[\m} b^{I_{10}}_{\n \r]}(x) Y_{[ab]}^{I_{10}}(y)
-2 b^{I_5}_{\m \n \r}(x) D_{[a} Y_{b]}^{I_5}(y)); \nonumber \\
f_{a b c \m \n}(x,y) &=& \sum
(2 D_{[\m} b_{\n]}^{I_5}(x) \e_{abc}{}^{de} D_d Y_e^{I_5}(y)
+ 3 b_{\m \n}^{I_{10}}(x) D_{[a} Y_{bc]}^{I_{10}}(y)); \nonumber \\
f_{abcd\m}(x,y)  &=& \sum
(D_\m b_{(s)}^{I_1}(x) \e_{abcd}{}^e D_e Y^{I_1}(y)
+ (\L^{I_5}-4) b_\m^{I_5}(x) \e_{abcd}{}^e Y_e^{I_5}(y)) \nn \\
f_{a b c d e}(x,y) &=& \sum b_{(s)}^{I_1}(x) \L^{I_1} \e_{abcde} Y^{I_1}(y); \nn
\eea
Numerical constants in these expressions are inserted so
as to match with the conventions of \cite{Kim:1985ez}.
Parentheses denote a symmetric traceless combination
(i.e. $A_{(ab)} = 1/2 (A_{ab}+A_{ba}) -1/5 g_{ab} A_a^a$).
$Y^{I_1}, Y_a^{I_5}, Y_{(ab)}^{I_{14}}$ and $Y_{[ab]}^{I_{10}}$ denote scalar,
vector and tensor harmonics whilst $\L^{I_1}$ and $\L^{I_5}$ are the
eigenvalues of the scalar and vector harmonics under (minus) the d'Alembertian.
The subscripts $t$, $v$ and $s$ denote whether the field is associated with
tensor, vector or scalar harmonics respectively, whilst the
superscript of the harmonic label $I_n$ derives from the number of
components $n$ of the harmonic. Relevant properties of the spherical
harmonics are summarized in appendix \ref{sph}.

In what follows it will be useful to label perturbations by both the degree $k$ of the
associated harmonic and by the degeneracy of such harmonics. For
example, $\pi^{kI}$ will denote the fluctuations associated with degree
$k$ scalar harmonics with $I$ labeling the $SO(6)$ quantum numbers.

\subsection{Gauge invariant quantities}

When computing the spectrum it is useful to impose the
de Donder-Lorentz gauge choice, as in \cite{Kim:1985ez}, which imposes
the following conditions on the metric fluctuations
\be
D^a h_{(ab)} = D^a h_{a \mu} = 0,
\ee
along with analogous conditions on the five-form fluctuations. These
gauge conditions remove terms involving gradients of spherical
harmonics.

As discussed in \cite{Skenderis:2006uy}, it is often the case that the
natural choice of coordinates for the asymptotic expansion takes the
fluctuations outside the de Donder gauge. Indeed, we will find here that
there is a distinguished coordinate choice which is outside de Donder
gauge. This issue may be dealt with using
gauge invariant combinations of the fluctuations; these
were derived up to quadratic order in the fluctuations in
\cite{Skenderis:2006uy}. For the purposes of this paper we will need
only certain combinations which are gauge invariant at linear order, namely:
\bea
\hat{\pi}^{kI_1} &=& {\pi}^{kI_1} - \L^{I_1} \f_{(s)}^{kI_1} \label{ginv}\\
\hat{B}^{kI_5}_{(v) \mu} & =& {B}^{kI_5}_{(v)\mu}
- \frac{1}{2} D_\mu \f^{kI_5}_{(v)}  \nn \\
%\hat{h}^{I_1}_{\m \n} &=& \tilde{h}_{\m \n}^{I_1}
%- D_{\m} \hat{B}^{I_1}_{(s)\n}
%-D_\n \hat{B}^{I_1}_{(s)\m},
%\qquad I_1 \neq 0. \nn \\
\hat{b}^{kI_1} &=& b^{kI_1}_{(s)} - \frac{1}{2} \f_{(s)}^{k I_1} \nn \\
%\hat{b}{}^{I_1}_{\m \n \r \s} &=& b_{\m \n \r \s}^{I_1}
%- \e_{\m \n \r \s}{}^\t \hat{B}_{(s)\t}^{I_1} \nn \\
\hat{b}{}^{kI_5}_{\m} &=& b_{\m}^{kI_5} - \frac{1}{2 (\L^{I_5}-4)} D_\m
\f_{(v)}^{k I_5}. \nn
\eea
Note also that ${h}^0_{\m \n}$ is
a deformation of the background metric and it indeed transforms
as a metric.

\subsection{The spectrum}

In this subsection we review the relevant parts of the spectrum of
fluctuations about $AdS_5 \times S^5$ computed in \cite{Kim:1985ez}.
As discussed in detail in \cite{Skenderis:2006uy},
one can relax the de Donder gauge fixing condition used in
\cite{Kim:1985ez} by replacing all fields by the gauge invariant
(hatted) versions given in the previous section.

The scalars relevant here satisfy the following linearized equations
\bea
\Box \hat{s}^{kI_1} &=& k (k-4) \hat{s}^{kI_1},  \quad k \geq 2, \nn \\
\Box \hat{t}^{kI_1} &=& (k+4) (k+8)\hat{t}^{kI_1}, \quad k \geq 0,
\eea
where we introduce the combinations
\be \la{l1}
\hat{s}^{kI_1} = \frac{1}{20 (k+2)} (\hat{\pi}^{kI_1}
- 10 (k+4) \hat{b}^{kI_1}), \qquad
\hat{t}^{kI_1} = \frac{1}{20 (k+2)} (\hat{\pi}^{kI_1} + 10 k \hat{b}^{kI_1}),
\ee
with inverse relations
$\hat{b}^{kI_1}=-\hat{s}^{kI_1}+\hat{t}^{kI_1},\
\hat{\pi}^{kI_1}=10k\hat{s}^{kI_1}+10(k+4)\hat{t}^{kI_1}$. The $s^I$
fields are dual to scalar chiral primary operators.

The relevant vector combinations are
\bea
a^{kI_5}_{\m} &=& (\hat{B}^{kI_5}_{(v)\m} - 4 (k + 3) \hat{b}^{kI_5}_{\mu}); \\
c^{kI_5}_{\m} &=& (\hat{B}^{kI_5}_{(v)\m} + 4 (k + 1) \hat{b}^{kI_5}_{\mu}), \nn
\eea
with the corresponding masses being
\be
m^2(a^k) = (k^2 -1); \hsp m^2 (c^k) = (k+3) (k+5).
\ee
Thus the $k=1$ modes of $a_{\mu}$ are massless and are dual to the R
symmetry currents.

The combination of $10d$ fields that satisfies the $5d$ linearized
Einstein equation is
\be \label{grav_lin}
\td{h}_{\m \n}^0 = ({h}_{\m \n}^0 + \frac{1}{3} g_{\m \n}^o {\pi}^0);
\ee
the shift by ${\pi}^0$ follows from the Weyl transformation required to
bring the $5d$ action into the Einstein frame.

\subsection{Kaluza-Klein reduction}

We are now ready to give the non-linear Kaluza-Klein map.
This map relates the coefficients appearing in
asymptotic expansion of the ten dimensional solution to the
coefficients in the asymptotic expansion of the five dimensional
solution obtained from the following action:
\be \label{5daction}
S = \frac{N^2}{2 \pi^2} \int d^5 x \sqrt{G}
\left(-\frac{1}{4} (R-2 \L)
+\frac{1}{2} G^{\mu \nu} \sum_{k=2}^4 \pa_\m S^{kI} \pa_\n S^{kI}
+\frac{1}{16} F_{\mu \nu} F^{\mu \nu}%+ \frac{1}{8} (k^2 -1) (A^{kI_5})^2
\right)
\ee
where $\L$ is the cosmological constant and
$F_{\mu \nu}= \pa_\m A^{1a}_\n-\pa_\n A^{1a}_\m$. In this expression
we have kept only the fields that are dual to operators
up to dimension 4, i.e. the metric $G_{\mu \nu}$ which is dual to the stress
energy tensor, the massless vector field $A_{\mu}^{1a}$
which is dual to the
R current and the scalars $S^{kI}$ dual which are dual
to chiral primaries of dimension $k$.
The reduction worked out in \cite{Skenderis:2006uy} included further
low lying KK modes. However only the terms listed in (\ref{5daction})
contribute to the formulae for the 1-point functions of
the operators up to dimension 4. In particular, the potential
terms for the scalars (derived in \cite{Skenderis:2006uy})
do not contribute.

The relation between the fields appearing in this action and the 10
dimensional fields reads,
\bea
S^{kI} &=& w(s^k) \hat{s}^{kI}, \quad k=2,3; \qquad
w(s^k) = \sqrt{8k (k-1) (k+2) z(k)/(k+1)}, \label{s2} \\
S^{4I} &=& %w(s^4)
\frac{2 \sqrt{3}}{5}\left(\hat{s}^{4I} - \frac{a_{4I, 2J, 2K}}{27 z(4)}
(83 \hat{s}^{2J} \hat{s}^{2K}
%+ L_{IJK}
%- \frac{7}{27 z(4)} a_{4I, 2J, 2K}
+7 D_{\m} \hat{s}^{2J} D^{\m} \hat{s}^{2K})\right); \label{s4} \\
G_{\m \n} &=& g^{o}_{\m\n} + \td{h}^{0}_{\m \n} + L_{\m\n};
\label{5dmet} \\
L_{\m \n} &=& - \frac{1}{12} \left(
\frac{2}{9} D_{\m} D^{\r} \hat{s}^{2I} D_{\n} D_{\r} \hat{s}^{2I}
- \frac{10}{3} \hat{s}^{2I} D_{\m} D_{\n} \hat{s}^{2I}
+(\frac{10}{9} (D \hat{s}^{2I})^2 - \frac{32}{9} (\hat{s}^{2I})^2) g^{o}_{\m
  \n} \right ); \nn \\
A^{1a}_{\m} &=& \frac{\sqrt{2}}{3} a^{1a}_{\m},  \label{A}
\eea
where  $z(k)$ is given in (\ref{nor_sc}) and
$a_{4I,2J,2K}$ is the triple overlap between scalar harmonics
(\ref{trip}).
In comparing these and subsequent formulae with those
given in \cite{Skenderis:2006uy} and \cite{Skenderis:2006di} note that
the formulae in these papers were for the case
of $SO(2) \times SO(4)$ singlet harmonics
for which $a_{40,20,20}/z(4) = 3/(2\sqrt{5})$.
%Similarly, in comparing the cubic coupling in (\ref{5daction}) and the
%coefficient $\l_{222}/3$ from Table 1 of \cite{Skenderis:2006di}
%one should use $a_{20,20,20}/z(2) = 1/(5\sqrt{3})$.
Here the hatted fields
are the gauge invariant combinations.
For computing the vevs of operators up to dimension
4, one needs in general gauge invariant combinations
quadratic in the fields and these are given in \cite{Skenderis:2006uy}.

For the bubbling solutions discussed in this paper, however,
linear gauge invariant combinations
suffice because, as explained later,
the solution is in the De Donder gauge to leading order
and deviates from it afterwards.
In particular, to the order we work, one can remove all hats from
all formulae in (\ref{s2})-(\ref{A}), except for $\hat{s}^{4I}$
for which one must use the linear gauge invariant combination
in (\ref{l1})-(\ref{ginv}).

The Kaluza-Klein relations in (\ref{s2})-(\ref{A})
result in an asymptotic expansion of the $5d$ fields to sufficiently
high order so that the vevs of dual operators can be
extracted, i.e. we obtain all terms up to order
$z^k$ for the fields\footnote{In \cite{Skenderis:2006di}
the KK map for $S^{2I}$ was computed up to terms of order $z^4$
but these higher order terms are not needed here.} $S^{4k}$, all
terms up to order
$z^2$ for $A^{1a}$ and all terms up to order $z^4$
for $G_{\m \n}$, where $z$ is the
Fefferman-Graham radial coordinate (see next section).

The reduction of gauge fields was not discussed in
\cite{Skenderis:2006uy} but can be determined from the results of
\cite{Arutyunov:1998hf} for the quadratic action.
Again, non-linear corrections to this reduction formula will not be needed
in what follows since they will not affect the vevs of the R symmetry
currents. The normalization of the $5d$ gauge fields is such that the
corresponding R symmetry currents have the standard normalization,
that is, their two point functions are given by \cite{Freedman:1998tz}
\be
\< J^a_i(x_1) J^b_j(x_2) \> = \frac{N^2}{2 (2 \pi)^4} \d^{ab} (\Box
\d_{ij} - \pa_i \pa_j) \frac{1}{(x_1 - x_2)^4},
\ee
where $4d$ coordinates are labelled by $x^i$ and $(a,b)$ label the $SO(6)$
indices.

\subsection{Holographic 1-point functions} \la{vev-form}

The final step is to use the method of holographic renormalization
to extract the vevs from the asymptotics of the $5d$ fields.
This is by now a standard procedure
except that here one needs to include additional terms to
accommodate extremal couplings (see section 5.4 of \cite{Skenderis:2006uy}).
The relation between field asymptotics and vevs is most transparent
in Hamiltonian variables where the radius plays the
role of time. The 1-point functions are then related to
the radial canonical momenta of the bulk fields, which are expressed
as (non-linear) functions of the field asymptotics \cite{PS1,PS2}.

So to obtain the vevs one should use the following steps:\\
1) expand the deviation of the ten dimensional solution from
$AdS_5 \times S^5$ in harmonics of $S^5$ and in the radial direction; \\
2) obtain the asymtpotic expansion of the five dimensional fields,
using (\ref{s2})-(\ref{A})); \\
3) use the expressions for the exact 1-point functions in terms of the
asymptotics of the $5d$ fields to obtain the vevs.\\
This is the route followed in  \cite{Skenderis:2006uy}.
Since we know the Kaluza-Klein map in general one can process the
1-point functions to express them directly in terms of the
asymptotic coefficients of the ten dimensional fields. Then the
final formulae can be used directly after step 1) without having to
know or use the Kaluza-Klein map. Such formulae were presented in
\cite{Skenderis:2006di} for a subset of operators and we present
the formulae for all operators up to dimension 4 here.

Let us first summarize the expressions for the holographic
1-point functions in terms of the asymptotics of the $5d$ fields.
The near-boundary expansion of the bulk metric $G_{\m \n}$, the gauge
field $A_\mu$  and
scalar fields $\Phi^k$, where $k$ is the dimension of the dual
operator, take the form
\bea \label{near-bdry}
ds_5^2 &=& %G_{\m \n} d x^\m d x^\n =
\frac{dz^2}{z^2}
+ \frac{1}{z^2}\left(G_{(0)ij}(x) + z^2 G_{(2)ij}(x)
+ z^{4} (G_{(4)ij}(x) + \log z^2 H_{(4)ij}(x))\right) dx^i dx^j;
\nn \\
A_{i}(x,z) &=& A_{(0)i}(x) + z^2 A_{(2)i}(x) + \cdots; \nonumber \\
S^2(x,z) &=&
z^2 \left(\log z^2 S^2_{(0)}(x) + \tilde{S}_{(0)}^2(x) + \cdots \right);
\nn \\
S^k(x,z) &=& z^{(4-k)} S^k_{(0)}(x) + \cdots +z^k S_{(2k-4)}^k(x) + \cdots,
\qquad k>2.
\eea
In these expressions the boundary fields
$G_{(0)ij}, A_{(0)i},
S^2_{(0)}, S^k_{(0)}$ parametrize the Dirichlet boundary
conditions and are also the field theory sources for the
QFT stress energy tensor, the R symmetry current
and operators of dimension 2 and $k$,
respectively. The gauge field is in radial axial gauge, $A_z = 0$.
The near-boundary
analysis determines all coefficients in these expansions except the
ones corresponding to the normalizable modes, namely
$G_{(4)ij}, A_{(2)i}, \tilde{S}^2_{(0)}, S^k_{(2k-4)}$.

Consider first the scalar operators $\cao_{S^{2I}}$ and
$\cao_{S^{3I}}$, where $I$ labels their degeneracy. We will be
interested in $SO(4)$ singlet operators which can be labeled by
their $SO(2)$ charge $m$, but we will express the holographic
relations in a more generally applicable way.
For these operators the holographic relations are \cite{Skenderis:2006uy}:
\be
\< \cao_{S^{kI}} \> = \frac{N^2}{2 \pi^2} (\pi_{(k)}^{kI}),
\ee
where $\pi_{(k)}^{km}$ indicates the part of the canonical momentum
of the field $S^{kI}$ that scales with weight $k$.
The relevant part of the canonical momenta can be expressed in terms
of the asymptotic expansion of the $5d$ fields as follows\footnote{Note
that we chose not to include in the definition of $\pi$ the
prefactor of $N^2/2 \pi^2$, in constrast to the conventions of
\cite{Skenderis:2006uy,Skenderis:2006di}.}
\be \label{mom}
\pi^{kI}_{(2 k-4)} =  (2 k- 4) [S^{kI}]_{k}
\ee
where the notation $[A]_k$ indicates the coefficient of the $z^k$
term in $A$ and $z$ is the Fefferman-Graham radial coordinate.
The relation (\ref{mom}) holds for $k \neq 2$;
when $k=2$ one should replaces the factor $(2 k-4)$ by 2.

As discussed in some detail in \cite{Skenderis:2006uy} the vevs of the
scalar operators of dimension four also involve quadratic terms; these
are necessary to accommodate extremal couplings. Thus the vevs in this
case are
\be
\< \cao_{S^{4I}} \> = \frac{N^2}{2 \pi^2} \left ( \pi_{(4)}^{4I}
+ 2 \sqrt{3} \frac{a_{4I,2J,2K}}{z(4)} \pi^{2J}_{(2)}
\pi^{2K}_{(2)} \right )
\label{o4_p}
\ee
Using the Kaluza-Klein map in (\ref{s2}-(\ref{s4}) we now express these vevs directly in terms of the
coefficients that appear in the $10d$ solution,
\vspace{2mm}

\framebox[\width]{
        \begin{minipage}{5.8in}
\bea
\< \cao_{S^{2I}} \> &=& \frac{N^2}{2 \pi^2} \frac{2 \sqrt{8}}{3}
[s^{2I}]_2; \hsp
\< \cao_{S^{3I}} \> = \frac{N^2}{2 \pi^2} \sqrt{3}
[s^{3I}]_3;
\la{ovev} \\
\< \cao_{S^{4I}} \> &=& \frac{N^2}{2 \pi^2} \frac{4 \sqrt{3}}{5}
[2 s^{4I}
+ \frac{74}{27 z(4)} a_{4I,2J,2K} s^{2J} s^{2K}
- \frac{14}{27 z(4)} a_{4I,2J,2K} (D_{\m}
s^{2J}) (D^{\m} s^{2K})]_4, \nn
\eea
\end{minipage}}

\vspace{2mm}
\noindent where $z(k)$ is the normalization of the degree $k$ spherical
harmonics, defined in \eqref{nor_sc} and $a_{4I,2J,2K}$ is the triple
overlap between scalar harmonics (\ref{trip}).
The expression for $\< \cao_{S^{4I}} \>$ can be further simplified
for solutions in which $s^2$ has vev (rather than source) behavior,
such as those under consideration in this paper. In such cases, the
asymptotics of (\ref{near-bdry}) imply that
\be
[(D_{\m} s^{2J} D^{\m} s^{2K}]_4 = [z^2 \pa_z s^{2J} \pa_z s^{2K}]_4 =
4 [s^{2J}s^{2K}]_4
\ee
and thus we obtain
\be \label{o4}
\< \cao_{S^{4I}}  \> = \frac{N^2}{2 \pi^2} \frac{4 \sqrt{3}}{5}
[2 s^{4I} + \frac{2}{3 z(4)} a_{4I,2J,2K} s^{2J} s^{2K}]_4.
\ee
Next consider the stress energy tensor; its vev can be obtained by
analyzing the coupled system of the metric and the scalar fields
$S^{2I}$. (The other $5d$ fields fall off too fast to contribute to
the stress energy tensor.) The part of the $5d$ action involving the
metric and one $S^2$ field
is same as the sector of gauged supergravity analyzed in
\cite{BFS1,BFS2}, where $S^2$ was called $\F$.
The result for the stress energy tensor can thus be carried over from
these works, with $S^{2} \rightarrow S^{2I}$. Thus one gets
\bea \label{tij}
\< T_{ij} \> &=& \frac{N^2}{2 \pi^2}\left(
G_{(4)ij} +\frac{1}{3} (\tilde{S}^{2I}_{(0)}
\tilde{S}^{2I}_{(0)})  G_{(0)ij}+
\frac{1}{8}[\Tr G_{(2)}^2 -(\Tr G_{(2)})^2] G_{(0)ij}\right. \\
&& \left. - \frac{1}{2} (G_{(2)}^2)_{ij} + \frac{1}{4} G_{(2)ij} \Tr G_{(2)}
+\frac{3}{2} H_{(4)ij} \right), \nn
\eea
where the summation over $I$ is implicit.

We now want to rewrite this expression in terms of the coefficients of the
ten-dimensional fields. To do this we should use (\ref{5dmet}) to express
the $5d$ coefficients in terms of the coefficients of the
ten-dimensional fields. To present a formula that is universally applicable  
we should start from a universal form of the $10d$ metric. 
One such form is the 
Fefferman-Graham form, which can always be reached by a coordinate 
transformation. Note however 
that the expression (\ref{tij}) presumes that the $5d$ metric $G_{\m \n}$ 
is in 
the Fefferman-Graham form, but because $L_{zz}, L_{zi}, \pi^0$ are 
generically non-zero, 
the $10d$ and  $5d$ metrics cannot simultaneously be in the Fefferman-Graham 
form.
This is not a problem since after the reduction to $5d$ we can always find
the appropriate coordinate transformation that brings $G_{\m \n}$ to the 
Fefferman-Graham form.
 
By definition the $10d$ metric is $(g^o+h^0)$ but it is more useful 
to consider instead the following combination,
\be \label{g10}
g_{\m \n} \equiv g^{o}_{\m\n} + \td{h}^{o}_{\m\n}% = G_{\m\n} - L_{\m\n}.
\ee
and use as a starting point this metric in the Fefferman-Graham form,
\be
ds^2 = g_{\m \n} d x^\m dx^\n = \frac{dz^2}{z^2}
+ \frac{1}{z^2}\left(g_{(0)ij}(x) + z^2 g_{(2)ij}(x)
+ z^{4} (g_{(4)ij}(x) + \log z^2 h_{(4)ij}(x))\right) dx^i dx^j;
\ee
The reason is that the $z$ dependence of $\pi^0$ it is not a priori known,
so had we started from $g^o + h^0$ we would not be able to work out 
explicitly the subsequent transformation that 
brings $G_{\mu \nu}$ to the Fefferman-Graham form. (Of course, in any 
given example one can always find this transformation, but here we
aim to give formulae that apply universally.)

Using 
\bea
L_{zz} &=& \frac{20}{27} z^2 \td{s}^{2I}_{(0)} \td{s}^{2I}_{(0)} +
\cdots; \hsp
L_{zi} =  - \frac{z^3}{2} \td{s}^{2I}_{(0)} \pa_i \td{s}^{2I}_{(0)} +
\cdots; \nn \\
L_{ij} &=& - \frac{19}{27} z^2 \td{s}^{2I}_{(0)} \td{s}^{2I}_{(0)} g_{(0)
  ij} + \cdots.
\eea
and (\ref{5dmet}) we find in particular that
\be
G_{zz} = \frac{1}{z^2} (1 + \frac{20}{27} z^4 
\td{s}^{2I}_{(0)} \td{s}^{2I}_{(0)}), \quad
G_{zi}= -\frac{1}{2} z^3 \td{s}^{2I}_{(0)} \pa_i \td{s}^{2I}_{(0)}
\ee
so the metric is indeed not in Fefferman-Graham form.
The coordinate transformation
\be
z = z' (1- \frac{5}{54} z'{}^4 \td{s}^{2I}_{(0)} \td{s}^{2I}_{(0)} 
+ \cdots ), \quad
x^i = {x^i}' + z'{}^6 \a \td{s}^{2I}_{(0)} \pa_i \td{s}^{2I}_{(0)} + \cdots,
\ee
with $\a$ a suitable numerical constant, brings the metric to the 
Feffreman-Graham form with the following coefficients:
\be
G_{(0)} = g_{(0)}, \quad  G_{(2)}= g_{(2)}, \quad 
G_{(4)}= g_{(4)} -\frac{14}{27} \td{s}^{2I}_{(0)} \td{s}^{2I}_{(0)} 
g_{(0)},\quad  H_{(4)}=h_{(4)}.
\ee
Inserting these expressions in (\ref{tij}) we get

\framebox[\width]{
        \begin{minipage}{5.6in}
\bea \label{tij2}
\< T_{ij} \> &=& \frac{N^2}{2 \pi^2}\left(
g_{(4)ij} - \frac{2}{9} (\tilde{s}^{2I}_{(0)}
\tilde{s}^{2 I}_{(0)})  g_{(0)ij}  \right . \\
&& \left . + \frac{1}{8}[\Tr g_{(2)}^2 -(\Tr g_{(2)})^2] g_{(0)ij}
- \frac{1}{2} (g_{(2)}^2)_{ij} + \frac{1}{4} g_{(2)ij} \Tr g_{(2)}
+\frac{3}{2} h_{(4)ij} \right), \nn
\eea
\end{minipage}}

\vspace{2mm}
\noindent where we emphasise that $g_{(k)ij}$ 
are the coefficients in the Fefferman-Graham
expansion of the $10d$ metric in (\ref{g10}). .

Let us finally consider the R symmetry currents; from the results of
\cite{BFS2} their vevs are
\vspace{2mm}

\framebox[\width]{
        \begin{minipage}{5.6in}
\be \la{rsym}
\< J_{i}^a \> = - \frac{N^2}{8 \pi^2} A_{(2)i}^{1a} \equiv
- \frac{\sqrt{2} N^2}{24 \pi^2} a_{(2)i}^{1a}
\ee
\end{minipage}}

\vspace{2mm}
\noindent
where again we rewrote the R symmetry current in terms of
ten-dimensional
fields to give the second equality in (\ref{rsym}).

\bigskip

Before leaving this section, let us comment on the wider applicability
of the highlighted expressions for the holographic vevs, \eqref{ovev},
\eqref{tij2} and \eqref{rsym}. In this paper we will analyse in detail
the LLM bubbling solutions, which preserve an $R \times SO(4) \times
SO(4)$ symmetry group, and are associated with $1/2$ BPS states of
${\cal N} = 4$ SYM on $R \times S^3$.

However $1/4$ and $1/8$ BPS states on $R
\times S^3$ which are built from operators involving only the six
scalars of ${\cal N} = 4$ also induce vevs only for the R-currents,
the stress energy tensor and the scalar chiral primaries. Thus the
expressions for the holographic vevs given here can be used to extract
such data from putative dual geometries, of the type constructed in
\cite{Chen:2007du}.
It would be straightforward to derive corresponding expressions
for more general
asymptotically $AdS_5 \times S^5$ solutions, which
involve more supergravity fields, such as the Janus solutions
recently derived in \cite{D'Hoker:2007xy}. Note in particular that the
expression given here for the stress energy tensor \eqref{tij2}
provides a rigorous way to extract the mass (including the Casimir term)
from the ten-dimensional solution.

\section{Bubbling solutions} \label{bub_sec}

The LLM bubbling solutions are\footnote{Note that
we use the notation $z$ with two completely different
meanings; as the function $z$ defined in (\ref{LLMsol})
and also as the Fefferman-Graham radial
coordinate, (\ref{near-bdry}). The meaning of $z$ should be clear from
the context.}
\bea
ds^2 &=& - h^{-2} (dt + V_i dx^i)^2 + h^2 (dy^2 + dx_i dx^i) + y e^{G}
d\Omega_3^2 + y e^{-G} d \td{\Omega}_3^2; \nn \\
h^{-2} &=& 2 y \cosh (G); \hsp
z = \half \tanh G; \label{LLMsol}\\
y \pa_y V_i &=& \ep_{ij} \pa_j z; \hsp
y (\pa_i V_j - \pa_j V_i) = \ep_{ij} \pa_y z; \nn \\
F_{5} &=& F_{\m \n} dx^{\m} \wedge dx^{\n} \wedge d \Omega_3 +
\td{F}_{\m\n} dx^{\mu} \wedge dx^{\nu} \wedge d \td{\Omega}_3; \nn \\
F &=& d B_t \wedge (dt + V) + B_t dV + d \hat{B}; \nn \\
\td{F} &=& d \td{B}_t \wedge (dt +V) + \td{B}_t dV + d \td{B}; \nn \\
B_{t} &=& - \qu y^2 e^{2G}; \hsp \td{B}_t = - \qu y^2 e^{-2 G}; \nn \\
d \hat{B} &=& - \qu y^3 \ast_{3} d \left ( \frac{z + \half}{y^2}
\right ); \hsp
d \td{B} = - \qu  y^3 \ast_3 d \left ( \frac{z - \half}{y^2} \right
), \nn
\eea
where $i =1,2$ and $\ast_3$ is the Hodge dual on the $R^3$
parameterized by $(y,x_1,x_2)$. The solutions are characterized by a
harmonic function on six dimensions, with sources on an $R^2$. That
is,
\be
\frac{z (x_1,x_2,y)}{y^2} = \frac{1}{\pi}
\int_{R^2} \frac{z(x'_1,x'_2,0) dx'_1 dx_2'}{( (x - x')^2 + y^2)^2},
\ee
where regularity of the solution
requires that $z(x'_1,x'_2,0)$ takes the values $\pm
\half$. The two-dimensional vector $V_i$ can be written as
\be
V_i (x_1,x_2,y) = \frac{\ep_{ij}}{\pi} \int_{R^2}
\frac{z(x'_1,x'_2,0) (x_j - x_j') dx'_1 dx_2'}{( (x - x')^2 + y^2)^2}.
\ee
In polar coordinates on $R^2$ this can be written as
\bea
V_{\td{\phi}} &=& - \frac{r}{\pi} \int_{R^2} \frac{ z(r',\td{\phi}',0)
  (r -r' \cos (\td{\phi} - \td{\phi}'))   r' dr' d \td{\phi}'}{ ((x
  -x')^2 + y^2 )^2}; \label{v-polar} \\
V_{r} &=& \frac{1}{\pi} \int_{R^2} \frac{ z(r',\td{\phi}',0)
  \sin( \td{\phi} - \td{\phi}')
 (r')^2 dr' d \td{\phi}'}{ ((x -x')^2 + y^2 )^2}. \nn
\eea

\subsection{$AdS_5 \times S^5$ solution}

The $AdS_5 \times S^5$ solution is obtained by taking sources for
$z$ on a disk of radius $r_0$. Then
\be
z^{o} = - \half \left ( \frac{(r^2 + y^2 - r_0^2)}{\sqrt{ (r^2 + r_0^2 + y^2)^2
    - 4 r^2 r_0^2}} \right )
\ee
where $r$ is a polar coordinate on $R^2$ such that $x_1 = r \cos
\td{\phi}$ and $x_2 = r \sin \td{\phi}$. Introducing the following coordinate change
on the $R^2$ parameterized by $y,x_1,x_2$
\be \la{coor}
y \equiv \td{R} \cos \td{\q} = R \cos \q; \hsp
r \equiv \td{R} \sin \td{\q} = \sqrt{R^2 + r_0^2} \sin \q; \hsp
\td{\phi} = \phi - t,
\ee
gives
\be
z^o = - \half + \frac{r_0^2 \cos^2 \q}{(R^2 + r_0^2 \cos^2 \q)}.
\ee
The coordinate shift (\ref{coor}) changes the flat metric on $R^3$
\be
ds_3^2 = d\td{R}^2 + \td{R}^2 (d\td{\q}^2 + \sin^2 \td{\q} d\td{\phi}^2)
\ee
to the following metric:
\be
ds^2_3 = (R^2 + r_0^2 \cos^2 \q) \left ( \frac{dR^2}{R^2 + r_0^2} + d
\q^2 \right ) + (R^2 + r_0^2) \sin^2 \q (d\phi - dt)^2.
\ee
The other functions in the metric take the values
\bea
(h^{-2})^o &=& r_0^{-1} (R^2 + r_0^2 \cos^2 \q); \hsp
(y e^{G})^o = r_0 \cos^2 \q; \\
V^o &=& - \frac{r_0^2 \sin^2 \q} {(R^2 + r_0^2 \cos^2 \q)} (d \phi - dt); \hsp
(y e^{-G})^{o} = \frac{R^2}{r_0}. \nn
\eea
The superscript $A^o$ denotes that these are the background $AdS_5
\times S^5$ functions, about which we will expand. Substituting these
values into the metric gives
\be
ds^2 = r_0 \left ( - (\hat{R}^2 + 1) dt^2 +  \frac{d\hat{R}^2}{(\hat{R}^2 + 1)} +
\hat{R}^2 d\td{\Omega}_3^2 + (d \q^2 + \sin^2 \q d\phi^2 +
\cos^2 \q d \Omega_3^2) \right ),
\ee
where $\hat{R} = R/r_0$. This is indeed the metric on $AdS_5 \times S^5$ with global
coordinates on $AdS_5$ and curvature radius
$\sqrt{r_0}$; henceforth $r_0$ will be set to one.

The five form field strength can be obtained in the following way. The
two form $d \hat{B}$ is given by
\bea
(d \hat{B})_{\q \phi} &=& - (d \hat{B})_{\q t} =
- \qu R^3 (R^2 + 1) \cos^3 \q \sin \q \pa_R \Phi; \\
(d \hat{B})_{R \phi} &=& - (d \hat{B})_{R t} =
\qu R^3 \cos^3 \q \sin \q \pa_{\q} \Phi; \nn \\
(d \hat{B})_{R \q} &=& - \qu R^3 \cos^3 \q  \frac{(R^2 + \cos^2 \q)}{
  \sin \q (R^2 + 1)} \pa_{\phi} \Phi, \nn
\eea
where
\be
\Phi = y^{-2} (z + \half),
\ee
and for $AdS_5 \times S^5$
\be \la{po}
\Phi^o = \frac{1}{R^2 (R^2 + \cos^2 \q)}.
\ee
The two form $d \td{B}$ is similarly given by
\bea
(d \td{B})_{\q \phi} &=& - (d \td{B})_{\q t} = - \half \cos \q \sin \q (R^2 + 1)
- \qu R^3 (R^2 + 1) \cos^3 \q \sin \q \pa_R \Phi; \nn \\
(d \td{B})_{R \phi} &=& - (d \td{B})_{R t} = - \half R \sin^2 \q
+ \qu R^3 \cos^3 \q \sin \q \pa_{\q} \Phi; \\
(d \td{B})_{R \q} &=& - \qu R^3 \cos^3 \q  \frac{(R^2 + \cos^2 \q)}{
  \sin \q (R^2 + 1)} \pa_{\phi} \Phi, \nn
\eea
Substituting into the expression for the five form then gives the following
expression for $AdS_5 \times S^5$:
\be
\td{F}^{o}_{tR} = R^3; \hsp
{F}^{o}_{\q \phi} = \cos^3 \q \sin \q,
\ee
as expected.

\subsection{Asymptotic expansion}

Now let us consider more general solutions which are asymptotic to
$AdS_5 \times S^5$. The field theory data will be extracted from their
asymptotic expansions around the $AdS_5 \times S^5$ boundary. This
expansion can be economically expressed as follows. Let the solution
be expressed in terms of the harmonic function $\Phi (x_1,x_2,y)$ with
\be
\Phi = \Phi^o + \Delta \Phi,
\ee
where $\Phi^o$ is the harmonic function of the $AdS_5 \times S^5$
background about which we perturb. $\Delta \Phi$ can be expressed as
\be
\Delta \Phi (x_1,x_2,y) = \frac{1}{\pi} \int_{R^2}
\frac{\Delta z(x'_1,x'_2,0) dx'_1 dx_2'}{( (x - x')^2 + y^2)^2},
\ee
where $\Delta z(x'_1,x'_2,0) = (z(x'_1,x'_2,0) - z^o(x'_1,x'_2,0))$.
Now note that $\Phi$ (and hence $\Delta
\Phi$) is a scalar harmonic function on $R^6$ which preserves $SO(4)$
rotational symmetry. The asymptotics can thus be expressed as
\be \la{e1}
\Delta \Phi (\td{R}, \td{\q}, \td{\phi}) = \sum_{k,m} (\Delta \Phi)_{km}
\frac{Y^m_k(\td{\q},\td{\phi})}{\td{R}^{k+4}},
\ee
where $Y^m_k(\td{\q},\td{\phi})$ are normalized $SO(4)$ singlet
spherical harmonics of degree $k$ with $m$ labeling their $SO(2)$
charge; the properties of such harmonics are discussed in appendix
\ref{sph}. By the addition theorem the coefficients in this expansion
are given by \cite{Skenderis:2006di}
\be
(\Delta \Phi)_{km} = 2^{k} (k+1) \pi^{-1}  \int_{R^2} \Delta z(x'_1,x'_2,0)
(C^{m}_{i_1 \cdots i_k} x^{i'_1} \cdots x^{i'_k}) dx_1' dx_2',
\ee
where $C^m_{i_1 \cdots i_k}$ are $SO(4)$ invariant symmetric traceless
tensors on $R^6$ of rank $k$ which are in one to one correspondence
with the $SO(4)$ singlet spherical harmonics. In particular
\be \la{mass}
(\Delta \Phi)_{20} = 4 \sqrt{3} \pi^{-1} \int_{R^2} \Delta z(x'_1,x'_2,0) (r')^3
dr' d\f',
\ee
where the explicit representation of the $SO(4) \times SO(2)$ singlet
tensor is used. In deriving this and subsequent expressions, one needs
to integrate the explicit form of the normalized traceless tensor on $R^6$, $C^m_{i_1
  \cdots i_k}$, over the source distribution in $R^2$. Since only the
coordinates $(x^1,x^2)$ are non-zero, terms in $C^m_{i_1
  \cdots i_k}$, for which $i \neq 1,2$ do not contribute.

Note that the expansion (\ref{e1}) begins at $k=2$. There is no $k=0$
term, since the leading asymptotics are those of $AdS_5 \times S^5$
and $k=1$ terms are unphysical since they can always be removed by
choosing the origin of the coordinate system to be at the centre of
mass. The centre of mass conditions imply that
\be \la{com}
\int_{R^2} z(r',\td{\phi}',0) r' e^{\pm i \phi'} (r' dr' d \td{\phi}' ) =
\int_{R^2} \Delta z(r',\td{\phi}',0) r' e^{\pm i\phi'} (r' dr' d \td{\phi}' ) = 0.
\ee
The harmonic function (\ref{e1}) is expressed in the usual
coordinates on $R^6$, but to perturb relative to $AdS_5 \times S^5$
the function needs to expressed in terms of the coordinates
$(R,\q,\phi)$. However, for $(R,\td{R}) \gg 1$, this change in
coordinates changes the form of (\ref{e1}) only at $k \ge 4$, that is,
\bea
\Delta \Phi ({R}, {\q}, {\phi},t) &=& (\Delta \Phi)_{2m}
\frac{Y^m_2({\q},{\phi-t})}{{R}^{6}} +
(\Delta \Phi)_{3m}
\frac{Y^m_3({\q},{\phi-t})}{{R}^{7}} +  \\
&& \hsp \frac{1}{R^8}
\left ( (\Delta \Phi)_{4m} Y^m_4({\q},{\phi-t}) + (\Delta \Phi)_{2m}
f^{m} (\q,\phi-t) \right )  +
\cdots \nn
\eea
where the functions $f^{m} (\q,\phi-t)$ are given by
\bea
f^{0} (\q,\phi-t) &=& f^0_0 Y_0 + f^0_2 Y^0_2(\q) + f^0_4 Y^{0}_4 (\q); \\
f^{\pm 2} (\q,\phi-t) &=& f^{\pm 2}_{2}  Y^{\pm 2}_{2}
(\q,\phi-t) + f^{\pm 2}_{4}  Y^{\pm 2}_{4} (\q,\phi-t). \nn
\eea
Note that $Y_{0} = 1$. These coefficients are obtained by expanding
$\Delta \Phi (\td{R}, \td{\q}, \td{\phi})$ using
\be
\frac{1}{\td{R}^2} = \frac{1}{R^2} \left (1 - \frac{\sin^2 \q}{R^2} +
\cdots \right ); \hsp
\sin\td{\q} = \sin \q \left (1 + \frac{\cos^2 \q}{2R^2} + \cdots
\right ),
\ee
and then projecting back onto the basis of spherical harmonics. In
what follows we will need only the following coefficients explicitly
\be
f^{0}_{0} = 0; \hsp f^{0}_4 =  - \frac{4 \sqrt{3}}{\sqrt{5}}; \hsp
f^{\pm 2}_{4} = - \frac{4 \sqrt{2}}{\sqrt{5}}.
\ee
The functions appearing in the metric can be expressed in terms of
$\Delta \Phi$ as follows
\bea
y e^{G} = \cos^2 \q (1 + R^2 (R^2 + \cos^2 \q) \Delta \Phi)^{\half}
(1 - \cos^2 \q (R^2 + \cos^2 \q) \Delta \Phi)^{- \half} \equiv \cos^2
\q (1 + \a); \nn \\
y e^{-G} = R^2 (1 + R^2 (R^2 + \cos^2 \q) \Delta \Phi)^{- \half}
(1 - \cos^2 \q (R^2 + \cos^2 \q) \Delta \Phi)^{\half} \equiv
R^2 (1 + \b); \nn \\
h^{-2} (R^2 + \cos^2 \q)^{-1} = (1 + (R^4 - \cos^2 \q) \Delta \Phi - R^2
\cos^2 \q (R^2 + \cos^2 \q)^2 \Delta \Phi^2)^{-\half} \equiv (1 + \g); \nn \\
h^2 (R^2 + \cos^2 \q) = (1 + (R^4 - \cos^2 \q) \Delta \Phi - R^2
\cos^2 \q (R^2 + \cos^2 \q)^2 \Delta \Phi^2)^{\half} \equiv (1 + \d). \nn
\eea
Note that the leading order terms in $(\a,\b\,\g,\d)$ are of order $1/R^2$;
to extract vevs of operators of dimension four and less it will be
sufficient to expand these quantities up to order $1/R^4$. Then
\bea
\a &=& (\half R^4 \Delta \Phi + R^2 \cos^2 \q \Delta \Phi - \frac{1}{8}
R^8 (\Delta \Phi)^2 + \cdots); \\
\b &=& (- \half R^4 \Delta \Phi - R^2 \cos^2 \q \Delta \Phi + \frac{3}{8}
R^8 (\Delta \Phi)^2 + \cdots); \nn \\
\g &=& (- \half R^4 \Delta \Phi + \frac{3}{8} R^8 (\Delta \Phi)^2 +
\cdots ); \nn \\
\d &=& (\half R^4 \Delta \Phi - \frac{1}{8} R^8 (\Delta \Phi)^2 +
\cdots ). \nn
\eea
Now consider the vector $V_i$: one needs to expand the perturbation of
the vector written in polar coordinates (\ref{v-polar}) from the background value. 
Using the centre of mass conditions (\ref{com}) one finds that
$(\Delta V) \equiv (V - V^o)$ is given by
\bea
\Delta V_{\td{\phi}} &=& - R^2 \sin^2 \td{\q} \Delta \Phi +
\frac{1}{6R^4} \left ( Y^{2,2} (\Delta \Phi)_{2 2} + Y^{2,-2} (\Delta
  \Phi)_{2 (-2)} + (\Delta\Phi)_{20} (2 Y^{2,0} + \frac{1}{\sqrt{3}}
  Y^{0})  \right ) \nn
\\
\Delta V_{r} &=& - \frac{i}{6 R^4}  \sin \td{\q}
\left (e^{2 i
  \td{\f}} \Delta \Phi_{22} - e^{-2 i \td{\f}} \Delta \Phi_{2(-2)}
\right ).
\eea
Changing coordinates to $(R,\q,\phi)$ gives
\bea
\Delta V_{\phi} &=& - \Delta V_{t} = \Delta V_{\td{\phi}}; \\
\Delta V_{R} &=& \sin \q \Delta V_{r} + \cdots; \hsp
\Delta V_{\q} = R \cos \q \Delta V_{r} + \cdots. \nn
\eea
It is convenient to use a gauge transformation to remove the radial
component of this vector. Shifting $\Delta V \rightarrow \Delta V + d
\Lambda$ is equivalent to shifting the coordinate $t \rightarrow t -
\Lambda$ whilst leaving $\tilde{\phi} = \phi - t$ unchanged. Choosing
\be
\Lambda = - \frac{i}{12 R^4} ( Y^{2,2} (\q,\td{\phi}) 
(\Delta \Phi)_{2 2} - Y^{2,-2} (\Delta \Phi)_{2 (-2)})
\ee
results in 
\be
\Delta V_{\phi} = \frac{v_{\phi} (\q ,\phi,t)}
{{R}^4} + \cdots; \hsp
\Delta V_{\q} = \frac{v_{\q}(\q,\phi,t)} {{R}^5} + \cdots,
\ee
with
\bea
v_{\phi} (\q ,\phi,t) &=& - R^6 \sin^2 \td{\q} \Delta \Phi -
\frac{i}{4} \pa_{\phi} \left ( Y^{2,2} (\Delta \Phi)_{2 2} -  Y^{2,-2} (\Delta
  \Phi)_{2 (-2)} \right )  \\ 
&& \qquad + \frac{1}{6} \left ( 
(\Delta\Phi)_{20} (2 Y^{2,0} + \frac{1}{\sqrt{3}} Y^{0})  \right );
  \nn \\ 
v_{\q} (\q ,\phi,t) &=& - \frac{i}{4}  
\pa_{\q} \left ( Y^{2,2} (\Delta \Phi)_{2 2} -  Y^{2,-2} (\Delta \Phi)_{2
  (-2)} \right ). \nn
\eea
Note that in these epxressions the spherical harmonics are functions
of $(\q,\phi,t)$. 

\subsection{Expansion of metric and five form}

The asymptotic expansion of the metric is given by
\bea
ds^2 &=& - dt^2 ( (R^2 + 1)(1 + \g) + \sin^2 \q (\g -\d)
- 2 \frac{v_{{\phi}}}{R^2}) + R^2 (1
+ \b) d \Omega_3^2 + (1 + \d) \frac{dR^2}{R^2 + 1} \nn \\
&& 
%-2 dt dR \frac{v_r \sin \q}{R^3} 
- 2dt d\q \frac{v_{\q}}{R^2}
+ 2 dt d\phi ( (\g- \d) \sin^2 \q - \frac{v_{{\phi}}}{R^2}) \la{metric} \\
&& + (1 + \d) d\q^2 + \cos^2 \q (1 + \a) d\Omega_3^2 + 2 \sin^2 \q d
\phi (
%\frac{dR \sin \q}{R^5} +
\frac{d\q  v_{\q} }{R^4}) \nn \\
&& \sin^2 \q d \phi^2 (1 + \d + \frac{\sin^2 \q}{R^2} (\d - \g) + 2
\frac{v_{{\phi}}}{R^4} ) +  \cdots, \nn
\eea
where the terms retained are sufficient to extract vevs of operators
of dimension four and less. That is, to extract the vevs of the
scalar operators with dimension less than or equal to four one
needs $(\pi,\f_{(s)})$ to order $1/R^4$. To extract the vev of the R symmetry
current one will need $B_{(v) \mu}$ with $\mu \ne R$ to order
$1/R^2$. For the vev of the stress energy tensor one needs $(\pi^0,
\td{h}^0_{\m \n})$ up to order $1/R^4$.

In actually extracting these fields there is considerable
simplification relative to the discussions of \cite{Skenderis:2006uy}.
Consider first the perturbations tangent to the sphere and note that
\be
h_{ab} dx^a dx^b = (1 + \a) (d \q^2 + \cos^2 \q d \Omega_3^2 + \sin^2
\q d \phi^2) + {\cal {O}}(\frac{1}{R^4}).
\ee
This implies that to order $1/R^4$ the metric perturbation along the
sphere is in de Donder gauge, with only the trace $\pi$
non-vanishing. Compared to \cite{Skenderis:2006uy} where
the fields $\phi_{(s)}$ (zero in de Donder gauge) were already excited
at order $1/R^2$ there is simplification. Gauge
invariant combinations of fluctuations are needed first at order
$1/R^4$ (or correspondingly when computing vevs of operators of
dimension four) and moreover only linearly gauge invariant quantities
will be needed at this order. One would only need to use gauge
invariant quantities at non-linear order for computing vevs of
operators with dimension greater than four.
Thus one can immediately extract
\be
\hat{\pi}^{km} = \pi^{km} = \frac{5}{2 R^{k}} (\Delta \Phi)_{km}
e^{- i m t}+ \cdots;
\hsp k = 2,3.
\ee
To obtain $(\pi^0, \pi^4, \f_{(s)}^4)$ we split the sphere
perturbation into its trace and traceless parts:
\bea
\pi &=& \left (5 \d + R^2 (3  - 2 \sin^2 \q) \Delta \Phi + \frac{2
  v_{{\f}}}{R^4} + \cdots \right ); \\
h_{(\q \q)} &=& \left (- \frac{1}{5} R^2 \Delta \Phi (3  - 2 \sin^2
\q) - \frac{2 v_{{\f}}}{5 R^4} + \cdots \right ); \nn \\
h_{(\phi \phi)} &=& \sin^2 \q \left (\frac{1}{5} R^2 \Delta \Phi (7 \sin^2
\q - 3) + \frac{8 v_{{\f}}}{5 R^4} + \cdots \right ); \nn \\
h_{(\chi^\a \chi^\b)} &=& \cos^2 \q \hat{g}_{\chi^\a \chi^\b}
\left ( \frac{1}{5} R^2 \Delta \Phi (2
- 3 \sin^2 \q) - \frac{2 v_{{\f}}}{5 R^4} + \cdots \right  ), \nn
\eea
with $h_{\q \phi}$ as given in (\ref{metric}), $\chi^\alpha$
are coordinates on $S^3$ and $\hat{g}_{\chi^\alpha \chi^\beta}$
is a unit radius metric on $S^3$. Projecting onto the
basis of spherical harmonics gives
\bea
\pi^{4 m} &=& e^{-i m t} \left ( \frac{5}{2 R^4} (\Delta \Phi)_{4 m } -
  \frac{5}{8 R^4 z(4)} a_{m p q} (\Delta \Phi_{2 n} \Delta \Phi_{2q})
  \right)  + 168  \f_{(s)}^{4m}  + \cdots ; \\
\f_{(s)}^{4 \pm 2} &=&  e^{\mp 2 i t}\left ( - \frac{\sqrt{2} }{12 \sqrt{5} R^4}
(\Delta \Phi )_{2 \pm 2} + \cdots \right );  \nn \\
\f_{(s)}^{4 0} &=& \left ( - \frac{\sqrt{3}}{12 \sqrt{5} R^4}
(\Delta \Phi )_{2 0} + \cdots \right );
\nn \\
 \pi^{0} &=& \left (-\frac{5 z(2) }{8 R^4} (\Delta \Phi_{2 n}
  \Delta \Phi_{2 (-n)}) + \cdots \right ), \nn
\eea
where here and in subsequent formulae $a_{mpq} \equiv
a_{4m,2p,2q}$. 
%In deriving these expressions it is useful to write the
%traceless part $h_{(ab)}$ of the perturbation as:
%\bea
%h_{(\q \q)} &=& \left (- \frac{1}{5} R^2 \Delta \Phi (3  - 2 \sin^2
%\q) - \frac{2 v_{{\f}}}{5 R^4} + \cdots \right ); \nn \\
%h_{(\phi \phi)} &=& \sin^2 \q \left (\frac{1}{5} R^2 \Delta \Phi (7 \sin^2
%\q - 3) + \frac{8 v_{{\f}}}{5 R^4} + \cdots \right ); \nn \\
%h_{(\chi^\a \chi^\b)} &=& \cos^2 \q \hat{g}_{\chi^\a \chi^\b}
%\left ( \frac{1}{5} R^2 \Delta \Phi (2
%- 3 \sin^2 \q) - \frac{2 v_{{\f}}}{5 R^4} + \cdots \right  ), \nn
%h_{(\q \phi)} &=& 
%\eea
The expressions for $D_{(a} D_{b)} Y^{km}$ given in appendix A are
used in extracting the values of $\phi_{(s)}$. For example, in the
case of the R charged harmonics the relevant perturbations are:
\bea
h_{(\q \q)} &=& \frac{2}{5 R^4} e^{\pm 2 i (\phi -t)}  
\left (\sin^4 \q - \sin^2 \q \right) (\Delta \Phi )_{2 \pm 2}; \nn \\
h_{(\phi \phi)} &=& \frac{1}{10 R^4} e^{\pm 2 i (\phi - t)} 
\sin^2 \q \left (- \sin^4 \q + \sin^2 \q \right) (\Delta \Phi )_{2 \pm 2}; \\
h_{(\chi^\a \chi^\b)} &=& \frac{1}{10 R^4} e^{\pm 2 i (\f-t)} 
\cos^2 \q \hat{g}_{\chi^\a \chi^\b}
\left ( - \sin^4 \q + \sin^2 \q \right  ) (\Delta \Phi )_{2 \pm 2}; \nn \\
h_{(\q \phi)} &=& \frac{\mp i }{4 R^4} e^{\pm 2 i (\f -t)} \sin^3 \q \cos \q
(\Delta \Phi )_{2 \pm 2}, \nn
\eea
which comparing with (\ref{dy1}) and (\ref{dy2}) gives
\be
\f_{(s)}^{4 \pm 2} =  e^{\mp 2 i t}\left ( - \frac{\sqrt{2} }{12
  \sqrt{5} R^4}(\Delta \Phi )_{2 \pm 2} \right ); \qquad
\f_{(s)}^{2 \pm 2} =  e^{\mp 2 i t} 
\left ( - \frac{1}{20 R^4} (\Delta \Phi )_{2 \pm 2} \right ). 
\ee
A similar analysis follows for the neutral harmonics.
Thus the gauge invariant combinations are
\be \la{pig}
\hat{\pi}^{4 m} = e^{-i m t} \left ( \frac{5}{2 R^4} (\Delta \Phi)_{4 m } -
  \frac{5}{8 R^4 z(4)} a_{m n p} (\Delta \Phi_{2 n} \Delta \Phi_{2p})
 + 200 \f_{(s)}^{4m} e^{imt} + \cdots \right );
\ee
Now consider the vector fields. The metric (non-zero)
fluctuations $h_{\m a}$ can be expressed as
\be
h_{t a} = \frac{i}{4 R^2 } D_{a} \left ( Y^{2,2} (\Delta \Phi)_{22} - Y^{2,-2}
(\Delta \Phi)_{2(-2)} \right ) - \frac{1}{\sqrt{6} R^2} (\Delta \Phi)_{20} Y^1_a.
\ee
The physical vector fields arise from
the projection of the $h_{\m a}$ terms onto vector harmonics to give
$B_{(v) \m}^{I_5}$. The
non-zero projection of $h_{\m a}$ onto scalar harmonics takes the
metric outside de Donder gauge, but the resulting vectors
$B^{I_1}_{(s) \mu}$ do not contribute to any gauge invariant
quantities computed here. Thus the only relevant vector term is
\be \la{vec1}
B^{1}_{(v) t} = - \frac{1}{\sqrt{6} R^2} (\Delta
  \Phi)_{20}.
\ee
Finally let us consider the metric perturbation. The perturbation
$\td{h}^{0}_{\m\n}$ receives
contributions only from the first line in (\ref{metric}).
Thus the metric
$g^{o}_{\m\n} + \td{h}^{0}_{\m\n}$ with
$\td{h}^{0}_{\m\n} = h^{0}_{\m\n} + \frac{1}{3} \pi^0 g^{o}_{\m\n}$
is given by
\bea
ds^2 &=& -dt^2 \left (R^2 + 1 - \frac{1}{4 \sqrt{3}R^2 } (\Delta \Phi)_{20}
+ \frac{z(2)}{6R^2} (\Delta \Phi_{2 n}
  \Delta \Phi_{2 (-n)})  \right ) \la{asymmet} \\
&&  + \frac{dR^2}{(R^2 +1)} \left (1 - \frac{z(2)}{3R^4} (\Delta \Phi_{2 n}
  \Delta \Phi_{2 (-n)}) \right ) \nn \\
&& + R^2 d\td{\Omega}_3^2 \left (1 +
\frac{1}{12 \sqrt{3}R^4 } (\Delta \Phi)_{20} +  \frac{z(2)}{6R^4} (\Delta \Phi_{2 n}
  \Delta \Phi_{2 (-n)})  \right ), \nn
\eea
where summation over $n = (-2,0,2)$ is implicit.

Next consider the five form field strength. To compute the vevs we need
only the modes $(b_{\mu}^{I_5}, b_{(s)}^{I})$ in the expansion. This
means that we need only expand $f_{\m a} \equiv (F_{\m \a})$ and
$f_{\q \phi} \equiv (F_{\q \phi} - F^{o}_{\q \phi})$ giving
\bea
f_{\q \phi} &=&  \sin\q \cos^3 \q \left ( - \qu R^3 (R^2+1) \pa_{R} \Delta
\Phi + \frac{\a}{R^2} (1 - 3 \sin^2 \q) + \sin \q \cos \q
\frac{\pa_{\q} \a}{2 R^2}  \right . \nn \\
&& \left .+ \frac{v_{{\phi}}}{R^4}  - \frac{\cos \q}{4 R^4 \sin \q}
\pa_\q v_{{\phi}} 
%- \cos \q \pa_{\phi} v_r) 
+ \cdots \right ); \nn \\
f_{R \q} &=& \sin\q \cos^3 \q \left ( - \qu R^3 \frac{\pa_{\phi} \Delta
  \Phi}{\sin^2 \q} + \cdots \right ); \\
f_{R \phi} &=& \sin\q \cos^3 \q \left (\qu R^3 \pa_{\q} \Delta \Phi +
\cdots \right ); \nn \\
f_{\q t} &=&  \sin\q \cos^3 \q \left (\qu R^5 \pa_{R} (\Delta \Phi) + 2 \a
- \frac{\cos \q}{2 \sin \q} \pa_{\q} \a + \cdots \right ); \nn \\
f_{\f t} &=& - \half  \cos^4 \q \pa_{\f} \a + \cdots. \nn
\eea
{}From the $f_{\q t}$ term one gets
\be
f_{\q t} = \sin \q \cos^3 \q ( - \frac{1}{2 \sqrt{3} R^2}
(\Delta\Phi)_{20} + \cdots),
\ee
{}from which one can extract
\be
b^1_t = - \frac{1}{8 \sqrt{6} R^2} (\Delta\Phi)_{20}.
\ee
Combining this with the vector (\ref{vec1}) extracted from the metric
one finds that
\be \la{vecasy}
a^1_t = - \frac{\sqrt{3}}{\sqrt{2} R^2} (\Delta\Phi)_{20}; \hsp
c^1_t = 0.
\ee
This is the anticipated result since the massive vector $c^1$ should
not be excited at this order.

{}From the $f_{\q \f}$ terms one finds
\bea
b_{(s)}^{km} &=& - \frac{1}{4 k R^k} e^{-i m t} (\Delta \Phi)_{km}; \hsp k=2,3 \\
b_{(s)}^{4m} &=& \left ( - \frac{1}{16R^4}
e^{-i mt } (\Delta \Phi)_{4m} - \frac{9}{2} \f_{(s)}^{4m} \right ), \nn
\eea
and therefore the gauge invariant quantities are
\be \la{bg}
\hat{b}_{(s)}^{km} = - \frac{1}{4 k R^k} e^{-i m t} (\Delta \Phi)_{km} -
5 \f_{(s)}^{4m}; \hsp k=2,3,4.
\ee
Putting together (\ref{pig}) and (\ref{bg}) gives
\bea \la{sgue}
\hat{s}^{km} &=&
e^{-i m t} (\frac{1}{4 k R^k} (\Delta \Phi)_{km} +
( 5 e^{im t} \f_{(s)}^{4m} - \frac{1}{192 R^4 z(4)} a_{mnp}
(\Delta \Phi_{2 n} \Delta \Phi_{2p})) \d_{k4} + \cdots). \hsp k=2,3,4;
\nn \\
\hat{t}^{4m} &=& - \frac{1}{192 R^4 z(4)} e^{-i m t} a_{mnp} (\Delta \Phi_{2 n}
\Delta \Phi_{2p}).
\eea
Note that there are no $\f_{(s)}^{4m}$ terms in the gauge invariant
fields $\hat{t}^{4m}$. This is a computational check: using
\cite{Skenderis:2006uy} these fields satisfy the field equations
\be
(\Box - 96) \hat{t}^{4m} = 32 z(4)^{-1} a_{mnp} \hat{s}^{2n} \hat{s}^{2p},
\ee
and thus at order $1/R^4$ can only receive contributions quadratic in
$\Delta \Phi_{2 n}$. The $\f_{(s)}^{4m}$ terms are linear in $\Delta
\Phi_{2 n}$ and thus cannot contribute to the fields  $\hat{t}^{4m}$
at this order.

\subsection{Holographic vevs}

Given the asymptotic expansions of the relevant fields we can extract
the values for the vevs using the formulae from section
\ref{vev-form}. The relation for the R symmetry current
vev (\ref{rsym}) along with (\ref{vecasy}) implies that
\be
\< J_t \> = \frac{N^2}{2 \pi^2} \frac{1}{4 \sqrt{3}} (\Delta
\Phi)_{20}.
\ee
To apply the formula (\ref{tij2})
for the vev of the stress energy tensor one must first bring the
metric (\ref{asymmet}) into Fefferman-Graham form, by the coordinate change
\be
z = \frac{1}{R} \left (1 - \frac{1}{4 R^2} + \frac{1}{8 R^4} - \frac{z(2)}{24
  R^4} (\Delta \Phi)_{2n} (\Delta \Phi)_{2(-n)} \right ).
\ee
Then
\bea
\< T_{tt} \> &=& \frac{N^2}{2 \pi^2} \left ( \frac{3}{16} + \frac{1}{4 \sqrt{3}} (\Delta
\Phi)_{20} \right ); \\
\< T_{\a\b} \> &=& \frac{N^2}{2 \pi^2} \left ( \frac{1}{16} + \frac{1}{12 \sqrt{3}} (\Delta
\Phi)_{20} \right ) g_{\a\b};
\eea
where $g_{\a\b}$ is the metric on the unit radius $S^3$. Using the
explicit form for $(\Delta
\Phi)_{20}$ from (\ref{mass}) and reinstating factors of $a$, the inverse radius of the
$S^3$, gives
\vspace{2mm}

\framebox[\width]{
        \begin{minipage}{5.6in}
\bea
\< J_t \> &=& \frac{N^2 }{2 \pi^2} a \int_{R^2} \rho (r^2 - \half a^2) r
dr d \f; \la{j-en} \\
\< T_{tt} \> &=& \frac{N^2}{2 \pi^2} \left ( \frac{3a^4}{16} +
a^2 \int_{R^2} \rho (r^2 - \half a^2 ) r dr d \f \right ) = \< T_{tt}
\>_{c} +  a \< J_t \>, \nn
\eea
\end{minipage}}

\vspace{2mm}
\noindent where $\< T_{tt}\>_{c}$ is the Casimir on $R \times S^3$ and the
density function $\rho (r,\f)$ satisfies
\be
\int_{R^2} \rho (r,\f) r dr d\f = 1; \hsp
\rho^{o} = \frac{1}{\pi a^2} \q (a - r).
\ee
We define $\q(x) = 1$ for $x \ge 0$ and $\q(x) = 0$ otherwise.
A general distribution is such that $\rho(r,\f)$ takes the value
$1/\pi a^2$ in a region of the plane with area $\pi a^2$, and is zero
everywhere else. The corresponding mass $E$ and R-charge $J$ are given by
integrating these expressions over the $S^3$, resulting in
\vspace{2mm}

\framebox[\width]{
        \begin{minipage}{5.6in}
\bea
J &=& N^2 a \int_{R^2} \rho (r^2 - \half a^2) r
dr d \f; \la{j-mass} \\
E  &=& N^2 \left ( \frac{3a^4}{16} +
a^2 \int_{R^2} \rho (r^2 - \half a^2 ) r dr d \f \right ) = E_{c}
+  a J. \nn
\eea
\end{minipage}}

\vspace{2mm}
\noindent These quantities have the expected behavior,
namely $J  = 0$ for $AdS$ with the Casimir energy $E_c$ taking
the expected value; the energy and angular momentum tend to zero
in the limit of a large $S^3$ and the BPS bound $(E -E_c) = aJ$ is saturated.

For the scalar operators
(\ref{o4}) along with (\ref{sgue}) implies the following result for the vevs:
\bea
\< \cao_{S^{km}}  \> &=& \frac{N^2}{\pi^2} \frac{(k-2)}{2^{\half k} (k+1)}
\sqrt{\frac{(k-1)}{k}}
e^{-i a m t} \left ( (\Delta \Phi)_{km} + 80 R^4 \f_{(s)}^{4m} \d_{k4}
\right ); \nn \\
80 R^4 \f_{(s)}^{4 \pm 2} &=& - \frac{4 \sqrt{10}}{3} (\Delta \Phi)_{2
\pm 2}; \hsp
80 R^4 \f_{(s)}^{4 0} = - \frac{4 \sqrt{5}}{\sqrt{3}} (\Delta \Phi)_{2
0},
\eea
with $(k-2) \rightarrow  1$  for $k=2$ and the scale $a$, the inverse
radius of the $S^3$, reinstated. For operators with $\left | m
\right | = k$ (and $k \neq 1$) the vevs are therefore
\vspace{2mm}

\framebox[\width]{
        \begin{minipage}{5.6in}
\be \la{ch-1}
\< \cao_{S^{k \pm k}}  \> = \frac{N^2}{\sqrt{k} \pi^2}
(k-2) \sqrt{k-1} e^{-i a k t} \int_{R^2} (r^k \rho)  e^{\pm i k \f}
r dr d \f.
\ee
\end{minipage}}

\vspace{2mm}
\noindent Recall that there is no $k=1$ operator in the $SU(N)$ theory; the
integral vanishes in this case because of the centre of mass condition
\eqref{com}.
%where $\Delta \rho \rightarrow \rho$ since
%\be
%\int_{R^2} r^l z^{o} e^{i m \f} r dr d\f = 0
%\ee
%for any $l$ when $m \neq 0$.
For the other operators with dimension less than four, one gets
\vspace{2mm}

\framebox[\width]{
        \begin{minipage}{5.6in}
\bea
\< \cao_{S^{2 0}}  \> &=& \frac{\sqrt{2} N^2 }{\sqrt{3} \pi^2}
\int_{R^2} (r^2 \Delta z) r dr d \f; \la{pofr} \\
&=& \frac{\sqrt{2} N^2 }{\sqrt{3} \pi^2}
\left ( \int_{R^2} \rho (r^2 - \half a^2) r dr d \f \right ); \nn \\
\< \cao_{S^{3 \pm 1}}  \> &=& \frac{N^2} {\pi^2}
e^{\mp i a t } \int_{R^2} (r^3 \rho) e^{\pm i  \f}
r dr d \f, \nn
\eea
\end{minipage}}

\vspace{2mm}
\noindent where in the second expression for the neutral operator
the explicit form of
$\rho^o$ is used.
For the operators with dimension four, again reinstating the inverse
radius of the $S^3$ one finds
\vspace{2mm}

\framebox[\width]{
        \begin{minipage}{5.6in}
\bea
\< \cao_{S^{4 0}}  \> &=& \frac{\sqrt{3} N^2 }{\sqrt{5} \pi^2}
\int_{R^2} \Delta z  (3 r^{4} - 4 a^2 r^2) r dr d \f; \label{vev4} \\
&=& \frac{\sqrt{3} N^2 }{\sqrt{5} \pi^2} \left ( \int_{R^2} \rho  (3
r^{4} - 4 a^2 r^2 + a^4) r dr d \f \right ); \nn \\
\< \cao_{S^{4 \pm 2}}  \> &=& \frac{4 \sqrt{3} N^2} {\sqrt{10} \pi^2}
e^{ \mp 2 i a t} \int_{R^2} \rho (r^{4} - a^2 r^2) e^{\pm 2 i  \f}
r dr d \f. \nn
\eea
\end{minipage}}

\vspace{2mm}
\noindent
The general structure of the vevs is thus
\be \la{generic1}
\< \cao_{S^{k m}}  \> = N^2 e^{- i m a t } \sum_{l=0}^{\half (k - \left |m \right
  |)} \a_l \int_{R^2} \rho (r^{k-2l} a^{2l}) e^{im \f} r dr d\f,
\ee
with certain coefficients $\a_l$. These vevs can also be written in the
form
\be
\< \cao_{S^{k m}}  \> = {\cal M}_{k} e^{- i a m t} (\Delta \Phi)_{km}
+ \a_{km} a^2 \< \cao_{S^{(k-2) m}}  \>
+ \b_{km} a^4 \< \cao_{S^{(k-4) m}}  \> + \cdots
\ee
where $({\cal M}_{k}, \a_{km})$ are appropriate constants. In the $a
\rightarrow 0$ limit only the first term survives, and as will discuss
below one recovers the Coulomb branch result. Since $k \ge |m|$
the vevs of maximally operators only receive contributions from the
first term. Our explicit computations go up to dimension four, but if
one assumes this structure persists in the vevs of higher dimension
operators then
the result \eqref{ch-1} holds for maximally charged operators of all
dimension. We will find that the field theory result does indeed
reproduce \eqref{ch-1} for all $k$, thus verifying this hypothesis.
By contrast the vevs of
non-maximally charged operators do receive other contributions and thus
one needs to calculate explicitly the appropriate coefficients
$(\a_{km},\b_{km},\cdots)$.

\bigskip

These expressions make manifest the
limiting behavior as $a \rightarrow 0$ and the theory passes to
that of the Coulomb branch of ${\cal N} = 4$ on $R^{3,1}$.
The R-charge and the energy as given in \eqref{j-en} vanish in this
limit, as expected for supersymmetric vacua of
${\cal N} = 4$ on $R^{3,1}$.
However, the scalar chiral primary vevs
remain non-trivial for appropriate density functions
$\rho(r,\phi)$. Each density function describing a regular bubbling geometry
consists of $N$ droplets $d_i$, such that $\rho(r,\phi)$ takes the value
$1/\pi a^2$ on the droplet, and the area of each droplet is $\pi
a^2$. Suppose the boundary of the droplet is described by $r = r_i + d_{i}
(\f)$, with $r_i$ constant and some suitable function $d_i (\phi)$. Then the density
function describing the droplet is
\be
\rho_{d_i}(r,\phi) = \frac{1}{\pi a^2} \q (r_i + d_{i}(\f) - r),
\ee
such that
\be \la{pob}
\int_{d_i} \rho_{d_i}(r,\phi) r dr d\f = \frac{1}{N}.
\ee
Coulomb branch solutions are then obtained in the limit that $r_i$
stays finite as $a \rightarrow 0$: the density function for each
droplet behaves as
\be
\rho_{d_i}(r,\phi) \rightarrow \frac{1}{N} \d (x^1 - x^1_{i}) \d(x^2 - x^2_i),
\ee
satisfying \eqref{pob}. Here $(x^1_i,x^2_i)$ describe the location of
the droplet in the 1-2 plane, and in this limit
each of the $N$ droplets is associated with
an eigenvalue of the matrices $(X^1,X^2)$. Clearly in the $a
\rightarrow 0$ limit the disc density function describing the
conformal vacuum becomes a delta function localized at the origin,
$\rho(r,\phi) \rightarrow \d (x^1) \d (x^2)$.

Now taking the $a \rightarrow 0$ limit in the vevs of the scalar
chiral primaries one gets
\bea
\< \cao_{S^{k m}}  \> &=& \frac{N^2}{\pi^2} \frac{(k-2)}{2^{k/2}
  (k+1)} \sqrt{\frac{(k-1)}{k}} (\Delta \Phi)_{km}; \la{Coulomb} \\
&=&  \frac{N^2}{\pi^2} 2^{k/2} (k-2)
\sqrt{\frac{(k-1)}{k}} \int_{R^2} dx^1 dx^2 \rho(x^1,x^2) (C^{km}_{i_1 \cdots i_k}
x^{i_1} \cdots x^{i_k}), \nn
\eea
in exact agreement with the Coulomb branch vevs
given in \cite{Skenderis:2006di}, restricting to an $SO(4)$
invariant distribution.

\section{Dual description} \label{dual_qft}

In this section we will consider half BPS states in ${\cal N} = 4$ SYM,
and their relation to free fermions. We discuss the
correspondence between an arbitrary half BPS state and a
two-dimensional density distribution, which in turn is to be
identified with the defining density function of the bubbling supergravity
solution. In particular, we show that the state is not completely
determined by this density distribution, but the density distribution
does determine uniquely the vevs of all single trace chiral primary
operators. These
in turn are precisely the information that is encoded in the
asymptotics of the LLM solutions. Thus one would anticipate that the
LLM solutions receive higher order corrections, involving information
beyond the density function, which capture the dual state uniquely.

\subsection{Half BPS states in ${\cal N} = 4$ SYM}

There is a one-to-one correspondence between half BPS
$SO(4)$ symmetric representations
of ${\cal N} = 4$ SYM and symmetric polynomials in the eigenvalues of
a complex matrix $Z$ or Schur polynomials. Here $Z$ is one
combination of the six Hermitian scalars $X^m$ of ${\cal N} = 4$ SYM, given
by $Z = X^1 + i X^2$.

There are several choices of basis for the gauge invariant multi-trace
polynomials of $Z$:
\begin{enumerate}
\item{The {\bf trace basis} of products of traces of $Z$ is an obvious
  gauge invariant basis. For the group $U(N)$ the multitraces can be
  labelled by $p(n)$ conjugacy classes of the permutation group $S_n$
  where $p(n)$ is the number of partitions of $n$. Labeling
  representatives of different conjugacy classes of $S_n$ by $\s_I$,
  the basis of multi-trace operators is given by ${\rm{Tr}}(\s_I Z)$:
\be
{\rm{Tr}}(\s_I Z) = \sum_{j_1 \cdots j_n} Z^{j_1}_{j_{\s_{I}(1)}}
  Z^{j_2}_{j_{\s_{I}(2)}} \cdots Z^{j_n}_{j_{\s_{I}(n)}}.
\ee
For $SU(N)$ $Z$ is traceless and one must therefore restrict to
elements of $S_n$ without $1$-cycles; the distinction between $U(N)$
and $SU(N)$ is however not important in the $N \rightarrow \infty$
limit relevant here.}

\item{The {\bf Schur polynomial basis} is, in the case of $U(N)$,
a sum over these trace operators, weighted by the characters of $\s$ in the representation
  $R$ of $S_n$, namely
\be
\chi_{R}(Z) = \frac{1}{n!} \sum_{\s \in S_n} \chi_{R}(\s) {\rm{Tr}}(\s
  Z).
\ee
The representations $R$ can be labeled by Young diagrams with $n$
  boxes, which correspond to partitions of $n$ and there are thus
  $p(n)$ Schur polynomials of degree $n$. An advantage of this basis is
  that the two-point functions are diagonal. Again modifications are
  needed for the case of $SU(N)$, but these give $1/N$ effects which
  will not be relevant here.}
\end{enumerate}
Note that another useful basis is the dual basis, dual to the
trace basis, but this will not play a role here.

An arbitrary half BPS state $|\Phi \rangle$
preserving $SO(4)$ R symmetry can
therefore be written as a superposition of states
\be
| \Phi \rangle = \sum_{R} a_{R} \chi_{R}(Z) | \Omega \rangle
= \sum_{I} b_{I} {\rm{Tr}}(\s_I Z)  | \Omega \rangle,
\ee
for suitable (complex) coefficients $a_R$ and $b_I$,
with $| \Omega \rangle$ being the conformal vacuum. Denoting by ${\cal
  O}^{\cal A} $ the set of gauge invariant operators, the vevs of these
operators in
the state $|\Phi \rangle$ are given by
\bea
\langle {\cal O^{\cal A}} \rangle_{\Phi} &=& \sum_{R,R'} a_{R}^{\ast} a_{R'}
\langle \Omega | (\chi_R (Z))^{\dagger} {\cal O}^{\cal A} \chi_{R'}(Z) | \Omega
\rangle; \la{tre} \\
&=& \sum_{I,J} b^{\ast}_{I} b_{J} \langle \Omega | ({\rm{Tr}}(\s_I
Z))^{\dagger} {\cal O}^{\cal A} {\rm{Tr}}(\s_J Z)  | \Omega \rangle. \nn
\eea
A state is an eigenstate of the dilatation operator and of
the R-symmetry (in the 1-2 directions) with eigenvalue $n$
if and only if the superposition involves only $S_n$. That is, only
operators involving $n$ fields $Z$ are included in the superposition.

\bigskip

It is also important to note that in the $N \rightarrow \infty$ limit
there are considerable simplifications in three point functions
appearing in \eqref{tre}. Let us consider first computations in the
trace basis, where the operators are normalized as
$C_{\s^{I}_n} {\rm Tr} (\s_I^n Z)$ with $n$ the dimension.
The normalization factors  $C_{\s^I_n}$ are such that the basis
is orthonormal in the large $N$ limit, namely
\be
C_{\s^{I}_n} C_{\s^{J}_m}
\langle \Omega | ({\rm{Tr}}(\s^n_I
Z))^{\dagger} {\rm{Tr}}(\s^m_J Z) | | \Omega \rangle = \d_{IJ} \d^{nm}
+ {\cal O}(1/N),
\ee
and the large $N$ scaling of $C^2_{\s^{I}_n}$ is $1/N^{n}$.

Now consider the three point functions \eqref{tre} in which
the operators ${\cal O}^{\cal A} $ are single trace operators built from the six scalar
fields $X^m$. These are clearly the relevant operators to compare with the
holographic results. As discussed in \cite{D'Hoker:1999ea}
three point functions which are extremal, so
that the conjugate operator has a dimension which is the sum
of the dimensions of the other operators, and those which are
non-extremal are known to have different large $N$ behavior.
Since the state $|\Phi \rangle$ is a sum of terms each of which
is maximally charged,
i.e. it has $j=\D$, it follows that the 3-point functions
are never extremal when ${\cal O}^{\cal A}$ is not maximally charged.
Moreover
the large $N$ behavior depends on whether the other operators in the
correlator are single or multi-trace. We discuss in appendix \ref{new} the large $N$
behavior of such correlators, and summarize here the relevant results:

{\bf Non-extremal correlators:} Non-extremal three point functions
for which ${\cal O}^{\cal A} $ are (orthonormal) single trace
operators scale as $1/N$ or smaller in the large $N$ limit. Note that this
assumes that the dimensions of all operators in the correlator are
small compared to $N$. In the
case that the operator ${\cal O}^{\cal A} $ is neutral under
 $SO(2) \times SO(4)$ R symmetry the correlators behave as $1/N$ only for the
diagonal terms, namely when $\s_{I}^{n} = \s_{J}^m$. When the operator
 ${\cal O}^{\cal A} $ has a non-maximal $SO(2)$ charge $m$ the correlators involving
single trace operators still behave as $1/N$; correlators involving
multi-trace operators are generically subleading in $N$, but for
special cases can also behave as $1/N$.

{\bf Extremal correlators:} Extremal three point functions scale as
one or smaller in the large $N$ limit. Correlators involving only
single trace operators behave as $1/N$, whilst multi-trace correlators
in which ${\rm{Tr}} (\s^n_I Z) = {\cal O}^{\cal A} {\rm {Tr}}(\s^m_J Z)$
(and therefore $\s^n_I$ is necessarily multi-trace) are of order one.

One can also rephrase these results in terms of the Schur polynomial
basis, as discussed in \cite{Corley:2001zk}. As we will show in section
\ref{dop1}, the vev of the $SO(2) \times SO(4)$ neutral
operators in the state built from a given Schur polynomial of
dimension $n$ is independent of the choice of Schur polynomial. To
leading order in $N$ it behaves as $n/N$. This result has an
immediate corollary: consider geometries dual to different
superpositions of the Schur polynomials, all of the same dimension $n$. Then
symmetry implies only the neutral operators acquire vevs but
these vevs differ only by $1/N$ effects, so
these geometries are not reliably distinguishable within supergravity.

Now consider a superposition of states of different dimension (and
thus R charge). In such a case $SO(2)$ charged single trace operators
acquire vevs, and the computation of the vevs of
maximally charged operators necessarily involves extremal
correlators. The $N$ scalings of these vevs depend crucially on the specific
Schur polynomials, or equivalently multi-trace operators, appearing in
the superposition. Superpositions involving single trace operators
will lead to vevs which are suppressed by $1/N$ relative to
multi-trace superpositions, and thus these are immediately distinguishable.
An explicit example illustrating this effect will be discussed in section \ref{dop2}.

\subsection{Relation to free fermions}

Consider irreducible representations of the symmetry group
$S_n$. These may be characterized by a sequence of non-negative
integers $\{\l\} = (\l_1,\cdots,\l_N)$ with $\l_{1} \ge \l_2 \ge
\cdots \ge \l_N \ge 0$ and $\sum_{i=1}^N \l_i = n$. The sequence
defines a Young tableau with the number of boxes in the $i$th row
being $\l_i$ and the total number of boxes being $n$; let $\chi^{n}_{\{\l
  \}} (Z)$ be the corresponding Schur polynomial. In this section we
will review the relationship between Schur polynomials and free
fermions.

We introduce a second quantized free fermion field
\be
\Psi(z,z^{\ast},t) = \sum_{l=0}^{\infty} \hat{C}_l e^{-i (l+1) t} \Phi_l(z,z^{\ast})
\ee
where $(\hat{C}_{l}, \hat{C}_l^{\dagger})$ satisfy
the anti-commutation relation $\{\hat{C}_l, \hat{C}_{m}^{\dagger} \} =
\d_{lm}$. Note that throughout this section we will set the inverse
radius $a$ of the $S^3$ to one; this sets the mass scale in the matrix
model, and thus of the fermions, to one.
The functions $\Phi_l(z,z^{\ast})$ are the orthonormal wavefunctions of
the lowest Landau level, and are given by
\be
\Phi_{l} (z,z^{\ast}) = \sqrt{\frac{2^{l + 1}}{\pi l!}} z^{l} e^{-  z z^{\ast}}.
\ee
The fermion field $\Psi(z,z^{\ast},t)$ satisfies the constraint that
the total number of fermions be $N$,
\be
\int dz dz^{\ast} \Psi^{\dagger} (z,z^{\ast},t) \Psi(z,z^{\ast},t) =
\sum_{l=0}^{\infty} \hat{C}^{\dagger}_l \hat{C}_l = N.
\ee
The ground state is denoted $| \Omega \rangle$ and is given by
\be
| \Omega \rangle = \hat{C}^{\dagger}_{N-1} \hat{C}^{\dagger}_{N-2}
\cdots \hat{C}^{\dagger}_1 \hat{C}^{\dagger}_0 | 0 \rangle,
\ee
where $| 0 \rangle$ is the Fock vacuum defined by $\hat{C}_l | 0
\rangle = 0$ for all $l$. We will denote by $| \Phi \rangle$ a generic state
containing N fermions; each such state can also be expressed as a
superposition of Schur polynomials. The Schur polynomial
$\chi^{n}_{\{\l \}} (Z)$ corresponds to the state
\be \la{schur}
\hat{C}^{\dagger}_{N-1+ \l_1} \hat{C}^{\dagger}_{N-2 + \l_2} \cdots
\hat{C}^{\dagger}_{1 + \l_{N-1}} \hat{C}^{\dagger}_{\l_N} | 0 \rangle.
\ee
Now consider the expectation value of the density function defined as
\be
\hat{U} (z,z^{\ast},t) = \Psi^{\dagger}(z,z^{\ast},t) \Psi(z,z^{\ast},t).
\ee
In the conformal vacuum
\bea
\langle \hat{U} (z,z^{\ast},t)
\rangle_{\Omega} &=& \sum_{l=0}^{N-1}
\Phi_{l}^{\ast} (z,z^{\ast}) \Phi_{l} (z,z^{\ast}); \\
 &=& \sum_{l=0}^{N-1}
\frac{2^{l+1}}{\pi l!} (z {z}^{\ast})^l e^{-2 z z^{\ast}}
\equiv  2 \pi^{-1} e^{-2 z z^{\ast}} E_{N-2} ( 2 z z^{\ast}), \nn
\eea
where by definition
\be
E_{N-1} (y) = \sum_{l=0}^{N} \frac{y^l}{l!}.
\ee
For $N \gg 1$,
\be
e^{-y} E_{N-1} (y) \rightarrow \q (N - y) + f(y),
\ee
where $\q(x) = 1$ for $x \ge 0$ and $\q(x) = 0$ for $x < 0$. The
function $f(y)$ describes the smearing of the step function; $f(y)$
has support only within a region around $y=N$ of width of order one
and is such that $f(y) < 0$ for $y < N$, $f(y) > 0$ for $y > N$ with
\be
\int_{0}^{\infty} dy f(y) = 0; \hsp \int_{0}^{\infty} dy | f(y) | =
    {\cal O}(1).
\ee
Thus to leading order as $N \rightarrow \infty$
\be
\langle \hat{U} (z,z^{\ast}) \rangle_{\Omega} = \frac{2}{\pi} \q (N - 2 |z|^2).
\ee
To compare with the supergravity results one therefore needs
\be \la{com1}
|z| = \sqrt \frac{N}{2} |w|; \hsp \langle \hat{U} \rangle_{\Omega} = 2 \rho,
\ee
with $|w|$ identified with the supergravity coordinate $r$.
In a generic state $| \Phi \rangle$ the density function is given by
\bea
\langle \hat{U} (z,z^{\ast},t) \rangle_{\Phi} &=& \sum_{l,m}
\Phi_{l}^{\ast}(z,z^{\ast}) \Phi_{m}(z,z^{\ast}) e^{i (l-m) t} \langle
\hat{C}^{\dagger}_{l} \hat{C}_m \rangle_{\Phi}; \la{genericity} \\
&=& \sum_{l,m} e^{i (l-m) t} (z^{\ast})^l z^m \sqrt{
  \frac{2^{2 + l+m}}{\pi^2 l! m!}} e^{-2 z^{\ast} z} U^{\Phi}_{lm},\nn
\eea
where we define
\be
U^{\Phi}_{lm} = \langle
\hat{C}^{\dagger}_{l} \hat{C}_m \rangle_{\Phi}.
\ee
The supergravity density function is related to this via
\be \la{com2}
\langle \hat{U} (t=0) \rangle_{\Phi} = 2 \rho.
\ee

\subsection{Extracting the state from a distribution}

In this section we consider how to derive
the specific superposition of Schur polynomials corresponding to a
given distribution. Recall that the expectation value of the
density function given in \eqref{genericity} is
\be
\langle \hat{U} (z,z^{\ast},t) \rangle_{\Phi} = \sum_{l,m}
\Phi_{l}^{\ast}(z,z^{\ast}) \Phi_{m}(z,z^{\ast}) e^{i (l-m) t} \langle
\hat{C}^{\dagger}_{l} \hat{C}_m \rangle_{\Phi}.
\ee
Thus by integrating a given density function with respect to a
suitable basis of orthonormal polynomials one can
extract the coefficients $\langle
\hat{C}^{\dagger}_{l} \hat{C}_m \rangle_{\Phi}$. Note however that
these expansion functions $\Phi_{l}^{\ast} \Phi_{m}$ are not
orthogonal, when integrated over the plane with unit measure.
So in practice it is actually more convenient to work with the
density function in phase space, $\hat{u}(p,q,t)$, which
is naturally expanded in a useful basis of orthonormal functions.
The explicit relationship between the density functions $\hat{U} (z,z^{\ast},t)$
and $\hat{u}(p,q,t)$ was given in \cite{Takayama:2005yq}:
\be
\hat{U} (z,z^{\ast},t) = \int \frac{d \Lambda d \Lambda^{\ast}}{4
  \pi^2} e^{- \Lambda^{\ast} z + \Lambda z^{\ast} - \qu \Lambda
  \Lambda^{\ast}} \int {dp dq} e^{- \Lambda (q +
  ip) +  \Lambda^{\ast} (q - ip)} \hat{u}(p,q,t).
\ee
Then, following \cite{Takayama:2005yq} one finds that
\be
\int dz dz^{\ast} {(-1)^k}
(z^{\ast})^j z^{k} \hat{U} =
\frac{\pa^{j+k}}{\pa \Lambda^j \pa \Lambda^{\ast k}}
\left (e^{-\qu \Lambda \Lambda^{\ast}} \int dp dq e^{-
  \Lambda (q + ip) + \Lambda^{\ast} (q - ip)}
  \hat{u} \right )_{\Lambda = \Lambda^{\ast} = 0}  \la{fff2}
\ee
Now to make manifest the behavior in the large $N$ limit one should
rescale these coordinates as in \eqref{com1} so that
\be
|z| = \sqrt {\frac{N}{2}} |w|; \qquad q = \sqrt{ \frac{N}{2}} x; \qquad
p = \sqrt{ \frac{N}{2}} y.
\ee
Retaining only the leading order terms in \eqref{fff2} as $N \rightarrow
\infty$ gives
\be
\int d^2w (w^{\ast})^j w^k \hat{U} = \int d^2x r^{j +k} e^{i (k-j)
  \f} \hat{u},
\ee
where $x = r \cos \f$, $y = r \sin \f$. As we have seen from the
holographic computations, and will discuss below,
all one point functions are expressed in terms of these integrals.
Thus at leading order in $N$ one can identify $|w| = r$ and
the difference between
the distributions $(\hat{U},\hat{u})$ is not visible.
Whilst it is more natural for the droplet distribution in the
bulk solution
to be identified with the phase space distribution, rather than the
$z$ space distribution, this is not distinguishable at leading order
in $N$.

In some calculations, such as that of one point functions which we
 will discuss below, it is more convenient to use the $z$ space
distribution, and exploit the simple form of its expansion in
exponentials. For the current purpose it is rather
more convenient to work with the phase
space distribution, since this is expanded in a natural basis of orthonormal
functions, the Laguerre polynomials. That is, the phase space distribution is given by
\cite{Takayama:2005yq}:
\bea
\pi \rho_{\Phi} &=& \sum_{m \le n} \sqrt{ \frac{m!}{n!}} (-1)^m \c^{\half
  (n-m)} e^{- \half \c} e^{i (m -n) \f} L_{m}^{n-m}(\c) \langle
\hat{C}^{\dagger}_m \hat{C}_n \rangle_{\Phi} \\
&& + \sum_{m >  n} \sqrt{ \frac{n!}{m!}} (-1)^n \c^{\half
  (m-n)} e^{- \half \c} e^{i (m -n) \f} L_{n}^{m-n}(\c) \langle
\hat{C}^{\dagger}_m \hat{C}_n \rangle_{\Phi}, \nn
\eea
where $\c = N r^2$ and $L_{m}^{n-m}(\c)$ is the Laguerre polynomial defined by
\be
L^{\a}_n(\c) = \sum_{p=0}^n (-1)^p \left (\begin{array}{c}
n + \a \\
n - p
\end{array} \right ) \frac{\c^p}{p!},
\ee
for which the orthogonality relation is
\be
\int_{0}^{\infty} d\c e^{-\c} \c^{\a} L^{\a}_{n}(\c) L^{\a}_{m}(\c) = \d_{mn}
\frac{(n+\a)!}{n!},
\ee
for integral $\a$.
In the conformal vacuum one gets
\be
\pi \rho_{\Omega} = \sum_{m=0}^{N-1} (-)^m e^{- \half \c} L_{m} (\c)
\ee
where $L_m(\c) \equiv L_m^0( \c)$. Using the identity
\be
\int_{0}^{\infty} d\c e^{-\half \c} L_{m}(\c) = 2 (-1)^m
\ee
one can show that this satisfies the normalization condition
$\int d^2x \rho_{\Omega} = 1$. Moreover
for large $N$ the distribution asymptotes as before to a disc,
$\pi \rho_{\Omega} \rightarrow \q (1 - r)$.

Using the orthogonality relation for the Laguerre polynomials
one can now extract the $\langle
\hat{C}^{\dagger}_m \hat{C}_n \rangle_{\Phi}$ via:
\bea \la{project}
\langle \hat{C}^{\dagger}_m \hat{C}_{m+p} \rangle_{\Phi}
&=& (-1)^m \qu \sqrt{\frac{(m+p)!}{m!}}
\int d \phi e^{i p \f} d \c \c^{p/2} e^{- \c/2} L^{p}_m (\c) \rho_{\Phi}; \\
\langle \hat{C}^{\dagger}_{m+p} \hat{C}_{m} \rangle_{\Phi}
&=& (-1)^m \qu \sqrt{\frac{(m+p)!}{m!}}
\int d \phi e^{- i p \f} d \c \c^{p/2} e^{- \c/2} L^{p}_m (\c) \rho_{\Phi}, \nn
\eea
where $(m,p) \ge 0$.

Thus from the distribution $\rho_{\Phi}$ one can extract the complete
set of $\langle \hat{C}^{\dagger}_m \hat{C}_n \rangle_{\Phi}$. Let us
now discuss whether knowledge of these is in principle
sufficient to determine the state $|
\Phi \rangle$\footnote {Related discussions appeared in the recent paper
  \cite{Balasubramanian:2007zt}.}.
Consider first the case where $| \Phi \rangle$ has
definite dimension $n$, so that
\be \la{phg}
| \Phi \rangle = \sum_{\{ \l \} } a_{ \{ \l \}} | n; \{ \l \}
\rangle.
\ee
Normalization of the state implies that $ \sum_{\{ \l \} } | a_{ \{ \l
  \}}|^2 = 1$. In such a state
$\langle \hat{C}^{\dagger}_m \hat{C}_p \rangle_{\Phi}$ is non-zero
only for $m = p$ and
\be
\langle \hat{C}^{\dagger}_m \hat{C}_m \rangle_{\Phi} = \sum_{\{ \l \}
} | a_{ \{ \l \}}|^2 \d_{ \{ \l \} m},
\ee
where $\d_{ \{ \l \} m} = 1$ iff the corresponding state contains a
fermion at level $m$. Therefore one cannot extract the phases of
$a_{ \{ \l \}}$ from this information: the
density function is not sufficient to completely determine the
state. There is one exception to this: when precisely $N$
of the $\langle \hat{C}^{\dagger}_m \hat{C}_m \rangle_{\Phi}$ are
non-zero and equal to one, the corresponding state is necessarily a
single Schur polynomial. In this case the summation in \eqref{phg}
collapses to one term, and the overall phase of the state plays no
role. Note that this is precisely the case that was discussed in
\cite{Balasubramanian:2006jt}, but for a general state of definite
dimension the distribution is not sufficient to determine the state.
To determine the phases in the general case one would need to know
in addition
\be \la{poty}
\langle \prod_{i} \hat{C}^{\dagger}_{m_i} \prod_{j} \hat{C}_{m_j}
\rangle_{\Phi}, \qquad \sum_{i} m_i = \sum_{j} m_j.
\ee
It will be made manifest in the next section
that $\langle \hat{C}^{\dagger}_m \hat{C}_p \rangle_{\Phi}$
determines the expectation values of single trace operators in the
state, see \eqref{single-trace}, whilst \eqref{poty} is related to the expectation values of
multi-trace (neutral) operators.

Note that the discussion so far has made no restriction on $N$. Even
if one can determine the  $\langle \hat{C}^{\dagger}_m \hat{C}_m
\rangle_{\Phi}$ exactly one can still not determine the state. Of
course when one takes the $N \rightarrow \infty$ limit and sharpens
the distribution such that it gives a regular supergravity solution
the situation will be worse. One will not be able to determine the
coefficients $\langle \hat{C}^{\dagger}_m \hat{C}_m
\rangle_{\Phi}$ exactly, and thence one can only determine the leading
behavior in $N$ of the $ | a_{ \{ \l \}}|^2 $.

Now consider a general state $|\Phi \rangle$ which does not have a
definite dimension, so that
\be
| \Phi \rangle = \sum_{n, \{ \l \} } a_{n, \{ \l \} } | n; \{ \l \} \rangle.
\ee
The neutral vevs $\langle \hat{C}^{\dagger}_m \hat{C}_m
\rangle_{\Phi}$ still do not determine the phases of the
$a_{n, \{ \l \} }$. However, some phase information is obtained via
the vevs of charged operators. That is,
\bea
\langle \hat{C}^{\dagger}_{m+p} \hat{C}_m
\rangle_{\Phi} &=& \sum_{n, \{ \l \}}  a_{n + p, \{ \l \}_p}^{\ast}
a_{n, \{ \l \} } \langle n + p; \{ \l \}_p |
\hat{C}^{\dagger}_{m+p} \hat{C}_m | n; \{ \l \} \rangle, \\
&=& \sum_{n, \{ \l \}}  a_{n + p, \{ \l \}_p}^{\ast}
a_{n, \{ \l \} }, \nn
\eea
where the Schur polynomial $  | n + p; \{ \l \}_p \rangle$ is
precisely $\hat{C}^{\dagger}_{m+p} \hat{C}_m | n; \{ \l \}
\rangle$. That is, the state $  | n + p; \{ \l \}_p \rangle$
differs from $ | n; \{ \l \} \rangle$ by only one fermion. So only a
subset of the phase information is obtained; this is sufficient to
determine the state when the superposition contains Schur polynomials
such that each of which differs by only one fermion
from at least one other polynomial in the superposition. If however the state
contains at least one Schur polynomial which differs by two fermions or more from
all other Schur polynomials in the superposition, then the phase of
the coefficients of these terms cannot be determined without vevs
of multi-trace operators. Thus for a general distribution
one would again need vevs of multi-trace operators to determine the coefficients in $| \Phi
\rangle$ uniquely.

The supergravity solutions are constructed entirely out of the density
function $\rho_{\Phi}$ and therefore contains information only about the
expectation values of single trace operators. To determine the state
$|\Phi \rangle$ one needs the expectation values of all other
operators, which are not determined by $\rho_{\Phi}$. Therefore, one
would expect that the higher order corrections to the
LLM bubbling solution, apart from correcting  the distribution
$\rho_{\Phi}$, would also involve additional information
so that the corrected solution captures the entire vacuum structure.
This is in line with the fact that the IIB supersymmetry rules
are expected to receive non-trivial higher derivative corrections.

\section{Computation of vevs}

In the previous section we set up the correspondence between a general
half BPS state and a density distribution. The information extracted
holographically is the vevs of chiral primary operators, and in this
section we will discuss how these vevs may be computed in an arbitrary
half BPS state. We compute explicitly vevs of all operators up to
dimension four, and the vevs of maximally charged operators of
arbitrary dimension, and find exact agreement with the holographic
results. These results are a detailed confirmation of the correspondence
between LLM bubbling solutions and 1/2 BPS states, and moreover
provide evidence that the vevs are not renormalized,
as one might anticipate given the sixteen preserved supercharges.

\subsection{Energy and R-charge}

In a generic state $| \Phi \rangle$ the energy and R-charge $J$
relative to that of the conformal vacuum is
\be
(E - E_c) = J = \sum_{m} m \langle \hat{C}^{\dagger}_m \hat{C}_m
\rangle_{\Phi} - \sum_{m =0}^{N-1} m =
\sum_{m} m \langle \hat{C}^{\dagger}_m \hat{C}_m
\rangle_{\Phi} - \half N (N-1) ,
\ee
where $E_c = 3N^2/16$ is the Casimir energy on $R \times S^3$. This
Casimir is clearly not reproduced correctly by the matrix model, since
there are contributions from all KK modes on the $S^3$ of all SYM
fields. Now note that
\be
\int d^2z |z|^{2k} \langle \hat{U}(z,z^{\ast},t) \rangle_{\Phi} =
\sum_{m} \frac{(m+k)!}{2^k m!} \langle \hat{C}^{\dagger}_m \hat{C}_m
\rangle_{\Phi}.
\ee
Thus
\be
\sum_{m} m \langle \hat{C}^{\dagger}_m \hat{C}_m
\rangle_{\Phi} = \int d^2z (2 |z|^{2} - 1)
\langle \hat{U}(z,z^{\ast},t) \rangle_{\Phi},
\ee
and hence
\bea
(E - E_c) = J & = & \int d^2z (2 |z|^{2} - \half (N+1))
\langle \hat{U}(z,z^{\ast},t) \rangle_{\Phi}; \\
&=& N^2 \int d^2 w ( |w|^2 - \half (1 + \frac{1}{N})) \rho, \nn
\eea
which agrees with the holographic result \eqref{j-en}, after taking the $N
\rightarrow \infty$ limit.

\subsection{Vevs of maximally charged operators}

We now consider how the vevs of single trace operators ${\cal
  O}^{\Delta,  k}$ of dimension $\Delta$ and $SO(2)$ charge $k$ may be computed.
Here we will use ${\cal O}^{\Delta,  k}$ to denote the operator in field theory,
  whose vevs we compute within free field theory, whilst
${\cal O}_{S^{\Delta k}}$ refers to the corresponding operator whose vevs at strong
  coupling were computed holographically.

Let us begin by computing the vev of the maximally charged single
trace scalar operator of dimension $k$, ${\cal O}^{k,k}$,
in a generic state.
The operators ${\cal O}^{k,k}$ are implemented as
follows:
\be \la{imp-1}
{\cal O}^{k,k}
= {\cal N}_{k} \l_{k,k} e^{i k t} \sum_{l=0}^{\infty}
\sqrt{\frac{(l+k)!}{l!}} \hat{C}^{\dagger}_{l+k} \hat{C}_l.
\ee
The factor $\l_{k,k}$ is chosen such that the two point function satisfies the
following normalization condition:
\be
\langle ({\cal O}^{k_1,k_1})^{\dagger} (t_1)
{\cal O}^{k_2,k_2} (t_2) \rangle =
{\cal N}_{k_1}^2 \d_{k_1 k_2}.
\ee
where ${\cal N}_{k}$ is defined in \eqref{no2}. ${\cal N}_{k}$ is
the appropriate
normalization for the two point functions extracted holographically.
The normalization condition implies that
\be
\l_{k,k}^{-2} = \sum_{N-k}^{N-1} \frac{ (l+k)!}{l!} = \frac{1}{(1+ k)}
\left (\frac{ (N+k)!}{ (N-1)!} - \frac{N!}{(N - k - 1)!} \right )
\stackrel{N \rightarrow \infty}{\rightarrow} N^{k} k + {\cal O}(N^{k-1}).
\ee
The corresponding integral representation of the operators is therefore
($k \neq 1$)
\bea
{\cal O}^{k,k} (t)
&=& {\cal N}_{k} \l_{k,k} 2 ^{\frac{k}{2}} \int dz dz^{\ast} z^{k}
\hat{U} (z,z^{\ast},t); \\
&=& \frac{N}{\pi^2 \sqrt{k}}  \sqrt{k-1} (k-2) 2^{\frac{k}{2}}
N^{-\frac{k}{2}} \int dz dz^{\ast} z^{k} \hat{U} (z,z^{\ast},t). \nn
\eea
The expectation value of this operator in a generic state $| \Phi
\rangle$ is then given by
\be
\langle {\cal O}^{k,k} (t)
\rangle_{\Phi} =
\frac{N}{\pi^2 \sqrt{k}}  \sqrt{k-1} (k-2) 2^{\frac{k}{2}}
N^{-\frac{k}{2}} \int dz dz^{\ast} z^{k} \langle \hat{U}
(z,z^{\ast},t) \rangle_{\Phi}.
\ee
The integral may be rewritten using
using (\ref{com1},\ref{com2}) as
\be
\langle {\cal O}^{k,k} (t) \rangle_{\Phi} =
\frac{N^2}{\pi^2 \sqrt{k}}  \sqrt{k-1} (k-2) e^{i k t} \int d^2w e^{i
  k \phi} |w|^{k} \rho,
\ee
in exact agreement with the holographic result \eqref{ch-1}.

Consider the $k=1$ operator. Here the distinction between
$U(N)$ and $SU(N)$ becomes important: the vanishing of the trace in the latter means
that there is no dimension one operator, ${\cal O}^{1,1}$.
This constraint can be incorporated here by restricting to
configurations in which
\be \la{11van}
\langle {\cal O}^{1,1} \rangle = 0 = \int d^2 w w \rho,
\ee
which is indeed the condition imposed on the holographic
distribution.

\subsection{Other scalar chiral primaries}

Now let us consider the remaining scalar chiral primary operators
${\cal O}^{\Delta k}$, which are not maximally charged, $\Delta > |
  k|$. The action of
such an operator on the conformal vacuum $| \Omega \rangle$ or any other
1/2 BPS state $| \Phi \rangle$ built from Schur polynomials
creates a state which cannot be described in terms of Schur polynomials.
The reason is that the other scalar operators contains not only the
  scalar fields $(Z,\bar{Z})$ but also the remaining four
${\cal N} = 4$ SYM scalar fields. The latter are not contained in the
  Schur polynomials, and thence not in the free fermion description.

Suppose however one wishes to compute one point functions of these scalar
operators in a state $| \Phi \rangle$. To do so, following \eqref{tre},
one needs to know three point functions between such an operator and two maximally charged
operators. Consider the computation of such a three point function in
free field theory; at tree level the computation actually
only involves the fields $(Z,\bar{Z})$. Take for example a three point
function such as
\be
\langle {\rm{Tr}} (\bar{Z})^k(x) {\cal O}^{2p,0} (y)  {\rm{Tr}} (Z)^k(z) \rangle,
\ee
for which the single trace $SO(2) \times SO(4)$ singlet operator
${\cal O}^{2p,0}$ has the structure
\be \la{ex1}
{\cal O}^{2p,0} = a_1 {\rm{Tr}} \left ( (Z \bar{Z})^p + \cdots
\right ) +
a_2 {\rm{Tr}} \left ( (Z \bar{Z})^{p-2} R^2 + \cdots \right ) + \cdots,
\ee
where the ellipses within the trace denote cyclic permutations,
$(a_1,a_2,\cdots)$ are constants and
$R^2 = \sum_{i=1}^4 (X_i)^2$ denotes collectively the
other scalars $X_i$ of ${\cal N} = 4$ SYM. Since the latter have no
propagators with the fields $(Z,\bar{Z})$ and cannot be
self-contracted, only the first term contributes in the three point
function \eqref{ex1}.

Therefore, one would anticipate being able to implement the scalar
operators with free fermions such that one can compute such one point
functions. One would not however expect to
be able to compute two point functions, or general higher point
functions of such operators, using the free fermion description.

\subsubsection{Neutral operators}

Suppose one implements the dimension two neutral operator as
\be
{\cal O}^{2,0} = {\cal N}_2 \l_{2,0} \sum_{m} \hat{C}_{m}^{\dagger} \hat{C}_{m} (m - \b).
\ee
Then the normalization factor $\l_{2,0}$ and the constant $\b$ should
be fixed such that the one point function of this operator vanishes in
the conformal vacuum; the three point function of the operator
with charged operators gives the correct ${\cal N} = 4$
results and the vev reduces to the Coulomb branch result as the radius
of the $S^3$ is increased.
Imposing the first constraint, $ \langle {\cal O}^{2,0} \rangle_{\Omega} = 0$,
implies that
\be
\b = N^{-1} \sum_{m=0}^{N-1} m =  \half (N-1).
\ee
Note that this value for $\b$ implies that the operator ${\cal
  O}^{2,0}$ annihilates the conformal vacuum,
${\cal O}^{2,0} | \Omega \rangle = 0$, and therefore the two point
function for this operator also vanishes. This makes manifest the point
made above, that one can only obtain the vevs of neutral operators
from the matrix model. The corresponding integral representation of
the operator is
\be
{\cal O}^{2,0} =  2
{\cal N}_2 \l_{2,0} \int dz dz^{\ast} (|z|^2 - \qu N)  \hat{U} (z,z^{\ast},t)
\ee
where subleading terms as $N \rightarrow \infty$ are dropped. The
normalization $\l_{2,0}$ is fixed by taking the flat space limit: only
the leading order term is retained, and comparison with the result
\eqref{Coulomb} gives
\be
\l_{2,0} = \frac{\sqrt{2}}{N \sqrt{3}}.
\ee
We have also checked explicitly that this result is consistent with
${\cal N} = 4$ three point functions, involving charged operators.
Using (\ref{com1}),(\ref{com2}) one can rewrite the vev as
\be
\langle {\cal O}^{2,0} \rangle = \frac{\sqrt{2}N^2}{\sqrt{3} \pi^2}
\int d^2 w (|w|^2 - \half) \rho,
\ee
in exact agreement with the holographic result (\ref{pofr}).

\bigskip

Now let us apply the same techniques to obtain expressions for the
vev of the dimension four neutral operator. For the operator
${\cal O}^{4,0}$ one gets
\be
{\cal O}^{4,0} = \frac{2 {\cal N}_4}{\sqrt{5} N^2} \sum \hat{C}_{m}^{\dagger}
\hat{C}_m (\frac{3 m^2}{4} + b_1 m + b_2).
\ee
Here the overall normalization is again fixed, so that the operator in the
integral representation gives the correct expression in the Coulomb
branch limit.
Imposing the vanishing of the vev in the conformal vacuum implies that
\be
2 N^3 + N^2 (4 b_1 -3) + N (8 b_2 - 4 b_1 + 1) = 0.
\ee
Calculating the three point function with single trace charged operators of
dimension two such that $\hat{s}_2 = {\cal N}_2^{-1} {\cal O}^{2,2}$ gives
\be
\langle {\hat{s}_2}^{\dagger} (t) {\cal O}^{4,0} \hat{s}_2 (t') \rangle
= \frac{2 {\cal N}_4}{\sqrt{5} N} \l_{2,2}^2 (6 N^2 + 4 N b_1 + \cdots ) =
\frac{2 {\cal N}_4}{ N \sqrt{5}},
\ee
where the ellipses denote terms which are subleading as $N \rightarrow
\infty$.
Then solving these equations to leading order in $N$ gives
\be
b_1 = - N; \hsp b_2 = \qu N^2.
\ee
These values can be shown to also be consistent with other three
point functions involving different charged operators.
In integral representation this implies that
\be
{\cal O}^{4,0} =  \frac{2 {\cal N}_4}{\sqrt{5} N^2} \int dz
dz^{\ast} (3 |z|^4 - 2 N |z|^2 + \qu N^2) \hat{U} (z,z^{\ast},t),
\ee
where again only leading terms as $N \rightarrow \infty$ are retained,
and using \eqref{com1} this gives
\be
\langle {\cal O}^{4,0} \rangle = \frac{\sqrt{3} N^2}{\sqrt{5} \pi^2}
\int d^2 w (3 |w|^4 - 4 |w|^2 + 1 ) \rho,
\ee
which agrees with the holographic result (\ref{pofr}).

\subsubsection{Charged operators}

Now let us treat the charged operators in a similar fashion.
For the dimension three operator
\be
{\cal O}^{3,1} = \frac{2 {\cal N}_3}{N^{3/2}} \int dz dz^{\ast} z
(|z|^2 + c_3) \hat{U}(z,z^{\ast},t),
\ee
which implies in the fermion representation
\be
{\cal O}^{3,1} = {\cal N}_3 \frac{1} {\sqrt{2} N^{3/2}} e^{i t} \sum_{m}
\left ( (m+2 + 2 c_3 ) (m+1)^{1/2} \right )
\hat{C}_{m+1}^{\dagger} \hat{C}_m.
\ee
Now we use the three point function (\ref{3pf-res}) to fix the
coefficient $c_3$; to leading order in $N$ this gives
\be
c_{3} = - \half N,
\ee
and thus
\be \la{31}
\langle {\cal O}^{3,1} \rangle = \frac{N^2}{\pi^2} e^{i t} \int d^2 w w (
|w|^2 - 1) \rho.
\ee
However the constraint \eqref{11van}
implies that the second term in (\ref{31}) vanishes, and thus
that the holographic result (\ref{pofr}) is reproduced.

For the dimension four operator
\be
{\cal O}^{4,2} = \frac{8 {\cal N}_4 }{\sqrt{10} N^{2}} \int dz dz^{\ast} z^2
(|z|^2 + c_4) \hat{U}(z,z^{\ast},t),
\ee
which implies in the fermion representation
\be
{\cal O}^{4,2} = e^{2 i t} \frac{2 {\cal N}_4 }{\sqrt{10} N^{2}}
\sum_{m} \left ( (m+3 + 2 c_4) \sqrt{(m+2)(m+1)} \right ) \hat{C}_{m+2}^{\dagger} \hat{C}_m.
\ee
Again we use a three point function (\ref{3pf-res}) to fix the
coefficient $c_4$; to leading order in $N$ this gives $c_{4} = -\half
N$, and thus the vev is
\be
\langle {\cal O}^{4,2} \rangle = \frac{4 \sqrt{3} N^2 }{\sqrt{10}
  \pi^2 } e^{2 i t} \int d^2w w^2 (|w|^2 - 1) \rho,
\ee
in agreement with (\ref{pofr}).

\bigskip

Thus, to summarize, we have implemented all single trace operators of
dimension $\Delta$ and $SO(2)$ charge $k$ quadratically in fermions as
\be \la{single-trace}
{\cal O}^{\Delta, k} = N^{-\half \Delta}
{\cal N}_{\Delta} e^{i k t} \sum_{m=0}^{\infty} P^{\Delta,k} (m) \hat{C}^{\dagger}_{m+k}
\hat{C}_m,
\ee
with $P^{\Delta,k} (m)$ fixed so as to give the correct
normalization and three point functions of ${\cal N} = 4$ SYM.
Up to dimension four, it was
sufficient to use three point functions involving only single trace operators,
but for higher dimension operators one might also have to use additional
three point functions involving multi-trace operators. Rewriting the vevs of these
operators as integrals over the distribution gives the general form
for the holographic vevs \eqref{generic1} and explicit agreement for all
operators up to dimension four and maximally charged operators of all dimension.

\section{Correspondence between supergravity solutions and states}

In this section we explore how much one can deduce about the dual
state from a given regular supergravity solution, using
the information about the vevs of chiral primaries. We have already argued that
even the exact
distribution function does not in general determine the state uniquely, and
in this section we will see how a given sharpened distribution (which
gives a regular supergravity solution) can correspond to a number of
distinct exact distributions.

We will also note that non-singular
supergravity solutions which break the $SO(2)$ rotational symmetry are
necessarily dual to infinite superpositions of Schur
polynomials. Superpositions of a small number of Schur polynomials
typically give rise to distributions which cannot be approximated by step
functions and thus do not correspond to regular geometries. Thus the
natural field theory bases for half BPS states, which use R charge
eigenstates, are not the natural bases for describing regular bubbling
geometries.

We illustrate this point by considering a disc distribution with a
ripple deformation of frequency $n$. Using the chiral primary vevs
we argue that such a distribution
is given by a coherent superposition of single trace operators. This
identification follows very naturally from earlier discussion of
quantum Hall liquids: area preserving deformations of a disc droplet
are naturally described by coherent superpositions of fermionic
excitations, or equivalently in terms of a collective chiral boson
description.

\subsection{Radially symmetric distributions} \la{dop1}

In this section we consider half BPS states associated with
superpositions of Schur polynomials of the same dimension. Such
states are eigenstates of the dilatation and R charge, so only $SO(2)$
neutral operators can acquire expectation values. The corresponding distributions
therefore preserve the rotational symmetry in the plane.

A given Schur polynomial $\chi^{n}_{\{\l \}} (Z)$ corresponds according
to \eqref{schur} to a distribution
\bea
\langle \hat{U} (z,z^{\ast},t) \rangle_{n,\l} &=& \sum_{p=1}^{N}
\Phi^{\ast}_{\l_p + N -p} (z,z^{\ast},t) \Phi_{\l_p + N - p}
(z,z^{\ast},t); \la{dchur} \\
&=& \frac{2}{\pi} e^{-2 |z|^2} \sum_{p=1}^{N} \frac{ (2 |z|^2)^{\l_p +
    N -p}}{ (\l_p + N -p)!}. \nn
\eea
Given such a distribution is radially symmetric, only the stress
energy tensor along with neutral operators acquire expectation
values. The expectation value of the former is clearly independent of
$\l$, and depends only on $n$.
One can now show that the vevs of neutral operators with dimensions
$2k \ll N$ also do not distinguish between different Schur polynomials, for
$N \rightarrow \infty$. First note that
\bea
\int d^2 z |z|^{2 k}  \langle \Delta \hat{U} (z,z^{\ast},t)
\rangle_{n,\l} = 2^{-k} \sum_{p=1}^{N} \left (
\frac{ (\l_p + N + k - p)!}{(\l_p + N - p)!} - \frac{ (N + k -
  p)!}{(N-p)!} \right ) \nn \\
 = 2^{-k} \sum_{p=1}^{N} \frac{ (N+k-p)!}{(N-p)!} (\psi (N + k -
p) - \psi (N - p)) \l_p + \cdots
%, \nn \\
 = 2^{-k} N^{k-1} k n  + \cdots \la{univ}
\eea
where $\Delta \hat{U}_{n,\l} = (\hat{U}_{n,\l} - \hat{U}_{\Omega})$ and
$\psi(x)$ is the Digamma function and ellipses denote terms
which are subleading as $N \rightarrow \infty$. Thus the leading term
depends only on $\sum_{p=1}^N \l_{p} = n$, and not the specific Schur polynomial.
The vevs of neutral operators
can be expressed in terms of such integrals as
\be
\langle {\cal O}^{2k,0} \rangle_{n,\l} =
\frac{{\cal N}_{2k}} {N^k} \sum_{l=0}^{k-1} d_l \int d^2z
|z|^{2(k-l)} N^{l} \langle \Delta \hat{U} (z,z^{\ast},t)
\rangle_{n,\l},
\ee
for certain coefficients $d_l$. (In the previous sections we gave the
$d_l$ explicitly for $k=1,2$.) Using \eqref{univ}, one finds that
\be \la{ind-result}
\langle {\cal O}^{2k,0} \rangle_{n,\l} = \frac{{\cal N}_{2k}
  n}{N} C_k, \qquad C_k = \sum_{l=0}^{k-1} d_l 2^{-(k-l)} (k-l),
\ee
regardless of the choice of $\l$. Note that this behavior concurs with the
explicit result for the vev in the state created by the single trace operator
${\rm Tr}(Z^n)$, given in \eqref{single}, and the latter result
determines that
\be
C_k = \frac{\sqrt{2k}}{2^{k-1} \sqrt{2k + 1}}.
\ee
Thus the vevs of neutral operators in an R-charge eigenstate are at leading order in $N$
independent of the specific choice of Schur polynomial superposition
creating that state.

\bigskip

Expressed in the coordinates appropriate for comparing with supergravity,
the density distribution takes the form
\be \la{exact}
\rho_{n,\l} = \frac{1}{\pi} e^{- N |w|^2} \sum_{p=1}^N \frac{
  (N|w|^2)^{\l_p + N - p}}{(\l_p + N - p)!},
\ee
and has the properties
\be
\int d^2w \rho_{n,\l} = 1; \hsp
\int d^2w |w|^2 (\rho_{n,\l} - \rho_{\Omega}) = \frac{n}{N^2}.
\ee
The latter implies that the excess energy relative to the conformal
vacuum is $n$, independently of $\l$. This density function is
such that $ 0 \le \pi \rho_{n,\l} \le 1$ everywhere; however, as for the
density function describing the conformal vacuum, $(\pi \rho_{n,\l})$
does not take the values $\{0,1\}$ everywhere, and therefore the
corresponding supergravity solution constructed from $\rho_{n,\l}$
would be singular. Just as for the
conformal vacuum, though, the density function can be written as a sum
of theta functions plus correction terms describing the smearing
which are subleading as $N \rightarrow \infty$. Retaining only the
former leads to a non-singular supergravity solution.

\begin{figure}
\begin{center}
\leavevmode \epsfxsize=.6\textwidth \epsfbox{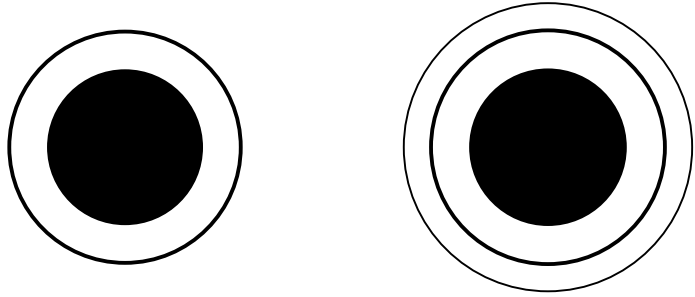}
\caption[]
{Sharpened distributions describing Schur polynomials of the same
dimension $n \ll N$ cannot be distinguished by the corresponding vevs
at leading order in $N$.}
\label{fig:figure1}
\end{center}
\end{figure}

For example, suppose one considers the Schur polynomial for which
$\l_1 = n+1$ with $\l_{p} = 0$ otherwise; this corresponds to the state
$\hat{C}^{\dagger}_{N + n}  \hat{C}_{N-1} | \Omega \rangle$. The
density function \eqref{exact} in this case consists of a smoothed
disc of radius one, along with a second peak localized around $|w|^2 =
(1 + \frac{n+1}{N})$. Now suppose one sharpens distribution so as to
get a regular supergravity solution, consisting of a disc plus an
annulus:
\bea
\rho &=& \frac{1}{\pi}: \hsp 0 \le |w|^2 \le (1 - \frac{1}{N}); \hsp
\left (1 + \frac{n}{N} \right ) \le |w|^2 \le \left (1 +
\frac{n+1}{N} \right ), \la{dis1} \\
\rho &=& 0: \hsp (1 - \frac{1}{N}) < |w|^2 < \left (1 + \frac{n}{N} \right );
\hsp |w|^2 > \left (1 + \frac{n+1}{N} \right ). \nn
\eea
The smearing of the disc and the annulus to obtain the exact density function
is described by correction terms which are subleading as $N \rightarrow \infty$.

As a second example, consider the Schur
polynomial for which $\l_1 = n_1$ and $\l_2 = n_2 = n + 1 - n_1$,
corresponding to $\hat{C}^{\dagger}_{N + n_1 - 1} \hat{C}^{\dagger}_{N +
  n_2 - 2} \hat{C}_{N-1} \hat{C}_{N-2} | \Omega \rangle$. Assuming
that $n_1$ and $n_2$ differ by a finite amount, the density function
\eqref{exact} consists of a smooth disc, along with two localized
peaks. (If $n_1$ and $n_2$ are comparable, these two peaks merge, to
look like the one.) One can then obtain a corresponding supergravity
solution by taking the density
distribution to consist of a disk plus two annuli:
\bea
\rho = \frac{1}{\pi}: && \hsp 0 \le |w|^2 \le (1 - \frac{2}{N}); \hsp
\left (1 + \frac{(n_2-2)}{N} \right ) \le |w|^2 \le \left (1 +
\frac{(n_2-1)}{N} \right ), \la{dis2} \\
&& \qquad \left (1 + \frac{(n_1-1)}{N} \right ) \le |w|^2 \le \left (1 +
\frac{n_1}{N} \right ). \nn
\eea
Both configurations we have described have the same energy $(n+1)$,
and by the arguments above also have the same one point functions at
leading order in $N$.

\begin{figure}
\begin{center}
 \begin{minipage}{0.45\hsize}
  \begin{center}
\epsfxsize=.9\textwidth
\epsfbox{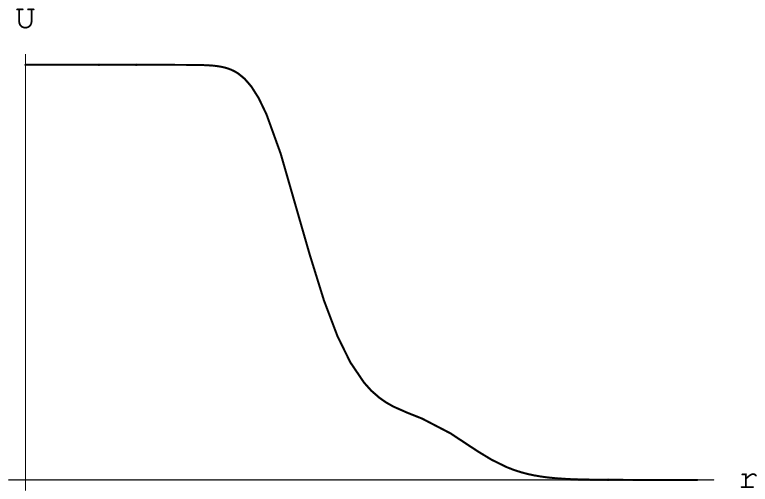}
  \end{center}
  \caption{Distribution for a typical Schur polynomial; there is no
  distinct peak. The figure shows $N=100$, $n=30$, with a random
  distribution of $\l_p$.}
  \label{fig:circular}
 \end{minipage}
\hspace*{1cm}
 \begin{minipage}{0.45\hsize}
  \begin{center}
\epsfxsize=.8\textwidth
\epsfbox{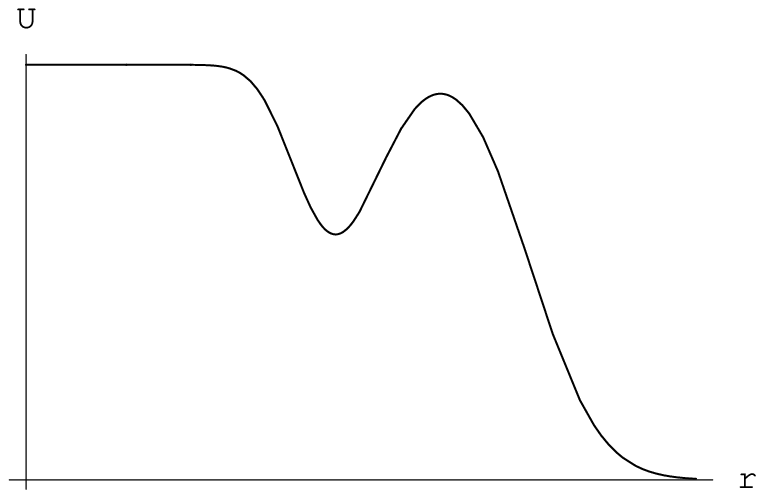}
  \end{center}
 \caption{Distribution for a Schur polynomial in which the $\l_p$ are
 equal. The figure shows $N=120$, $n=30$ with $\l_p =2$ for $p \le 15$.}
  \label{fig:equal}
\end{minipage}
\end{center}
\end{figure}

More generally, for an arbitrary Schur polynomial one can obtain a
corresponding supergravity solution by sharpening the distribution
into a set of annuli, as illustrated in Figure~\ref{fig:figure1}.
However, there is clearly not a unique map
from such a set of annuli to a given Schur polynomial: Schur
polynomials which are very similar to each other, and superpositions
of similar Schur polynomials, give density distributions which can only
be distinguished at subleading order in $N$. That is, the associated
sharpened supergravity distributions, which consist only of annuli,
can be the same. To give an example, consider a specific superposition of Schur
polynomials of the same dimension
\be
| \Phi \rangle = \sum_{k=0}^{n} a_{k} \hat{C}^{\dagger}_{N+ n -k}
\hat{C}^{\dagger}_{N - 1 -k} | \Omega \rangle,
\ee
with $\sum_{k=0}^{n} |a_{k}|^2 = 1$. When only one coefficient $a_{k}$
is non-zero this
collapses to the case discussed above of a single Schur polynomial,
for which the sharpened distribution is a disc plus annulus. However, many other
superpositions, such as those for which one $a_k$ is much greater than
the rest, will give precisely the same sharpened distribution.
Moreover, a typical Schur polynomial for which $1 \ll n \ll N$ in which many
of the $\l_p$ are non-zero and different can give rise to a distribution which does not
approximate a disk plus annuli: generically there are no strong peaks at $|w| >
1$, as illustrated in Figure \ref{fig:circular}. By contrast a Schur
polynomial for which the $\l_p$ are equal does give rise to a strong
peak, see Figure \ref{fig:equal}. For more discussions on these
issues, see \cite{Balasubramanian:2005mg},
\cite{Balasubramanian:2006jt} along with the recent paper \cite{Balasubramanian:2007zt}.

\begin{figure}
\begin{center}
\leavevmode \epsfxsize=.25\textwidth \epsfbox{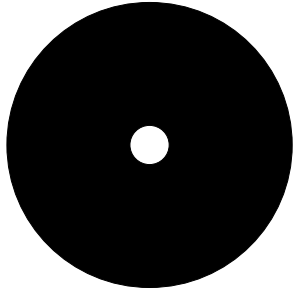}
\caption[]
{A maximal giant graviton corresponds to a disc with a small hole at
  the centre.}
\label{fig:figure2}
\end{center}
\end{figure}

\begin{figure}
\begin{center}
\leavevmode \epsfxsize=.4\textwidth \epsfbox{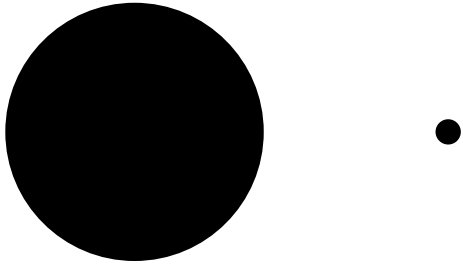}
\caption[]
{A disc plus droplet corresponds to a superposition of Schur
  polynomials of all dimension.}
\label{fig:figure3}
\end{center}
\end{figure}

Note furthermore that a giant graviton, which is also a Schur polynomial
of given dimension, necessarily corresponds to a radially symmetric
distribution and not a disc with an excited droplet illustrated
in Figure \ref{fig:figure3}, as suggested in some earlier papers.
Indeed the maximal giant graviton, which has dimension $N$ and is a
Schur polynomial for which all $\l_{p} = 1$, corresponds to a
distribution which approximates a disc with a hole in the middle,
as illustrated in Figure \ref{fig:figure2}.

\bigskip

In the above discussion we have focused on Schur polynomials with dimension less
than $N$. A Schur polynomial with dimension greater than or equal to $N$
can never be represented by a single trace component, since there is
of course no single trace operator with such a dimension. Indeed if
one considers
the two distributions \eqref{dis1} and \eqref{dis2} but now with $n \ge
N$, one finds that the vevs for the dimension four neutral operators
are respectively
\be \la{ponb}
\langle {\cal O}_{S^{40}} \rangle = \frac{\sqrt{3} N}{\sqrt{5}
  \pi^2} ( 2 \td{n} + 3 \td{n}^2);  \qquad
\langle {\cal O}_{S^{40}} \rangle = \frac{\sqrt{3} N}{\sqrt{5}
  \pi^2} (2 \td{n} + 3 (\td{n}^2 - 2 \td{n}_2 \td{n}_1)),
\ee
where $\td{n} = n/N$, $\td{n}_1 = n_1/N$ and $\td{n}_2 = n_2/N$ with
$\td{n} = \td{n}_1 + \td{n}_2 \ge 1$. The leading order terms as $N
\rightarrow \infty$ for $(n_1,n_2) \gg 1$ have been retained.
These vevs are indeed
distinguishable even as $N \rightarrow \infty$; note that the vev for
the single annulus is greater than that of the other distribution.
Of course, following
the general discussion given earlier, this distinguishability will
still not be
sufficient to determine the dual state $|\Phi \rangle$ uniquely:
the single trace operator vevs cannot provide enough
information. Moreover, again a typical Schur polynomial will not give
rise to a distribution with strong peaks which is well approximated by
annuli.

\bigskip

Consider a disc plus annulus configuration with fixed energy $n$. For
given $n$ one might think that the radius of the annulus $|w|^2 = 1 +
\b$ can be taken to be arbitrarily large provided that its width is
decreased accordingly. However, one has to take into the flux
quantization condition: as discussed in \cite{Lin:2004nb} one can
integrate the five-form flux over a sphere surrounding the annulus,
and the flux must be quantized. This quantization requires that the
total area of the annulus (and indeed of any isolated droplet)
must be a multiple of $\pi/N$. Thus the width of the
annulus defined as $\delta = |w|^2_{max} - |w|^2_{min}$ is a multiple of $1/N$.
The total energy of the configuration of the disc plus annulus, relative to the conformal
vacuum, is given by $n = N \b + 1$
and therefore for given $n$ one gets $\b = (n-1)/N$, as in
\eqref{dis1}. This is the maximal radius for a given energy: multiple
annuli or thicker annuli necessarily have lower radii.

This same bound is also visible directly from the density
distribution. First note that the function
$ e^{-\chi} \chi^{\a}/\a! $
for large $\a$ has support only around $\chi = \a$. Thus the
distribution \eqref{dchur} extends to a maximum radius
\be
|w|^2_{max} = 1 + \frac{\l_1 -1}{N},
\ee
which is maximal for the Schur polynomial in which $\l_1 = n$ and
$\l_i = 0$ otherwise. This implies an upper bound
on the magnitudes of vevs of the scalar operators, for a given
energy.

\subsection{Non symmetric distributions}

A distribution which is not radially symmetric corresponds
to a superposition of R-eigenstates. The generic state $|\Phi \rangle$
can be written in terms of Schur polynomials as
\be \la{po4}
| \Phi \rangle = \sum_{n,\l} a_{n,\l} \chi^{n}_{\{\l \}} | \Omega
\rangle \equiv \sum_{n,\l} a_{n,\l} |n,\{\l \} \rangle.
\ee
with the Schur polynomials being orthonormal, and normalization of the
state implies $\sum_{n,\l} | a_{n,\l} |^2 = 1$.
Now consider an operator ${\cal O}_{\Delta,j}$
of dimension $\Delta$, R-charge $j$.
%with $\m$ labelling its degeneracy.
The vev of such an operator in this state is
\be
\langle {\cal O}_{\Delta,j} \rangle_{\Phi} =
\sum_{n,\l,\l'}  a_{n+j,\l'}^{\ast} a_{n,\l}
\langle  (n+j), \{ \l' \} | {\cal O}_{\Delta,j} | n, \{ \l \} \rangle.
\ee
The three point functions appearing in this sum are non-extremal,
except when $\Delta = j$, i.e. the operator is maximally charged. As
discussed previously, the leading order contribution to non-extremal
correlators as $N \rightarrow \infty$ is independent of the specific
choices of Schur polynomial. Thus, the leading order contributions to
the vevs of non-maximally charged operators can be computed using
\be \la{po3}
| \Phi \rangle \approx \sum_{n} \td{a}_n | n \rangle,
\ee
where $| n \rangle$ is any state of R-charge $n$, and the
coefficients $\td{a}_n$ are such that
\be
\sum_{ \{ \l \} } | a_{n,\l} |^2 = | \td{a}_n |^2.
\ee
However, the vevs of the maximally charged operators do depend on the
specific Schur polynomials appearing in the state $| \Phi \rangle$,
because the corresponding three point functions are extremal. Thus
these vevs can be used to distinguish between different distributions
with the same $\td{a}_n$, even at leading order in $N$.
For example, consider the states
\be
| \Phi \rangle = a_{0} | \Omega \rangle + a_{n,\{ \l \} } | n, \l \rangle,
\ee
where $|a_0|^2 + |a_{n,\{ \l \} }|^2 = 1$. To leading order in $N$,
the vevs of the energy, R-charge and neutral operators are independent
of the specific choice of $\{ \l \}$. However, the vevs of the
  maximally charged operators with dimension $n$ clearly do
  distinguish them. Let us compare the state created by the single
  trace operator, namely $|n, \l \rangle \rightarrow {\cal O}^{n,n} |
  \Omega \rangle$ where the single trace operator is defined in
\eqref{imp-1}, and that associated with the Schur polynomial for
  which $\l_1 = n$, namely  $|n, \l \rangle \rightarrow
  \hat{C}^{\dagger}_{N+n-1} \hat{C}_{N-1} | \Omega \rangle$. Then the
  expectation values of the single trace operator of charge $-n$ are
  respectively
\bea
&& |\Phi \rangle = a_{0} | \Omega \rangle + a_{n} {\cal O}^{n,n} |
  \Omega \rangle : \hsp \langle {\cal O}^{n,-n} \rangle_{\Phi} = N_n
  a_{0}^{\ast} a_n e^{-i
    nt}; \\
&& |\Phi \rangle = a_{0} | \Omega \rangle + a_{n}
\hat{C}^{\dagger}_{N+n-1} \hat{C}_{N-1} | \Omega \rangle  : \hsp
\langle {\cal O}^{n,-n} \rangle_{\Phi} = \frac{N_n}{\sqrt{n}}
a_{0}^{\ast} a_n e^{-i nt}, \nn
\eea
which differ by a factor of $\sqrt{n}$.

We should note here, however, that such superpositions which involve
only a small number of Schur polynomials do not generically give rise
to smooth supergravity solutions. In the example just given, taking
$(a_{0},a_n)$ to be real, the corresponding distribution takes the form
\be
\rho(w,\phi) = \rho(|w|) + \td{\rho}(|w|) \cos (n \phi), \la{proof}
\ee
with the functions $(\rho(|w|),\td{\rho}(|w|))$ dependent on the
  specific Schur polynomial. In the case that
$|n, \l \rangle \rightarrow
  \hat{C}^{\dagger}_{N+n-1} \hat{C}_{N-1} | \Omega \rangle$ the
  functions are
\bea
\pi \rho(|w|)&=&\theta(1{-}N^{-1}{-}|w|^2){+}|w|^{2(N-1)}
e^{-N |w|^2} \left (
|a_0|^2 \frac{N^{N-1}}{(N-1)!}{+}|a_n|^2 \frac{N^{N+n-1}}{(N+n-1)!}
|w|^{2n} \right  ); \nn \\
\pi \td{\rho}(|w|) &=& a_{0} a_n |w|^{2(N-1) + n} e^{-N
  |w|^2} \frac{N^{N+ n/2 -1}}{\sqrt{(N-1)! (N+n-1)!}}.
\eea
Regularity of the supergravity solution requires that $\pi
\rho(w,\phi)$ takes the values $\{0,1\}$ everywhere. This condition can
however never satisfied by a function of the form (\ref{proof}), with
$\td{\rho}(|w|)$ non-zero. One can see this easily as follows. Suppose
the function $\rho(w,\phi)$ satisfies this condition at $\phi = 0$, so
that
\be
\pi \rho(w, 0) = \rho(|w|) + \td{\rho}(|w|) = \{0,1 \},
\ee
for $\td{\rho}(|w|) \ge 0$. Then
\be
\pi \rho(w, \d \f) - \pi \rho(w, 0) = - \half \pi n^2 \td{\rho}(|w|) \d \f^2
+ \cdots
\ee
Regularity would require that the right hand side takes only the values
$\{-1,0,1 \}$ for all $|w|$ and $\d \f$, but this is not possible given
that $\d \f^2$ is a continuous function of $\d \f$.
Smoothing the distribution so that $\pi
\rho(w,\phi)$ does take the values $\{0,1 \}$ everywhere introduces
additional terms (with small coefficients) into the state $| \Phi
\rangle$, and such terms may not be distinguishable at leading order
in $N$.

\bigskip

Conversely, a droplet distribution which is not radially symmetric but
which gives rise to a regular supergravity solution is always
associated with a superposition of an infinite number of Schur
polynomials. Such a distribution can be written in terms of step
functions whose arguments are the boundaries of the droplets:
\be
\theta(|w|^2 - f(\phi)); \hsp
f(\phi) = f_{0} + \td{f}(\phi),
\ee
where $f_0$ is a zero mode whilst
$\td{f}(\phi)$ has no zero mode and can be expanded in Fourier
modes. Now even when the droplet boundary
$\td{f}(\phi)$ contains only one frequency $n$,
the distribution will contain all multiples of this
frequency. One can see this by mode expanding the distribution, using
\eqref{project}, or by computing the multipole moments of the
distribution as
\be
\int d \f d|w|^2 |w|^{2l} e^{i m \f} \rho = (1 + l)^{-1} \int d\f
e^{im \f}  (f_{0} + \td{f}(\phi))^{1 + l}.
\ee
The latter integral is clearly non-zero when the frequency $m$ is
contained in $\td{f}(\phi)$ or its products, demonstrating that
the distribution contains products of the frequencies contained in
$\td{f}(\phi)$. Therefore, in the case
that the droplet boundary contains only
frequency $n$  the corresponding state is a
superposition of Schur polynomials of dimension $kn$ involving all $k
\ge 0$.

\subsection{An example: a disc with a ripple} \la{dop2}

Consider a disc with a ripple such that the boundary of the
distribution is at
\be \label{rip_dis}
|w|^2 = 1 + \a \cos (n\f).
\ee
\begin{figure}
\begin{center}
\leavevmode \epsfxsize=.3\textwidth \epsfbox{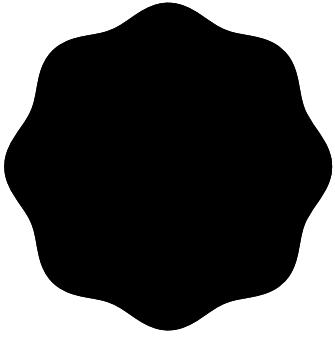}
\caption[]
{A disc with a ripple of frequency $n$ corresponds to a superposition
  of Schur polynomials whose dimensions are multiples of $n$. The
  actual figure has $n=8, \a=0.1$.}
\label{fig:figure4}
\end{center}
\end{figure}
By the arguments above, such a distribution corresponds to a state $|
\Phi \rangle$ which is a superposition of Schur polynomials of
dimension $kn$, namely
\be
| \Phi \rangle = \sum_{k,\{ \l \} } a_{k, \{ \l \} } | kn; \{ \l \}
\rangle.
\ee
In general, one will need to compute all one-point functions of all
single and multitrace operators in order
to deduce the coefficients $a_{k, \{ \l \}}$. However, given that this
distribution is highly symmetric, with only one defining parameter,
let us try to use the vevs of the lowest dimension single trace
operators to deduce the
superposition. Evaluating (\ref{j-mass}) for the distribution
with boundary at (\ref{rip_dis}),
the energy and R-charge in this state are
\be
E - E_c = J = \qu N^2 \a^2,
\ee
whilst from (\ref{pofr})-(\ref{vev4}) the vevs of the neutral operators are
\be
\langle {\cal O}_{S^{20}} \rangle = \frac{N^2 \sqrt{2}}{4 \sqrt{3}
  \pi^2} \a^2; \hsp
\langle {\cal O}_{S^{40}} \rangle = \frac{N^2 \sqrt{3}}{2 \sqrt{5}
  \pi^2} \a^2,
\ee
and from (\ref{ch-1}) the vevs of the maximally charged operators
with dimension $n$
are
\be
\langle {\cal O}_{S^{n, \pm n}} \rangle = \frac{N^2}{2 \pi^2 \sqrt{n}}
(n-2) \sqrt{n-1} e^{\mp 2 i nt} \a. \la{chg1}
\ee
First note that the energy and the vevs of neutral operators are reproduced by
a superposition of the form (\ref{po3}) with
\be
| \Phi \rangle = e^{-\frac{N^2
  \a^2}{4 n}} \sum_{k=0}^{\infty} \frac{ (N \a)^k}{ (2\sqrt{n})^{k}
  \sqrt{k!}} | kn \rangle.
\ee
To compute the energy and the neutral vevs at leading order in $N$ we may
use any representative orthonormal state $|kn \rangle$, using the
result of \eqref{ind-result}. Thus the energy is given by
\be
\langle E - E _c \rangle_{\Phi} = e^{-\frac{N^2 \a^2}{2 n}}
\sum_{k=1}^{\infty} \frac{ (N \a)^{2k}}{2^{2k} n^{k} k!} (kn) = \qu N^2 \a^2.
\ee
The vev of the dimension four operator can be computed using the
result given in \eqref{single}:
\bea
\langle {\cal O}^{40} \rangle_{\Phi} &=&
e^{-\frac{N^2 \a^2}{2 n}}
\sum_{k=0}^{\infty} \frac{ (N \a)^{2k}} {2^{2k} n^{k} k!}
\langle {\cal O}^{40} \rangle_{kn}  \\
&=& \frac{\sqrt{12} e^{-\frac{N^2 \a^2}{2 n}}}
{\sqrt{5} \pi^2} \sum_{k=1}^{\infty} \frac{ (N
  \a)^{2k}} {2^{2k} n^{k} k!}  (kn) = \frac{N^2 \sqrt{3}}{2 \sqrt{5}
  \pi^2} \a^2. \nn
\eea
Now consider the vev of the maximally charged operator: as we have
emphasized this vev is sensitive to the
specific Schur polynomials contained in the state $|kn \rangle$.
Let the single trace operator of dimension $n$ be ${\cal O}^{n,n} =
{\cal N}_n \hat{s}_n$, where by construction  $\hat{s}_n | \Omega
\rangle$ is orthonormal. Now suppose that $|kn \rangle$
is the state created by products of this operator such that
\be
|kn \rangle = \frac{1}{\sqrt{k!}} (\hat{s}_n)^{k} | \Omega \rangle.
\ee
The states $|k n \rangle$ are orthonormal to each other in the
large $N$ limit. The easiest way to check this is to go back to
SYM language where
$|kn \rangle =  \frac{1}{\sqrt{k!}}
\left((\Tr Z^n)/\sqrt{N^n n}\right)^k |\Omega \rangle$
and then use standard large $N$ counting. The leading contribution
in $\langle kn|kn\rangle$ comes from disconnected diagrams where
each $\Tr Z^n$ is contracted with a corresponding $\Tr \bar{Z}^n$.

Then by construction
\be
\langle n (k+1) | \hat{s}_n | kn \rangle = \sqrt{k+1},
\ee
which implies that
\bea
\langle {\cal O}_{S^{n, \pm n}} \rangle_{\Phi} &=& {\cal N}_{n} e^{\mp 2 i nt}
e^{-\frac{N^2 \a^2}{2 n}}
\sum_{k=0}^{\infty} \frac{ (N \a)^{2k+1}} {2^{2k+1}
n^{k} k! \sqrt{n(k+1)}} \langle n (k+1) | \hat{s}_n | kn \rangle
\\
&=& \frac{{\cal N}_{n} N \a}{2
  \sqrt{n}} e^{\mp 2 i nt} =
\frac{N^2}{2 \pi^2 \sqrt{n}}
(n-2) \sqrt{n-1} e^{\mp 2 i nt} \a, \nn
\eea
in exact agreement with the holographic result \eqref{chg1}.
By contrast, if $|kn \rangle$ were instead the state
created by the single trace operator ${\rm Tr} (Z^{kn})$, then using
the appropriate single trace operator three point functions one can
show that the holographic result for the charged operator vev would not
be reproduced: the vev would be down by a factor of $N$.

Note that when $\a$ is infinitesimal the state $|\Phi \rangle
$ reduces to a perturbation of the conformal vacuum by the operator
$\hat{s}_{n} |\Omega \rangle$:
\be
|\Phi \rangle \rightarrow (1 + \frac{N \a}{2  \sqrt{n}} \hat{s}_n ) |
\Omega \rangle,
\ee
in agreement with the identification of infinitesimal perturbations
made in \cite{Grant:2005qc}. Thus the disc with a ripple is consistent
with being
dual to a state created by coherent superpositions of states
created by powers of the single trace
operators $\hat{s}_n$ acting on $|\Omega\rangle$.
Indeed, the state $|\Phi \rangle$ can be written as
a coherent state,
\be
| \Phi \rangle = | \d ) \equiv e^{-\half \d^2} \sum_{k=0}^{\infty}
\frac{\d^k}{\sqrt{k!}} | k n\rangle,
\ee
for $\d = N \a/2 \sqrt{n}$ with $| k n\rangle $ the state containing
$k$ quanta of $\hat{s}_n$.

\subsection{Ripplon deformations and the chiral boson description}

The identification of the ripple deformed disc with a state built from
a coherent superposition of the operators $\hat{s}_n$ follows
naturally from earlier discussions of edge excitations in quantum Hall
liquids, see for example \cite{Stone, Stone2}.
Consider a distribution consisting of a single droplet, whose boundary
can be parametrized as
\be
|w|^2 = 1 + X(\f),
\ee
where $X(\f)$ is an arbitrary function with no zero modes. This
function $X(\f)$ describes area preserving deformations of the
disc. Now so far we have used the fermion picture, that is, the
Schur polynomials, to describe excitations relative to the conformal
vacuum. This is a natural basis to describe droplets which are
separated from the disc, but it is not the natural basis for
describing coherent ripplon deformations in the shape of the
droplet.

Such ripplons can best be described by quantizing the chiral boson
field $X(\f)$; quantizing its Fourier mode expansion gives rise to a Hilbert
space associated with bosonic creation and annihilation operators.
As discussed in \cite{Stone2}, these operators can
in turn be identified with elements in the symmetric polynomial (or
equivalently, the trace) basis. Thus edge waves and deformations in
the shape of the droplet are most naturally described within the trace
(chiral boson) description rather than the Schur polynomial (fermion)
description. This extends the identification
made for the single frequency ripple discussed above to more general ripples.

It is interesting to note that the algebra of area preserving
diffeomorphisms of the droplet is actually $W_{\infty}$. It emerges in
both the fermionic and bosonic formulations as the algebra of unitary
transformations of physical states. It would be interesting to understand
the meaning and implications of  $W_{\infty}$ for holography.

\section{Discussion} \la{disc}

In this paper we have discussed holography for bubbling solutions.
Solutions that are asymptotically $AdS_p \times S^q$ contain an
infinite amount of holographic data that can be extracted by
{\it algebraic manipulations}, namely the holographic 1-point functions
that characterize the vacuum of the dual QFT. This is the
simplest information one can extract from a given supergravity
solution. Conversely, knowledge
of the 1-point functions is in principle sufficient in order to
reconstruct bulk solutions from QFT data.
Two-point and higher-point functions can also in principle be
extracted, but this requires solving at least the linearized equations
around the solution (for 2-point functions) whilst for $n$-point functions
one needs to solve the $(n{-}1)$-th order equations. Explicitly
solving such equations is an intractable
problem, except for very symmetric solutions. Moreover, in the case of
interest vevs are protected, given the non-renormalization
of 3-point functions of chiral primaries at the conformal vacuum,
whilst there are no
such non-renormalization theorems protecting generic two point
functions, and corresponding four point functions in the conformal vacuum.

In the first part of this paper we reviewed the holographic 1-point functions
derived in \cite{Skenderis:2006uy}
for the stress energy tensor, the R-currents and all chiral primaries
up to dimension four for asymptotically $AdS_5 \times S^5$
solutions of IIB supergravity that involve the metric and the
5-form. These results hold generally when the solution is dual
to a state (rather than describing a deformation), i.e. they do not dependent
on the amount of supersymmetry preserved by the solution or its
bosonic isometries (except that the solution should be asymptotic
to $AdS_5 \times S^5$), and thus these 1-point functions can be used to extract
holographic data from any such solution. The 1-point functions are
given in terms of the asymptotic coefficients of the ten dimensional solution
and are presented in (\ref{ovev}), (\ref{tij2}) (\ref{rsym})
(which for the reader's convenience were boxed). Note that the expressions are
non-linear in the asymptotic coefficients.

The next step was the evaluation of these holographic formulas
on the LLM solutions. The holographic 1-point functions are by construction
diffeomorphism covariant, so the asymptotic coefficients can be
extracted in any coordinate system. A clever choice of coordinates
however can reduce the required labor significantly. Recall
that the LLM solution is determined by a harmonic function $\F$ in six
dimensions with sources on a 2-plane. The asymptotic expansion
around $AdS_5 \times S^5$ can be efficiently performed by
writing $\F = \F^o + \Delta \F$, where $\F^o$ is the harmonic
function that leads to $AdS_5 \times S^5$,  and then expanding in
$\Delta \F$. One still needs to convert these expansions
into radial expansions. This is done by first using flat
coordinates on $R^6$ so that the asymptotic expansion of $\D \F$
takes a standard form and then transforming to the
coordinates most natural for the $AdS_5 \times S^5$ solution.
This procedure minimizes the number of non-linear terms entering the
computation of the 1-point functions.

We obtained explicit
expressions for the  vevs in terms of integrals over the 2-plane
defining the solution; these are given in
(\ref{j-en})-(\ref{j-mass})-(\ref{ch-1})-(\ref{pofr})-(\ref{vev4})
(which again are boxed). Note that in
(\ref{ch-1}) we give the vevs of all maximally charged operators,
i.e. operators with dimension equal to R-charge, despite the fact that
the general analysis in the previous section was done for
operators up to dimension four. The ability to produce
such a result is due to special properties of the LLM solution
combined with the previous Coulomb branch
results of \cite{Skenderis:2006di}. Given the
large amount of supersymmetry preserved by the LLM solutions,
one would expect that these vevs should not renormalize
and thus that they must be reproduced by a weak coupling
computation. Put differently, these vevs provide checks for
the correct identification of the dual theory.

The vevs satisfy a number of non-trivial consistency checks.
Firstly, the vev of the energy is proportional to the vev of the
R-charge (up to the the Casimir energy of SYM on $S^3$) as is required by
supersymmetry. Secondly, all vevs, except for the
energy which should become equal to the Casimir energy,
should vanish for the theory at the conformal vacuum and this is
indeed the case for the vevs we derive. Thirdly, the LLM solutions in the
decompactification limit of $S^3$ go over to $SO(4)$ symmetric
distributions of D3 branes, with the sources on the 2-plane
now describing the distribution of D3 branes. These solutions are dual to
$N=4$ SYM on the Coulomb branch. The LLM vevs indeed
correctly reduce to the Coulomb branch vevs given in \cite{Skenderis:2006di} in this
limit.

Before proceeding we should make a comment about the
mass of these spacetimes\footnote{An earlier discussion about the
mass of the LLM solutions can be found in \cite{Giombi:2005zq}.}.
Had the solutions been asymptotically flat, one would have obtained
their mass from the $g_{tt}$ component of the metric. In our case however
the solution is asymptotically $AdS_5 \times S^5$ so such
a prescription is in general not valid. There are two issues
here. One is that the solution involves in a non-trivial
fashion a compact part of the geometry and the other is that the
non-compact part is asymptotically AdS. For asymptotically AdS spacetimes
the issue of mass has been revisited and thoroughly analyzed in recent years
\cite{Balasubramanian:1999re,deHaro:2000xn,Skenderis:2000in,Hollands:2005wt,
Papadimitriou:2005ii} resulting in holographic formulas which
relate the mass to the asymptotics of the metric and other matter
fields.

Taking into account the compact part is also non-trivial
since none of the existing consistent truncation formulas from
ten to five dimensions is directly
applicable. One can however reduce the solution to five dimensions
without truncating, keeping all fields relevant for the computation of
the mass.
This is essentially the method of KK holography \cite{Skenderis:2006uy}
and results in the rigorous formula for the holographic stress energy tensor
given in (\ref{tij2}). One can then obtain the mass from the
$T_{tt}$ component, as usual.

The field theory dual to the LLM solutions is expected to be $N=4$ SYM
on $R \times S^3$ in a half supersymmetric state.
A general way of analyzing this theory would be to carry out
path integral quantization. The requisite
supersymmetry is preserved by quantizing around 1/2 supersymmetric
solutions of $N=4$ SYM on $R \times S^3$. Examples of
such solutions were discussed in \cite{Hashimoto:2000zp}
and more recently in \cite{Ishiki:2006rt}. These
solutions are time-dependent and in correspondence
with the Coulomb branch of $N=4$ SYM on $R^{(3,1)}$.
In particular, the curvature coupling implies the scalar $Z$
satisfies an equation of the form $\dot{Z} \sim i Z$.
This implies that the R-symmetry current
$j_\mu \sim \Tr \bar{Z} \stackrel{\leftrightarrow}{\pa_\mu} Z$
and the operator $\co^{2,0} \sim \Tr \bar{Z} Z$
are proportional to the each other when evaluated on
these solutions (and similarly for related higher dimension operators).
This provides an additional consistency check for the holographic vevs,
which the vevs in (\ref{j-en})-(\ref{pofr}) indeed satisfy.

Due to the extended supersymmetry one might expect that
the exact values of the vevs of 1/2 supersymmetric
gauge invariant operators could be computed by a semiclassical computation.
This would provide a rigorous computation of the
vevs from first principles. Actually carrying out this computation is
not so easy in practice, though, because of subtleties associated with
correctly treating the integration measure. The issue is the
following: for the computations of interest one will want to integrate
out most of the SYM fields, including the other four scalars $X^i$ and
off-diagonal degrees of freedom of the complex matrix $Z$, leaving
only an integral over the eigenvalues of this matrix. This in turn involves
correctly parametrizing the path integral measure as given in
\be
{\cal Z} = \int [dZ dZ^{\dagger} \prod_i dX^i \cdots] e^{iS_{YM} [Z,Z^{\dagger},X^{i},\cdots]},
\ee
and then integrating out appropriately. Now for a general computation
one does not expect to be able to integrate out exactly all these
degrees of freedom. Integrating first over the $S^3$ would
lead to a complicated interacting multi-matrix model which will not
in general be solvable. However, the
holographic computations for the vevs along with the fact that we can reproduce
them by the holomorphic matrix model, imply that at least for these
computations one {\it can} explicitly integrate out the other degrees
of freedom. Demonstrating this by a first principles computation would
be useful as it would explain the
regime of validity and the limitations of the free fermion description.
Moreover, one may in this way show how certain computations
can be carried out in multi-matrix models, even when they are not
exactly solvable.

In the absence of a rigorous derivation of the free fermion description from
first principles, we proceeded by using it as a working assumption.
On symmetry grounds the state that any given bubbling solution
is dual to is a superposition of states obtained from the conformal
vacuum by the action of a 1/2 BPS operator. The question is then
whether one can uniquely determine the precise superposition from the data
encoded in the solution. Using the identification of the coloring
of the 2-plane with the phase space distribution of the free fermions
we show that this information alone does not completely determine
the state in the large $N$ limit. It does determine it enough however
so that the vevs of all single trace 1/2 BPS operators in that state
are uniquely determined. This is precisely the information encoded
holographically in the asymptotics of the solution. The missing
information is related to vevs of multi-trace operators.

A general single trace 1/2 BPS operator depends on fields other than
the complex $Z$ field. This implies that these operators cannot not in general
be implemented with free fermions. Nevertheless, we showed that
for the purpose of the computation of the vevs and to leading order in $N$
such an implementation is possible and all such operators are expressed
in terms of bilinears of fermion creation and annihilation operators.
Using these expressions we show that {\it all} vevs computed holographically
agree {\it exactly} with the field theory computation in the large $N$ limit
for any distribution.

To illustrate our discussion we analyzed a number of examples.
In accordance with our general discussion, we showed that all
vevs associated with any symmetric distributions are degenerate
to leading order $N$. For non-symmetric distributions,
the vevs of charged operators (which by symmetry considerations are zero in
symmetric distributions) can (partly) distinguish between different states.
However, an infinite superposition of states of definite R-charges
is required to obtain a regular geometry. We also analyzed in detail
the case of the distribution being a ripple on a disc.
This case has been analyzed previously for an infinitesimal ripple
in \cite{Grant:2005qc}. We showed here that a finite ripple corresponds
to a coherent state of single trace operators.

We should also comment on the striking parallels between this system
and the 2-charge D1-D5 fuzzball solutions.
Both systems can be characterized by a set of curves: in the LLM case
these are curves in $R^2$
describing the droplet boundaries, whilst in the D1-D5 case these are
curves in an auxiliary space describing the supertube shape and its internal
excitations. The holographic
analysis for this system has recently been done in \cite{KST}.
In both cases, only when the curves are circular and
preserve rotational symmetry do the geometries correspond to vacua
built from a single operator (in the R-charge basis). Regular geometries
in which the rotational symmetry is broken correspond to infinite
superpositions of states in the R-charge basis, with the coefficients
of the superpositions related to the Fourier expansions of the
curves. Thus the natural bases in the dual field theory, which are
labeled by their R charges, are not the natural bases
for regular geometries.

It would be interesting to use the holographic anatomy techniques
discussed in this paper to analyze $1/4$ and $1/8$ BPS bubbling
solutions \cite{Chen:2007du}.
One would expect that these include
both geometries dual to states and those dual to deformations. A
holographic analysis should determine how the boundary conditions
which ensure regularity in the interior of these geometries are related to the
vevs/deformations in the dual theory. More generally one may hope
that combining supersymmetric classification techniques with
holographic anatomy might lead to more efficient holographic
engineering of geometries dual to supersymmetric field theory
states and deformations.

\section*{Acknowledgments}

The authors are supported by NWO, KS via
the Vernieuwingsimpuls grant ``Quantum gravity and particle physics''
and MMT  via the Vidi grant ``Holography,
duality and time dependence in string theory''. This work was also
supported in part by the EU contract MRTN-CT-2004-512194. We would like
to thank both the 2006 Simons Workshop and the theoretical physics
group at the University of Crete, where some of this work was
completed. We also thank Shunji Matsuura and Diego Trancanelli for
pointing out typos and other minor errors in the earlier versions of this paper.  

\appendix

\section{Properties of spherical harmonics} \la{sph}

The defining equations for the spherical harmonics are
\bea
\Box_y Y^{I_1} &=& \L^{I_1} Y^{I_1}, \qquad \L^{I_1} = - k (k+4),
\quad k=0,1,2,... \label{sc_h} \\ %\nonumber \\
\Box_y Y_a^{I_5} &=& \L^{I_5} Y_a^{I_5}, \qquad \L^{I_5} =
-(k^2 +4k -1), \quad k=1,2,... \nonumber \\
\Box_y Y_{(ab)}^{I_{14}} &=& \L^{I_{14}} Y_{(ab)}^{I_{14}}, \qquad
\L^{I_{14}} = -(k^2 + 4k -2), \quad k=2,3,... \nonumber \\
\Box_y Y_{[ab]}^{I_{10} } &\equiv & \L^{I_{10}}  Y_{[ab]}^{I_{10}},
\qquad  \L^{I_{10}} = -(k^2 + 4k -2),
\quad k=1,2,... \nonumber \\
D^a Y_a^{I_5} &=& D^a Y_{(ab)}^{I_{14}} =D^{a} Y_{[ab]}^{I_{10} }=0.
%\qquad
%D_{[a} Y_{bc]}^{I_{10}\pm} = \pm \frac{i}{3} \e_{abc}{}^{de} Y_{[de]}^
%{I_{10}\pm}
\nonumber
\eea
The overall normalization is chosen so
that the harmonics are normalized as
\be \label{nor_sc}
\int Y^{I_1} Y^{I_2} = \pi^3 z(k) \d^{I_1 I_2}, \qquad
z(k) = \frac{1}{2^{k-1} (k+1) (k+2)}
\ee
The triple overlap between spherical harmonics is defined as
\be \la{trip}
\int Y^{(k_1, I_1)} Y^{(k_2, I_2)} Y^{(k_3, I_3)}
= \pi^3 a_{k_1 I_1, k_2 I_2, k_3 I_3},
\ee
where $(k,I)$ is the degree of the scalar harmonic and $I$ labels the
$SO(6)$ quantum numbers.
Recall that the scalar harmonics can be represented as
\be \label{harm_C}
Y^{(k_1,I_1)} = C^{I_1}_{i_1 \cdots i_k} x^{i_1} \cdots x^{i_k}
\ee
where $x^{i_n}$ are Cartesian coordinates on $S^5$ and
$C^I_{i_1 \cdots i_k}$ is a totally
symmetric traceless rank $k$ tensor of $SO(6)$.
The normalization in
(\ref{nor_sc}) corresponds to delta function normalization
for the $C^I$'s, i.e.
\be \la{c-norm}
\langle C^{I_1} C^{I_2} \rangle \equiv
C^{I_1}_{i_1 \cdots i_k}C^{I_2 i_1 \cdots i_k}
= \d^{I_1 I_2}.
\ee
Note that
\be
a_{k_1 I_1, k_2 I_2, k_3 I_3}  =\frac{1}{(\frac{1}{2}\Sigma+2)! 2^{\frac{1}{2}(\Sigma-2)}}
\frac{k_1! k_2! k_3!}{\a_1! \a_2! \a_3!} \langle C^{I_1} C^{I_2}
C^{I_3} \rangle.
\ee
where $\Sigma=k_1+k_2+k_3,\ \a_1{=}\frac{1}{2}(k_2+k_3-k_1)$ etc.
Useful identities for the scalar harmonics include
\bea
D^{a} D_{(a} D_{b)} Y^{I} &=& 4 (1 + \frac{\Lambda^{I}}{5}) D_a Y^{I};
\\
\Box_y D_{(a} D_{b)} Y^{I} &=& (10 + \Lambda^{I}) D_{(a} D_{b)} Y^I; \nn
\\
\Box_y D_a Y^{I} &=& (\Lambda^{I} + 4) D_a Y^I. \nn
\eea
Vector harmonics are normalized so that
\be \la{nor_v}
\int Y^{I_1}_a Y^{I_2 a} = \pi^3 z(k) \d^{I_1 I_2},
%\qquad z(k) = \frac{\pi^3}{2^{k-1} (k+1) (k+2)}.
\ee
where $z(k)$ is as given in \eqref{nor_sc}.
We introduce the following coordinates on $S^5$
\be
ds^2= d \q^2  + \cos^2 \q  d \W_3^2 + \sin^2 \q  d \f^2.
\ee
The differential equation (\ref{sc_h}) for the scalar harmonics
is separable. Imposing $SO(4)$ symmetry implies that the spherical
harmonics depend only on $\q$ and $\f$. The general
solution can then be expressed in terms of a hypergeometric functions,
\be
Y^{(k,m)}(\q,\f)= c_{(n,m)} y^k_m(\q) e^{i m \f}
\ee
where $c_{(n,m)}$ is a normalization constant and
the function $y^k_m(\q)$ is given by
\be
y^k_m(x)=x^{|m|}
{}_1F_2(-\frac{1}{2}(k-|m|),2+\frac{1}{2}(k+|m|),1 + |m|;x^2)
\ee
with $x=\sin \q$ (there are also a second solution
with leading behavior $x^{-|m|}$ but this solution does not
reduces to a finite polynomial for any choice of the quantum numbers).
The hypergeometric function reduces to a finite polynomial
when either the first or second argument is zero or a negative
integer. This leads to the following cases
\be
(k=2l, \quad m=2 n), \qquad (k=2l+1, \quad m=2 n +1)
\qquad n\in [-l,l], \ l \in Z^+
\ee
with
\bea
y^{2l}_{2n}(x)&=&x^{2 |n|} {}_1F_2(-l+|n|,2+l+|n|,1 + 2 |n|;x^2) \\
y^{2l+1}_{2n+1}(x)&=& x^{|2 n+1|} {}_1F_2(-l+|n|,3+l+|n|,2 + 2 |n|;x^2) \nn
\eea
The harmonics that are also $SO(2)$ symmetric are given by
\be \label{m0}
Y^{(2l,0)}(\q,\f) = \frac{ (-)^{l} }{2^{l} \sqrt{2l+1}}
    \left( \sum_{m=0}^{l} (-)^m
\left (
\begin{array}{c}
l \\
m
\end{array}
\right )
\left (
\begin{array}{c}
l + m + 1 \\
l + 1
\end{array}
\right )
(\sin\q)^{2m} \right ).
\ee
The lowest harmonics are therefore
\bea \label{y_expl}
Y^{(2,0)} &=& \frac{1}{2 \sqrt{3}}(3 \sin^2 \q -1), \\
Y^{(4,0)} &=& \frac{1}{4 \sqrt{5}} (10 \sin^4 \q - 8 \sin^2 \q +1), \nn
%\\ Y^{(6,0)} &=& \frac{1}{8 \sqrt{7}}
%(35 \cos^5 \q - 45 \cos^4 \q + 15 \cos^2 \q -1) \nn
\eea
We will also need the following normalized charged scalar harmonics
\bea
Y^{(k,\pm k)} &=& \frac{1}{2^{k/2}} \sin^k \q e^{\pm i k \f}; \la{max}
\\
Y^{(3, \pm 1)} &=& \frac{\sqrt{3}}{4} \sin \q (2 \sin^2 \q - 1) e^{
  \pm i \f}; \\
Y^{(4, \pm 2)} &=& \frac{1}{2 \sqrt{10}} \sin^2 \q (5 \sin^2 \q - 3)
e^{\pm 2 i \f};
%Y^{(5, \pm 3)} &=& \frac{\sqrt{5}}{4 \sqrt{6}} \sin^3 \q (3 \sin^2 \q -
%2) e^{\pm 3 i \f}; \\
%Y^{(5, \pm 1)} &=& \frac{1}{4} \sin \q (5 \sin^4 \q - 5 \sin^2 \q + 1)
%e^{\pm i \f}.
\eea
Note that the triple overlap between charged and neutral harmonics is
given by
\be \la{tri-3}
%\int Y^{(k,k)} Y^{(k,-k)} Y^{(2p,0)} &=& \frac{\pi^3}{(k + p +2)! 2^{k + p -1}}
%\frac{ (k!)^2 (2p!)}{ (k-p)! (p!)^2} \langle C^{k,-k} C^{2p,0}
%C^{k,k} \rangle; \nn \\
\langle C^{(k,-k)} C^{(k,k)}
C^{(2p,0)} \rangle = \frac{1}{2^{p-1} \sqrt{2 p +1}},
\ee
where $C^{(p,q)}$ denotes the symmetric tensor corresponding to the
degree $p$, $SO(2)$ charge $q$ spherical harmonic.
The relevant vector harmonics are those with only components along
$\phi$:
\bea
Y^{1} &=& \frac{1}{\sqrt{2}} \sin^2 \q d \phi; \\
Y^{3} &=& \frac{\sqrt{3}}{2} \sin^2 \q (2 \sin^2 \q - 1) d \phi.
\eea
In extracting the $\phi^{4m}_{(s)}$ perturbations the following are useful:
\bea
D_{(\q} D_{\q)} Y^{4,\pm 2} &=&  \frac{1}{2 \sqrt{10}}
e^{\pm 2 i \f} \left ( -48 \sin^4 \q + \frac{264}{5} \sin^2 \q - 6
\right );
\la{dy1} \\
D_{(\phi} D_{\phi)} Y^{4,\pm 2} &=&  \frac{1}{2 \sqrt{10}}
e^{\pm 2 i \f} \sin^2 \q \left (12 \sin^4 \q - \frac{66}{5} \sin^2 \q
+ 6 \right ); \nn
\\
D_{(\chi} D_{\chi)} Y^{4,\pm 2} &=&  \frac{1}{2 \sqrt{10}}
e^{\pm 2 i \f} \cos^2 \q \left (12 \sin^4 \q - \frac{66}{5} \sin^2 \q
\right ); \nn
\\
D_{\q} D_{\phi} Y^{4,\pm 2} &=&  \pm i \frac{1}{ \sqrt{10}}
e^{\pm 2 i \f} (15 \sin^3 \q \cos \q - 3 \sin \q \cos \q), \nn
\eea
and
\bea
D_{(\q} D_{\q)} Y^{2,\pm 2} &=&  
e^{\pm 2 i \f} \left ( - \frac{4}{5} \sin^2 \q + 1 \right ); \la{dy2} \\
D_{(\phi} D_{\phi)} Y^{2,\pm 2} &=&  
e^{\pm 2 i \f} \sin^2 \q \left (\frac{1}{5} \sin^2 \q - 1 \right ); \nn
\\
D_{(\chi} D_{\chi)} Y^{2,\pm 2} &=& 
e^{\pm 2 i \f} \cos^2 \q \left (\frac{1}{5} \sin^2 \q \right ); \nn
\\
D_{\q} D_{\phi} Y^{2,\pm 2} &=&  \pm i 
e^{\pm 2 i \f} \sin \q \cos \q, \nn
\eea
along with the corresponding expression for the neutral harmonics:
\bea
D_{(\q} D_{\q)} Y^{4 0} &=& \frac{1}{4 \sqrt{5}} \left (-96 \sin^4 \q +
\frac{504}{5} \sin^2 \q - \frac{48}{5} \right ); \\
D_{(\phi} D_{\phi)} Y^{4 0} &=& \frac{1}{4 \sqrt{5}} \left (24\sin^4 \q +
\frac{24}{5} \sin^2 \q - \frac{48}{5} \right ); \nn \\
D_{(\chi} D_{\chi)} Y^{4 0} &=&  \frac{1}{4 \sqrt{5}} \left (24\sin^4 \q -
\frac{176}{5} \sin^2 \q + \frac{32}{5} \right ). \nn
\eea

\section{Scalar chiral primaries}

The single trace scalar chiral primary operators of dimension $k$
are defined as
\be
{\cal O}_{S^{kI}} = \frac{{\cal N}_k}{N^{k/2} \sqrt{k}}
C^{I}_{i_1 \cdots i_k} {\rm Tr} (X^{m_{i_1}}
\cdots X^{m_{i_k}})
\ee
where the properties of the degree $k$
symmetric traceless tensors $C^{I}_{I_1 \cdots i_k}$ are given in
appendix \ref{sph}. The operators are normalized such that
\be
\langle {\cal O}_{S^{k_1 I_1}} (x) {\cal O}_{S^{k_2 I_2}} (y) \rangle
= {\cal N}^2_{k_1}
\d^{k_1 k_2} \frac{\d^{I_1 I_2}}{|x - y|^{2k_1}},
\ee
where the scalar fields $X^{m}$ are normalized such that
\be \la{norm-ee}
\langle X^{m}_{a} (x) X^{n}_b (y) \rangle = \frac{\d_{a b} \d^{mn}}{ |
  x - y |^2},
\ee
where $(a,b)$ are color indices.
The appropriate normalization of the dimension $k$
chiral primaries to match with supergravity is
\be \la{no2}
{\cal N}^2_{k} = \frac{N^2}{\pi^4} (k -1) (k -2)^2,
\ee
for $k \neq 2$ with ${\cal N}^2_2 = N^2/\pi^4$.

The planar three point function for such scalar chiral primaries is given by
\be
\langle {\cal O}_{S^{k_1 I_1}} (x) {\cal O}_{S^{k_2 I_2}} (y) {\cal
  O}_{S^{k_3 I_3}} (z)
\rangle = \frac{ {\cal N}_{I_1} {\cal N}_{I_2} {\cal N}_{I_3}}{N} \frac{\sqrt{k_1 k_2 k_3}
  \langle C^{I_1} C^{I_2} C^{I_3} \rangle} {|x - y|^{2 \a_3} |y - z|^{2
  \a_1} |x - z|^{2 \a_2} }.
\ee
Here $2 \a_{3} = k_1 + k_2 - k_3$
and $(\a_1,\a_2)$ are defined analogously. The triple overlap of the
symmetric traceless tensors is denoted $\langle C^{I_1} C^{I_2}
C^{I_3} \rangle$; recall that these tensors are orthonormal \eqref{c-norm}.

Now let us consider the specific case of three point functions between
one neutral ($SO(4) \times SO(2)$ singlet) operator ${\cal O}^{2k,0}$
with dimension $2k$ and two conjugate $SO(2)$ charged operators
$ {\cal O}^{n,n}$. The
corresponding spherical harmonics are given in \eqref{m0} and
\eqref{max} respectively, with the triple overlap being given in
\eqref{tri-3}. The three point function implies that the vev of the neutral
operator in the (unit normalized) state created by ${\cal O}^{n,n}$ is
\be
\langle {\cal O}^{2k,0} \rangle = {\cal N}_{2k} \frac{n
  \sqrt{2k}}{2^{k-1} N \sqrt{2k+1}}, \la{single}
\ee
Therefore the vevs of the
neutral operators in these states are given by
\be
\langle {\cal O}^{2,0} \rangle_n = \frac{\sqrt{2} n}{\pi^2
  \sqrt{3}}; \qquad
\langle
{\cal O}^{2k,0} \rangle_n = \frac{ n }{\pi^2} \frac{
  (k-1)}{2^{k-2}} \sqrt{ \frac{2k (2k-1)}{2k+1}}.
\ee
We will make use of several other three point functions, involving two
maximally charged operators:
\bea
\langle {\cal O}^{3,-3} {\cal O}^{3,1} {\cal O}^{2,2} \rangle &=&
\frac{{\cal N}_3}{N} 3 \sqrt{2} \langle C^{(3,-3)} C^{(3,1)} C^{(2,2)}
\rangle = \sqrt{3} \frac{{\cal N}_3}{N} ; \la{3pf-res} \\
\langle {\cal O}^{4,-4} {\cal O}^{4,2} {\cal O}^{2,2} \rangle
&=& \frac{{\cal N}_4}{N} 4 \sqrt{2} \langle C^{(4,-4)} C^{(4,2)} C^{(2,2)}
  \rangle = 4 \frac{{\cal N}_4}{\sqrt{5} N}, \nn
\eea
where $C^{(p,q)}$ denotes the symmetric tensor corresponding to the
degree $p$, $SO(2)$ charge $q$ spherical harmonic.

\section{Large $N$ behavior of three point functions} \la{new}

To compute vevs of single trace chiral primary operators
in generic half BPS states we use the corresponding three point
functions. To determine the dominant effects in the large $N$ limit we
thus need to know the $N$ dependence of correlators of the form
\be
C_{\s_n^I} C_{\s_m^J} \langle ({\rm Tr} (\s_n^I Z))^{\ast} {\cal O}^{\cal A}
{\rm Tr} (\s_m^J Z) \rangle
\ee
where ${\cal O}^{\cal A}$ is a single trace chiral primary,
$\s_n^I$ denotes a conjugacy class of $S_n$ and the
normalization factors $C_{\s_n^I}$ are such that the operators
are orthonormal in the large $N$ limit:
\be
C_{\s_n^I} C_{\s_m^J}
\langle ({\rm Tr} (\s_n^I Z))^{\ast}  {\rm Tr} (\s_m^J Z) \rangle =
\d_{nm} \d^{IJ} + {\cal O}(1/N).
\ee
Using the propagators given in \eqref{norm-ee} one finds that
$C_{\s_n^I}^2 \sim 1/N^n$; note that throughout this section we will suppress
factors of order one. It is convenient to introduce the notation
\be
{\cal O}^{\s[m]}_{n} = \frac{1}{N^{n/2}} \prod_{i} {\rm Tr}(Z^{n_i}); \qquad
\sum_{i} n_i = n; \qquad \sum_{i} 1 = m,
\ee
for an operator of dimension $n$ involving $m$ traces with a
permutation labeled by $\s[m]$. As discussed in \cite{D'Hoker:1999ea} there are two
distinct cases of correlators to consider, the extremal correlators in
which the dimension of the conjugate operator is equal to the sum of
the dimensions of the other operators and non-extremal correlators.

Let us consider first non-extremal correlators, focusing on the
case where $ {\cal O}^{\cal A}$ is an $SO(2)$ neutral operator, namely
it is an operator ${\cal O}^{2p,0}$ of dimension $2p$ such that
\be
{\cal O}^{2p,0} = \frac{1}{N^p} {\rm Tr} ( \bar{Z}^p Z^p + \cdots),
\ee
where the ellipses denote cyclic permutations. Now charge conservation
implies that the correlator
\be
\langle ({\cal O}^{\s[m_1]}_{n_1})^{\dagger} {\cal O}^{2p,0}  ({\cal
  O}^{\s[m_2]}_{n_2}) \rangle
\ee
is only non-zero when $n_1 = n_2$. The $N$ dependence varies according
to the specific choices of $(\s[m_1],\s[m_2])$. As discussed in
\cite{D'Hoker:1999ea}, for a generic choice
the correlator will have the same $N$ scaling as the related $(m_1 +
m_2 + 1)$-point correlator of single trace operators, namely as
$1/N^{m_1 + m_2 -1}$; thus for single trace operators the scaling is
$1/N$. Recall that an $n$-point correlator of single trace
operators behaves as
\be
\langle {\cal O}^{k_1} {\cal O}^{k_2} \cdots {\cal O}^{k_n} \rangle
\sim \frac{1}{N^{n-2}}, \qquad n \ge 2.
\ee
However, for specific choices of  $(\s[m_1],\s[m_2])$ the $N$ scaling
can be enhanced, because there are disconnected components to the
diagrams. In particular, large $N$ counting gives
\be \la{pov}
\langle ({\cal O}^{\s[m]}_{n})^{\dagger} {\cal O}^{2p,0}  ({\cal
  O}^{\s[m]}_{n}) \rangle \sim \frac{1}{N},
\ee
for any $m$, whilst for $\s[m_1] \neq \s[m_2]$ one always finds a subleading
$N$ dependence, with the 3-point function being at most of order
\be
\langle ({\cal O}^{\s[m_1]}_{n})^{\dagger} {\cal O}^{2p,0}  ({\cal
  O}^{\s[m_2]}_{n}) \rangle \sim \frac{1}{N^2}.
\ee
(This result for $m_1 = 1$ was given in \cite{D'Hoker:1999ea}.)
Thus vevs of neutral operators are thus dominated by diagonal
three point functions of the type \eqref{pov}. For the vevs of non-maximally
charged operators, the relevant correlators are also non-extremal;
the leading terms scale as $1/N$ and arise from single trace
correlators and specific multi-trace correlators. We will not however
need detailed results for the latter.

Now let us turn to the extremal correlators in which
$ {\cal O}^{\cal A}$ is a maximally charged single trace
operator. Again the correlator involving single trace operators
behaves as $1/N$, but in this case correlators involving multi trace
operators can dominate, since they can grow as $1$. In particular,
\be
\langle ({\cal O}^{\s[m + 1]}_{n+k})^{\dagger} {\cal O}^{k,k}  ({\cal
  O}^{\s[m + 1]}_{n}) \rangle \sim 1, \qquad {\cal O}^{\s[m + 1 ]}_{n+k}
=  {\cal O}^{k,k}  ({\cal O}^{\s[m]}_{n}). \la{ex-22}
\ee
Note that analogous results are obtained in the Schur polynomial
basis; see \cite{Corley:2001zk} for related discussions.

\section{Killing spinors for LLM solutions}

We discuss in this appendix the computation of the Killing spinors
of the LLM solutions. This computation was carried out in
appendix A of \cite{Lin:2004nb} but only half of Killing
spinors were correctly identified, even though the
projection operators were given correctly, and
furthermore the spacetime dependence is not given correctly.
These corrections do not affect the final answer for
the supergravity solution (although some intermediate steps in the
derivation are affected). They may have a real effect however
in similar computations for less supersymmetric solutions.
Furthermore the correct Killing spinors may be needed
for other purposes, for example for analyzing supersymmetric probe
branes in this background.

We use the notation of \cite{Lin:2004nb} and choose the same basis of gamma matrices
\be
\G_\m = \g_\m \otimes 1 \otimes 1 \otimes 1, \quad
\G_a = \g_5 \otimes \s_a \otimes 1 \otimes \hat{\s}_1,   \quad
\G_{\tilde{a}} = \g_5 \otimes 1 \otimes \tilde{\s}_a \otimes \hat{\s}_2,
\ee
where $\s_a, \tilde{\s}_a, \hat{\s}_a$ are the Pauli matrices.

The ten dimensional spinor is decomposed as
\be \label{spinor}
\eta = \e_{a} \otimes \chi_a \otimes \tilde{\chi}_a
\ee
where $\chi_a, \tilde{\chi}_a$ are geometric Killing spinors
of $S^3$, i.e. they obey
\be
\nabla_c \chi_a = a \frac{i}{2} \g_c \chi_a, \qquad a=\pm 1,
\ee
and a similar equation for $\tilde{\chi}_a$, where $\nabla_c$ is the
standard connection on a unit 3-sphere. We normalize these
spinors as $\chi_a^\dagger \chi_a= \tilde{\chi}_a^\dagger \tilde{\chi}_a=1$.
The fact that the
spinors are correlated as in (\ref{spinor}) follows from
the analysis in \cite{Lin:2004nb}.

The Killing spinor equation then reduces to \cite{Lin:2004nb}
\bea
&&(i a e^{-\half (H+G)} \g_5 \hat{\s}_1 + \half \g^\m \pa_\m (H+G)) \e
+ 2 M \e =0, \la{first} \\
&&(i a e^{-\half (H-G)} \g_5 \hat{\s}_2 + \half \g^\m \pa_\m (H-G)) \e
- 2 M \e =0, \la{second} \\
&&\nabla_\m \e + M \g_\m \e = 0 \la{third}
\eea
where
\be
M= - \frac{1}{4} e^{-\frac{3}{2} (H+G)} \g^{\m \n} F_{\m \n} \g^5 \hat{\s}_1
\ee
Processing these equations one finds that the spinor $\e_a$ should
satisfy the following equations \cite{Lin:2004nb}
\be \label{proj}
P_a^- \e_a = R_a^+ \e_a=0,
\ee
where we introduce the commuting projection operators
\be
P_a^\pm = \half
\left(1 \pm (i e^{-G} \g_5 + a \sqrt{1 + e^{-2 G}} \G_3 \hat{\s}_1)\right),
\qquad R_a^\pm = \half (1 \pm i a \G_1 \G_2)
\ee
satisfying
\be
(P_a^\pm)^2 = P_a^\pm, \quad P_a^+ P_a^- =0, \quad
(R_a^\pm)^2 = R_a^\pm,  \quad R_a^+ R_a^- =0
\quad [P_a^\pm, R_a^\pm]=0
\ee
Each of this projection cuts the number of spinors by 1/2, so we have
a total of 8 Killing spinors for $a=+1$ and 8 Killing spinors for
$a=-1$. The most general solution of (\ref{proj}) is
\be \label{kill}
\e_a = R_a^- P_a^+ \tilde{\e}_a
\ee
where $\tilde{\e}_a$ are (at this point) unconstrained spinors.

In \cite{Lin:2004nb} the following solution of (\ref{proj})
was given,
\be \label{LLMkil}
\e=e^{i \d \g^5 \G^3 \hat{\s}^1} \e_1, \qquad
\G^3 \hat{\s}^1 \e_1 = a \e_1, \qquad \sinh 2 \d = a e^{-G}
\ee
These are in fact only half of the Killing spinors in (\ref{kill}).
To see this introduce a new projector,
\be
S_a^{\pm} = \half (1 \pm a \G^3 \hat{\s}^1), \qquad
(S_a^{\pm})^2 = S_a^{\pm}, \quad S_a^+ S_a^-=0,
\ee
and decompose $\tilde{\e}_a$ as
\be
\tilde{\e}_a = \tilde{\e}_a^+ +\tilde{\e}_a^-, \qquad
S_a^\pm  \tilde{\e}_a^\pm = \tilde{\e}_a^\pm, \quad
S_a^\pm  \tilde{\e}_a^\mp =0
\ee
A short computation yields,
\be
P_a^+ \tilde{\e}_a^+
= \cosh \d  e^{i \d \g^5 \G^3 \hat{\s}^1} \tilde{\e}_a^+
\ee
which is the spinor in (\ref{LLMkil}).
Upon multiplication by $R_a^-$ one obtains half of the Killing spinors
in (\ref{kill}), namely we miss the ones based on $\td{\e}_a^-$.

To specify the Killing spinor we need to specify $\tilde{\e}_a$.
To this end we consider the fermion bilinear
$f_2 = i \bar{\e} \hat{\s}_2 \e$. It was shown in
\cite{Lin:2004nb} that $f_2$ equals
\be \label{f2}
f_2 = e^{\half (H+G)}
\ee
Inserting the spinors in (\ref{kill}) and defining
\be
\tilde{\e}_a^\pm = c^{\pm} e_a^\pm
\ee
we find that (\ref{f2}) implies
\be
c^\pm = \frac{e^{\frac{1}{4}(H+G)}}{\sqrt{\sqrt{1 + e^{-2 G}} \pm 1 }}
\ee
and
\be \la{con3}
i \bar{e}_a \hat{\s}_2 R_a^- e_a = 2
\ee
where
\be
e_a=e_a^+ + i \g_5 e_a^-,
\ee
The Killing spinor becomes
\bea
\e_a &=& \frac{1}{\sqrt{2}} e^{\frac{1}{4}(H+G)} R_a^- \left(
(\cosh \d + i a \g_5 \sinh \d) e_a^+ + (-\sinh \d + i a \g_5 \cosh \d) e_a^-
\right) \nonumber \\
&=& \frac{1}{\sqrt{2}} e^{\frac{1}{4}(H+G)} R_a^-
\left(e^{i \d \g^5 \G^3 \hat{\s}^1} e_a^+
+i a \g_5  e^{-i \d \g^5 \G^3 \hat{\s}^1} e_a^- \right)
\eea
{}From these spinors one can construct appropriate fermion bilinears and
determine the supergravity solution as in \cite{Lin:2004nb}; the
supergravity solution is exactly as given in \cite{Lin:2004nb}. Note
that the functions $(G,H)$ are such that
\be
e^{H} = y; \qquad z = \half \tanh G,
\ee
where $z$ is the defining function of the supergravity solution.

There is however a further issue in constructing the actual Killing spinors:
the spinors by construction satisfy \eqref{first} and \eqref{second}
since it is these equations which were processed. One still needs to
check explicitly that all components of \eqref{third} are
satisfied. Now the spinors as given in \cite{Lin:2004nb} do not depend
at all on the time coordinate $t$. This is however inconsistent with
the $t$ component of \eqref{third}; one can show that
\be
\nabla_{t} \e_a + M \g_{t} \e_a \neq 0
\ee
for constant $(e_a^+,e_a^-)$. Another way to see that the Killing
spinor solution is not quite correct is by considering the limiting
case of $AdS_5 \times S^5$. The known explicit
expressions for the Killing spinors of $AdS_5 \times S^5$ do depend explicitly
on the time coordinate; this remains true for the specific
combinations of spinors which form the set of sixteen discussed above.

The resolution of this issue is straightforward: $(e_a^+,e_a^-)$ are not
constant,
but must contain suitable $t$ (and indeed also $\phi$) dependent phase
factors so that \eqref{third} is satisfied. These phase factors drop
out of \eqref{con3} and all other fermion bilinears used to construct
the supergravity solution.

\end{document}